\documentclass{aa}
\usepackage[hidelinks,colorlinks=true,linkcolor=blue,citecolor=blue,urlcolor=black]{hyperref}
\usepackage{natbib}
\usepackage{bigints}
\usepackage{frcursive}
\begin{document}

\title{Understanding entropy in massive halos: \\
	The role of baryon decoupling\\
	}
\author{S. Molendi\inst{1}, M. Balboni\inst{1,2,3}, I. Bartalucci\inst{1}, S. De Grandi\inst{4}, M. Gaspari\inst{5}, F. Gastaldello\inst{1}, S. Ghizzardi\inst{1}, 
  L. Lovisari\inst{1}, G. Riva\inst{1,6}, M. Rossetti\inst{1}  and P. Tozzi\inst{7}
	} 
\offprints{S. Molendi \email{silvano.molendi@inaf.it}}

\institute{
INAF - IASF Milano, via A. Corti 12 I-20133 Milano, Italy 
\and
Dipartimento di Scienza e Alta Tecnologia, Universit\`a dell’Insubria, Via Valleggio 11, I-22100 Como, Italy
\and
DIFA - Universit\`a di Bologna, via Gobetti 93/2, I-40129 Bologna, Italy
\and
INAF - Osservatorio Astronomico di Brera, via E. Bianchi 46, I-23807 Merate (LC), Italy
\and
Department of Physics, Informatics and Mathematics, University of Modena and Reggio Emilia, I-41125 Modena, Italy
\and
Dipartimento di Fisica, Universit\`a degli Studi di Milano, Via G. Celoria 16, I-20133 Milano, Italy
\and
INAF – Osservatorio Astrofisico di Arcetri, Largo E. Fermi, I-50122 Firenze, Italy
}
\date{\today}
\abstract
{}
{The goal of the work presented in this paper is to use observed entropy profiles to infer constraints on the accretion process in massive halos. 
}
{We compare entropy profiles from various observational samples with those generated by an updated version of the semi-analytical models developed in the early 2000s, modified to reflect recent advancements in our understanding of large-structure formation.} 
{Our model reproduces the growing departure from self-similarity observed in data as we move inward in individual profiles and down in mass across different profiles.  These deviations stem from a phase of extremely low gas content centered around $10^{13}$M$_\odot$. According to our model, halos at this mass scale are missing between 50\% and 90\% of their baryons, corresponding to a gas fraction ranging between 2\% and 8\%.
 }
{Baryon decoupling, the mechanism at the heart of our model, proves effective in explaining much of the behavior we sought to understand.	
}

\titlerunning{Entropy in Massive Halos}

\authorrunning{Molendi et al.}

\maketitle
\section{Introduction}\label{sec:intro}

It has long been understood that the high entropy of gas in massive halos ($M_{\rm h} \gtrsim 10^{13}$M$_\odot$) is a direct consequence of the virialization process.
In simple terms, the gas falls onto the halo and, around the virial radius, converts its kinetic energy into thermal energy by impacting on other gas, that is at rest in the potential well of the system. Since the sound speed in the infalling gas is much smaller than the infall velocity, the impact generates a shock that raises the entropy of the gas. In stark contrast, once inside the halo, the processes operating on the gas are largely adiabatic, both conduction and mixing are highly suppressed \citep[see][and refs. therein]{Molendi:2023}, leaving the entropy unaltered.  Thus, by measuring the entropy of the gas in massive halos, we acquire valuable information on shock heating  at the virial radius and, ultimately, on the accretion process. 

The goal of the work presented in this paper is to use observed entropy profiles to infer constraints on the accretion process. 
In pursuing this objective, we adopt  a semi-analytical approach. As discussed by a few authors two decades ago \citep{Tozzi_Norman:2001,Voit:2003,Voit:2005}, the process by which entropy is generated is a simple one.
The bulk of the entropy is generated at the virial shock. 
Early semi-analytical work \citep{Tozzi_Norman:2001,Voit:2003,Voit:2005} met with some success, the predicted slopes broadly aligned with observations of massive clusters. 
However, with the advent of more precise measurements \citep[see][]{Pratt_entropy:2006,Pratt:2010}, it became clear that even beyond the core region, where processes such as radiative cooling and heating from the Active Galactic Nucleus (AGN) heavily influence gas thermodynamics, these simple models fell short of explaining critical features. Specifically, lower-mass systems exhibited entropy profiles that, when rescaled self-similarly, were offset relative to those of more massive systems. This discrepancy indicates that the self-similar scaling inherent in the models fails to capture some essential physical processes.
To address these limitations, semi-analytical studies were quickly supplemented \citep{Borgani:2005,Voit_entropy:2005}, and later largely supplanted, by simulation-based approaches \citep[see][and references therein]{Altamura:2023}. While simulations have successfully overcome several obvious shortcomings of earlier models, they introduce their own set of challenges, such as computational complexity, resolution limitations, and uncertainties in modeling sub-grid physics like feedback and cooling processes.
Here, we aim to determine whether, and to what extent, an updated version of the semi-analytical models from the early 2000s, incorporating significant advancements in our understanding of large-structure formation, can account for the extensive observational evidence gathered over the past two decades.

As mentioned earlier, reproducing observed entropy profiles is a step toward a broader goal: gaining deeper insights into the accretion process. This focus distinguishes our approach from much of the simulation work in recent years \citep[e.g.,][]{Barnes:2017,Altamura:2023,Braspenning:2024}. Simulations typically employ detailed descriptions of numerous processes governing the formation and evolution of large-scale structures in the Universe, aiming to predict a wide array of baryonic properties across different states and scales, with entropy profiles being just one among many.
The two approaches are complementary: findings based on simulations offer a framework for conducting targeted studies of specific aspects of structure formation, while semi-analytical efforts, like the one presented in this paper, provide  insights into particular processes that can subsequently be incorporated into simulations.

The paper is organized as follows. Sect. \ref{sec:engen} reviews the process of entropy generation as outlined in the semi-analytical studies of the early 2000s. In Sect. \ref{sec:radprof} we derive radial entropy profiles by combining the entropy-halo mass relation obtained in Sect. \ref{sec:engen} with the radial mass distribution. We also show that these  profiles align reasonably well with observations at large radii, though they fall significantly below observed values at smaller radii. 
In Sect. \ref{sec:pre}, we illustrate that incorporating pre-heating into the entropy generation process  reduces the discrepancy between the model and observed data, but does not resolve it.
In Sect. \ref{sec:decoup}, we modify the self-similar accretion model by decoupling the accretion of baryons from that of dark matter. We demonstrate that the ensuing break in self-similarity can be recovered in part by rescaling entropy by the gas fraction. In Sect. \ref{sec:obs}, we compare entropy and density profiles derived assuming baryon decoupling with observed profiles from three different samples of galaxy clusters and one sample of X-ray bright galaxy groups, finding overall good  agreement. We also argue that the difference between observed profiles of other, less X-ray bright groups and those predicted from our model can be resolved by invoking heating from the central AGN. In Sect. \ref{sec:theo}, we compare predictions from our model with those from other theoretical work. In Sects. \ref{sec:discussion} and \ref{sec:summary} we respectively discuss and summarize our findings. 

Throughout the paper we assume a $\Lambda$ cold dark matter cosmology with $H_0 = $70 km s$^{-1}$ Mpc$^{-1}$, $\Omega_M = 0.3$ and $\Omega_\Lambda = 0.7$
Furthermore, we define $R_\Delta$ as the radius inside which the cluster mass density is $\Delta$
times the critical density of the Universe, $M_\Delta$ is the halo mass
within $R_\Delta$.

\section{Entropy generation}\label{sec:engen}

We start by reviewing the spherical accretion model discussed by several authors \citep[e.g.][]{Tozzi_Norman:2001,Voit:2005} two decades ago. 
In simple terms, the accretion of gas in massive halos can be divided into three phases: a first phase where gas falls onto the halo, a second where it is shock heated and a third where it settles into near hydrostatic equilibrium within the halo.  An important point is that the processes occurring in the first and third phase are assumed to be adiabatic and therefore leave the entropy of the gas unaltered. The variation in entropy occurs only in the second phase, this crucial aspect is nicely represented in Fig. 2 of \cite{Tozzi_Norman:2001}.
Furthermore, the whole process is assumed to be radially symmetric and can be described in terms of smooth accretion of spherical shells that are shocked at a single radius, namely the virial radius. 
  
\subsection{Baseline model}

In the simplest case, the gas starts off cold and acquires all its entropy by going through a shock at the virial radius.  Under these conditions the equations describing the shock can be reduced to a set of four, \citep[see Eqs. 67 to 70 in][]{Voit:2005}. We re-propose them here.

Mass flow through the accretion radius:
\begin{equation}
	\dot{M}_{\rm g} = 4 \pi R_{\rm v}^2 \rho_{\rm g,u}  v_{\rm v} ,
	\label{eq:mdot}
\end{equation}
where $\dot{M}_{\rm g}$ is the gas accretion rate, $R_{\rm v}$ is the virial radius which coincides with the shock radius, $\rho_{\rm g,u}$ is the upstream gas density and $v_{\rm v}$ is the gas velocity at the virial radius. 

Velocity of the accreted gas:
\begin{equation}
	v_{\rm v}^2 =  {GM_{\rm h} \over R_{\rm v} } ,
	\label{eq:vac}
\end{equation}
where $M_{\rm h}$ is the halo mass.
Eq. \ref{eq:vac} is derived by equating the kinetic energy at the virial radius to the variation in potential energy between the turn-around radius, where velocity is zero, and  the accretion radius, under the assumption the turn around radius is, $R_{\rm ta} = 2 R_{\rm v}$.

Energy conservation:
\begin{equation}
	k_{\rm b} T =  {1 \over 3} \mu m_{\rm p}   v_{\rm v}^2 ,
	\label{eq:kt2}
\end{equation}
where $k_{\rm b}$ is the Boltzmann constant, $\mu$ is the mean molecular mass of the gas and $m_{\rm p}$ is the proton mass. Eq. \ref{eq:kt2} describes the conversion of the upstream kinetic energy (right hand side) into downstream thermal energy (left hand side).

Gas compression:
\begin{equation}
	\rho_{\rm g,d} = 4  \rho_{\rm g,u}  ,
	\label{eq:rho2}
\end{equation}
where $\rho_{\rm g,d}$ and $\rho_{\rm g,u}$  are respectively the gas density downstream and upstream  of the shock. Eq. \ref{eq:rho2} relates the upstream to the downstream density of the gas under the assumption of high Mach number, in which case the downstream density is 4 times the upstream one.
Note that this set of equations provides an approximate description of the process valid for strong shocks. For instance,
it does not account for the compression and adiabatic heating undergone by the infalling gas before being shocked \citep[see][for details]{Tozzi_Norman:2001}. 
We will nonetheless work within the framework provided by Eqs. \ref{eq:mdot} to  \ref{eq:rho2} and evaluate the impact of the approximations at the appropriate time.

From Eqs. \ref{eq:mdot}, \ref{eq:vac} and \ref{eq:rho2} one can rewrite the downstream density  as a function of halo mass, gas accretion rate and accretion radius:
\begin{equation}
	\rho_{\rm g,d} \, = \, { 1 \over \pi G^{1/2} } \, R_{\rm v}^{-3/2} \,  M_{\rm h}^{-1/2} \dot{M}_{\rm g}  \, .
	\label{eq:rho3}
\end{equation}
Similarly, from Eqs. \ref{eq:vac} and \ref{eq:kt2} the gas temperature may be expressed as a function of halo mass and accretion radius:
\begin{equation}
	k_{\rm b} T_{\rm v}  \, = \, {1 \over 3} \mu m_{\rm p}  {GM_{\rm h} \over R_{\rm v}}\, .
	\label{eq:ktb}
\end{equation}
Making use of  Eqs. \ref{eq:rho3} and \ref{eq:ktb} one can write the entropy\footnote{
It is important to note that in studies of the intra-cluster medium (ICM), the term "entropy" refers to a quantity related to the thermodynamic specific entropy,  $s$: $s =  3/2 \, k_{\rm b} \ln K  + const$ \citep[see Eqs.9 and 10 in][]{Donahue:2022} .Additionally, specific entropy is an intensive variable.}, which, following \cite{Balogh:1999}, we define as:

\begin{equation}
	K \equiv {k_{\rm b}T \over \mu m_{\rm p} \rho_{\rm g}^{2/3}} ,
	\label{eq:entro_def}
\end{equation}
in the form:
\begin{equation}
	K_{\rm v} =  {1 \over 3} \, (\pi G^2)^{2/3} \, { M_{\rm h}^{4/3} \over \dot{M}_{\rm g}^{2/3} } .
	\label{eq:entro_sm}
\end{equation}
Assuming that the gas fraction of the shells undergoing accretion is constant and equal to the cosmic baryon value,  we may write:
\begin{equation}
	\dot{M}_{\rm g} = f_{\rm b}  \dot{M}_{\rm h} \, ,
    \label{eq:const_acc}
\end{equation}
where $ f_{\rm b}$ is the cosmic baryon fraction and $\dot{M}_{\rm h}$  is the matter accretion rate on the halo.  Thus, Eq. \ref{eq:entro_sm} can be rewritten in the form:

\begin{equation}
	K_{\rm v} =  {1 \over 3} \, (\pi G^2)^{2/3} \,  f_{\rm b}^{-2/3} \, 
	\Biggl({M_{\rm h} \over \dot{M}_{\rm h}}\Biggr)^{2/3} \, M_{\rm h}^{2/3}  .
	\label{eq:entro_sm2}
\end{equation}



\subsection{Entropy versus Mass relation}

\begin{figure}
	\hspace{-0.5cm}
	\centerline{\includegraphics[angle=0,width=9.2cm]{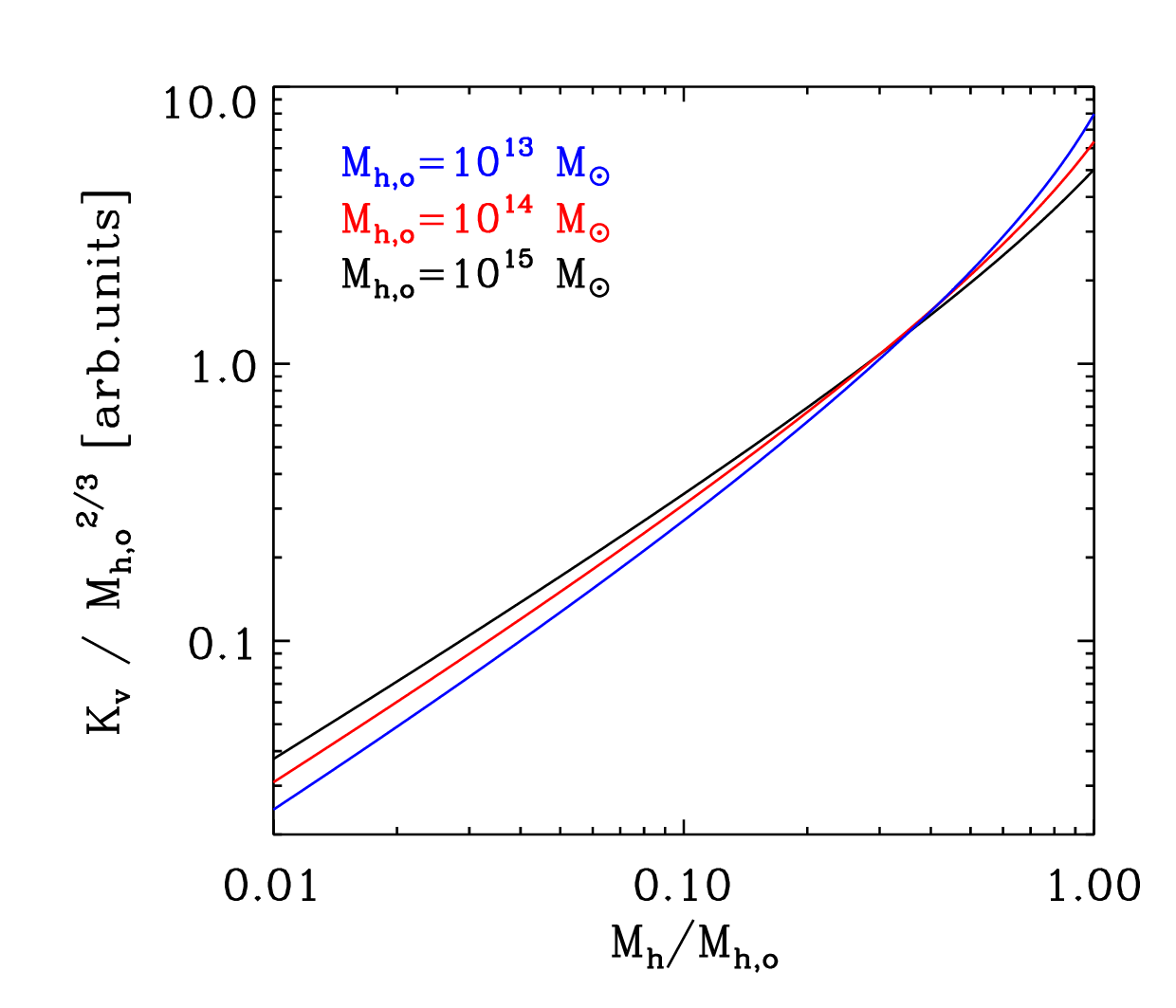}}
	\caption{Scaled entropy versus halo mass relation for three halos of $10^{13}$ M$_\odot$ , $10^{14}$ M$_\odot$ and $10^{15}$ M$_\odot$ at $z=0$ respectively.
	}
	\label{fig:kvsmh}
\end{figure}

Within the framework we have explored, the relationship between entropy and halo mass is described by Eq. \ref{eq:entro_sm2}. To estimate the halo mass, $M_{\rm h}$, and its growth rate, $\dot{M}_{\rm h}$, we make use of the analytical model described in \cite{Salcido:2020}, see their Eq. 12. and Fig. 3.
In Fig. \ref{fig:kvsmh} we show the case of three halos of $10^{13}$ M$_\odot$, $10^{14}$ M$_\odot$ and $10^{15}$ M$_\odot$ respectively, at $z=0$. The entropy has been rescaled by dividing by $M_{\rm h,o}^{2/3}$, where $M_{\rm h,o} \equiv M_{\rm h}(z=0)$. 

\begin{figure}
    \hspace{-0.5cm}
	\centerline{\includegraphics[angle=0,width=9.2cm]{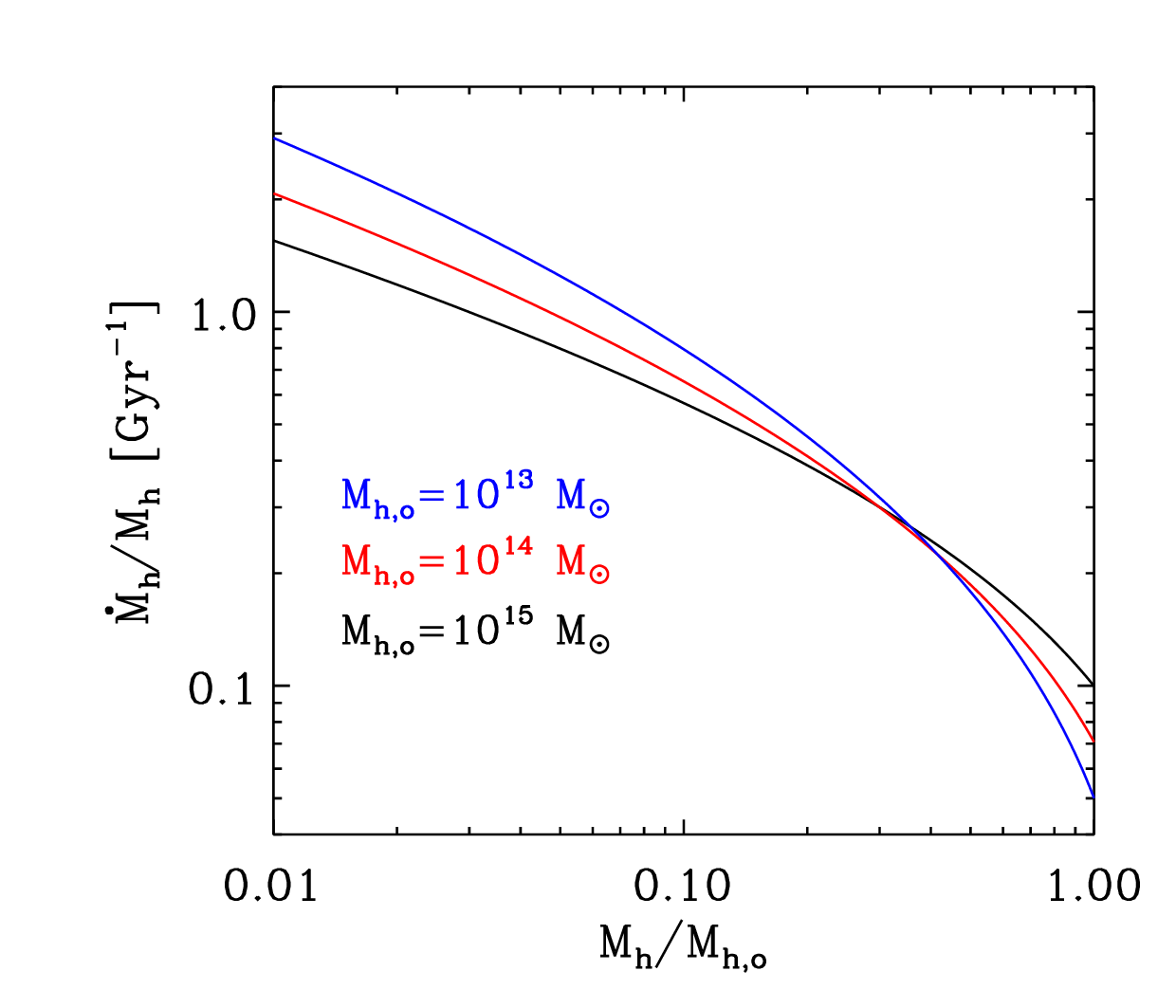}}
	\caption{Specific accretion rate versus halo mass relation for three halos of $10^{13}$ M$_\odot$ , $10^{14}$ M$_\odot$ and $10^{15}$ M$_\odot$ at $z=0$ respectively.
	}
	\label{fig:mdotvsm}
\end{figure}

There are two key points to highlight.
First: the growth of entropy with halo mass for the three systems is similar. This arises from the weak dependency of the specific accretion rate, $\dot{M}_{\rm h}/M_{\rm h}$, on $M_{\rm h,o}$, as illustrated in Fig. \ref{fig:mdotvsm}.
Second: entropy increases more rapidly with decreasing $M_{\rm h,o}$. This behavior can again be attributed to the relationship between entropy and the specific accretion rate. As shown in Fig. \ref{fig:mdotvsm}, the specific accretion rate decreases more significantly with decreasing $M_{\rm h,o}$. 

An interesting way of characterizing the entropy versus halo mass relationship is through the logarithmic slope ${\rm d}\ln K_{\rm v}/{\rm d}\ln M_{\rm h }$ \citep[see][]{Voit:2003}. It can be shown, see App. \ref{sec:app} for a derivation, that the following relation holds:
 
\begin{equation}
	 {{\rm d}\ln K_{\rm v} \over {\rm d}\ln M_{\rm h}} =  {2 \over 3} \, \left( 1 + {1 \over \alpha} - { {\rm d}\ln \alpha \over {\rm d}\ln M_{\rm h }} \right) ,
	\label{eq:slope}
\end{equation}
where  $\alpha \equiv {\rm d}\ln M_{\rm h } /{\rm d}\ln t $.

\begin{figure}
    \hspace{-0.5cm}
	\centerline{\includegraphics[angle=0,width=9.2cm]{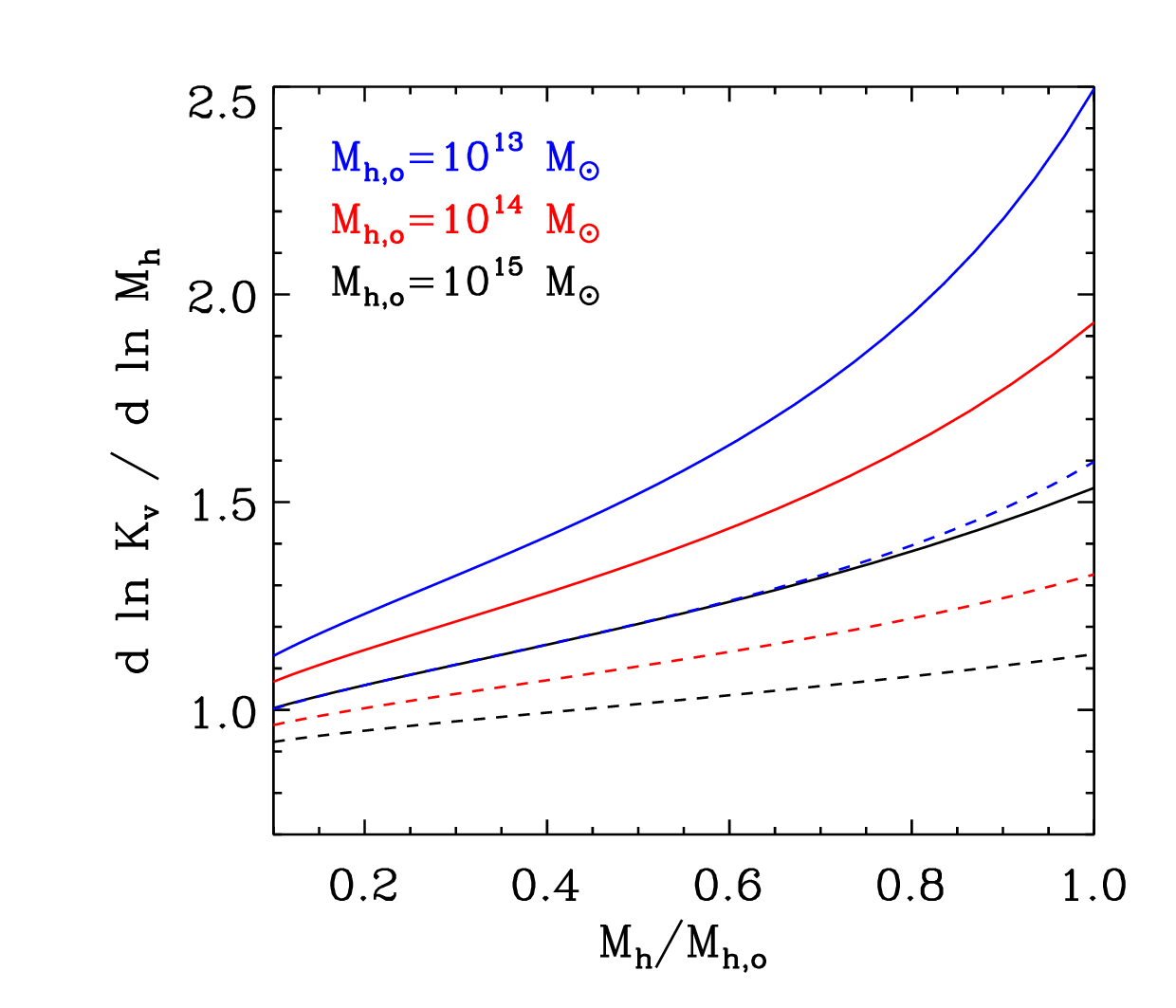}}
	\caption{Logarithmic slope, ${\rm d}\ln K_{\rm v}/{\rm d}\ln M_{\rm h}$, versus halo mass relation for three halos of $10^{13}$ M$_\odot$, $10^{14}$ M$_\odot$ and $10^{15}$ M$_\odot$ at $z=0$ respectively, as estimated from Eq. \ref{eq:slope}. Dashed lines are computed neglecting the ${\rm d}\ln \alpha / {\rm d}\ln M_{\rm h}$ term.
	}
	\label{fig:slope}
\end{figure}

The logarithmic slope ${\rm d}\ln K_{\rm v}/{\rm d}\ln M_{\rm h }$ is reported in Fig. \ref{fig:slope} for three halos of $10^{13}$ M$_\odot$, $10^{14}$ M$_\odot$ and $10^{15}$ M$_\odot$ at $z=0$ respectively.   
As pointed out in \cite{Voit:2003}, the increase in slope at large values  of $M_{\rm h}/M_{\rm h,o}$  is related to a decrease in accretion rate, see Fig. \ref{fig:mdotvsm}. 

\section{Radial entropy profiles} \label{sec:radprof}
In this section we derive radial entropy profiles from  the model presented in Sect. \ref{sec:engen}. We do this in three steps.
First we make use of Eq. \ref{eq:entro_sm2}  that relates entropy to halo mass, then we assume a radial mass distribution, finally we combine the two to derive the radial entropy profile. 

As we shall see later, in the last step we consider two distinct approaches to derive entropy profiles.
The first is a straightforward combination of the functional forms: $K(R)= K_{\rm v}(M_{\rm h}(R))$\footnote{The ${\rm v}$ subscript is present on the right hand side of the equation to highlight that this is the entropy generated at a time when the virial radius was going through the radius $R$, it is dropped on the left hand side of the equation because what we wish to emphasize in this case is that $K(R)$ is the entropy at a generic radius $R$.}, the second imposes that gas within the halo follow hydrostatic equilibrium.

\subsection{Mass versus radius relation}

Several relations have been proposed in the literature to describe the distribution of mass within a halo \citep[e.g.][and refs. therein]{Ettori:XCOP2019}, we assume one of the mostly commonly used, the NFW profile \cite{Navarro:1997}.
We write the mass within a radius R as:

\begin{equation}
	M_{\rm h}(<R) =  4 \pi \rho_o R_{\rm s}^3 \left[ \ln (1+x) - {x \over 1+x}\right] ,
	\label{eq:nfw}
\end{equation}
where  $\rho_o$ is a reference density, $x=R/R_{\rm s}$, and $R_{\rm s}$ is the scale radius.
Furthermore, we fix the shape of the profile by assuming that $c_{200}$, defined as: $c_{200} \equiv R_{200}/R_{\rm s}$,
assumes the value $c_{200} = 4$.
We point out that variation of $c_{200}$ within reasonable limits \citep[e.g.][]{Ettori:XCOP2019} do not change results presented later.  
Furthermore, for a $\Lambda$CDM Universe, with $\Omega_M = 0.3$ and $\Omega_\Lambda = 0.7$, the virial overdensity at $z=0$ is $\Delta_{\rm v}(z=0) \sim 100$ \citep[see Eq. 44 in][]{Voit:2005} and the virial radius is: $R_{\rm v} \sim 1.25\, R_{200}$. Thus, the concentration computed using the virial radius, which is defined as: $c_{\rm v} \equiv R_{\rm v}/R_{\rm s} $, turns out to be: $c_{\rm v} \sim 5$.

\subsection{Analytical derivation of the slope of the entropy profile}

\begin{figure}
    \hspace{-0.5cm}
	\centerline{\includegraphics[angle=0,width=9.2cm]{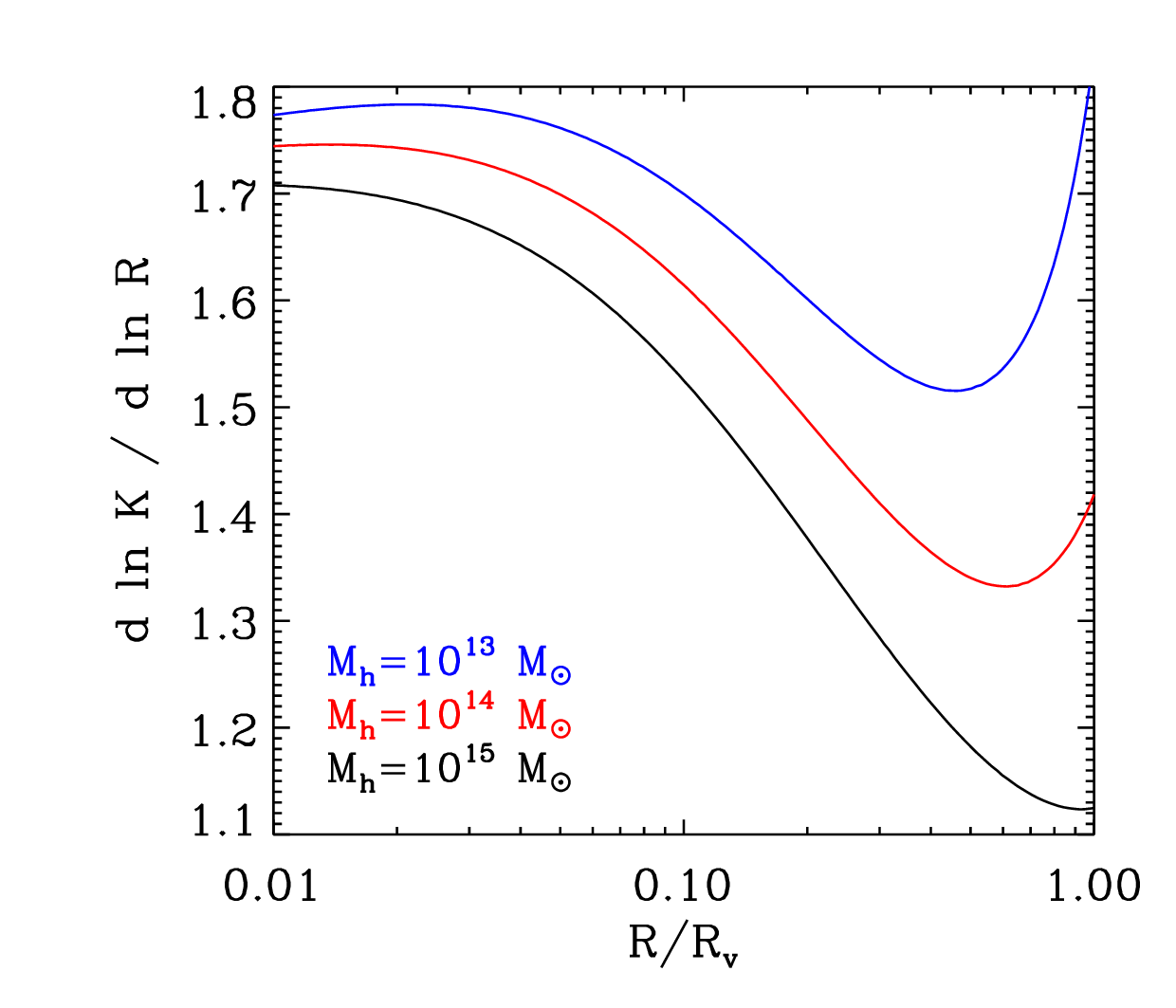}}
	\caption{Logarithmic slope of the radial entropy profile  for three halos of $10^{13}$ M$_\odot$, $10^{14}$ M$_\odot$ and $10^{15}$ M$_\odot$ at $z=0$ respectively, as estimated from Eq. \ref{eq:dlnk_dlnr}. The radial coordinate is normalized to the virial radius.
	}
	\label{fig:dlnk_dlnr}
\end{figure}

Interestingly, the logarithmic slope of the NFW mass profile can be expressed as:

\begin{equation}
		{{\rm d}\ln M_{\rm h} \over {\rm d}\ln R } \; =  \; {x^2 \over (1+x)^2} \, {1 \over   \ln (1+x) - {x \over 1+x}}    \, .
	\label{eq:dlnm_dlnr}
\end{equation}
By writing: 
\begin{equation}
	{ {\rm d}\ln K \over {\rm d}\ln R} \; = \; {{\rm d}\ln K_{\rm v} \over {\rm d}\ln M_{\rm h}} \; \times \; {{\rm d}\ln M_{\rm h} \over {\rm d}\ln R}
	\label{eq:dlnk_dlnr0}
\end{equation}
and substituting Eqs. \ref{eq:slope} and \ref{eq:dlnm_dlnr} into Eq. \ref{eq:dlnk_dlnr0} we get:

\begin{multline}
	 { {\rm d}\ln K \over {\rm d}\ln R} \; = \; {2 \over 3} \, \left( 1 + {1 \over \alpha} - { {\rm d}\ln \alpha \over {\rm d}\ln M_{\rm h}} \right) \times \\ 
	{x^2 \over (1+x)^2} \, {1 \over   \ln (1+x) - {x \over 1+x}}  \, .
	\label{eq:dlnk_dlnr}
\end{multline}
This expression allows us to perform a direct computation of the logarithmic slope of the radial entropy profile, which we report in Fig. \ref{fig:dlnk_dlnr}, for three halos of $10^{13}$ M$_\odot$, $10^{14}$ M$_\odot$ and $10^{15}$ M$_\odot$ at $z=0$ respectively. 

\begin{figure*}
	\centering
    \hspace{-0.5cm}
   	\includegraphics[width = 0.35\textwidth]{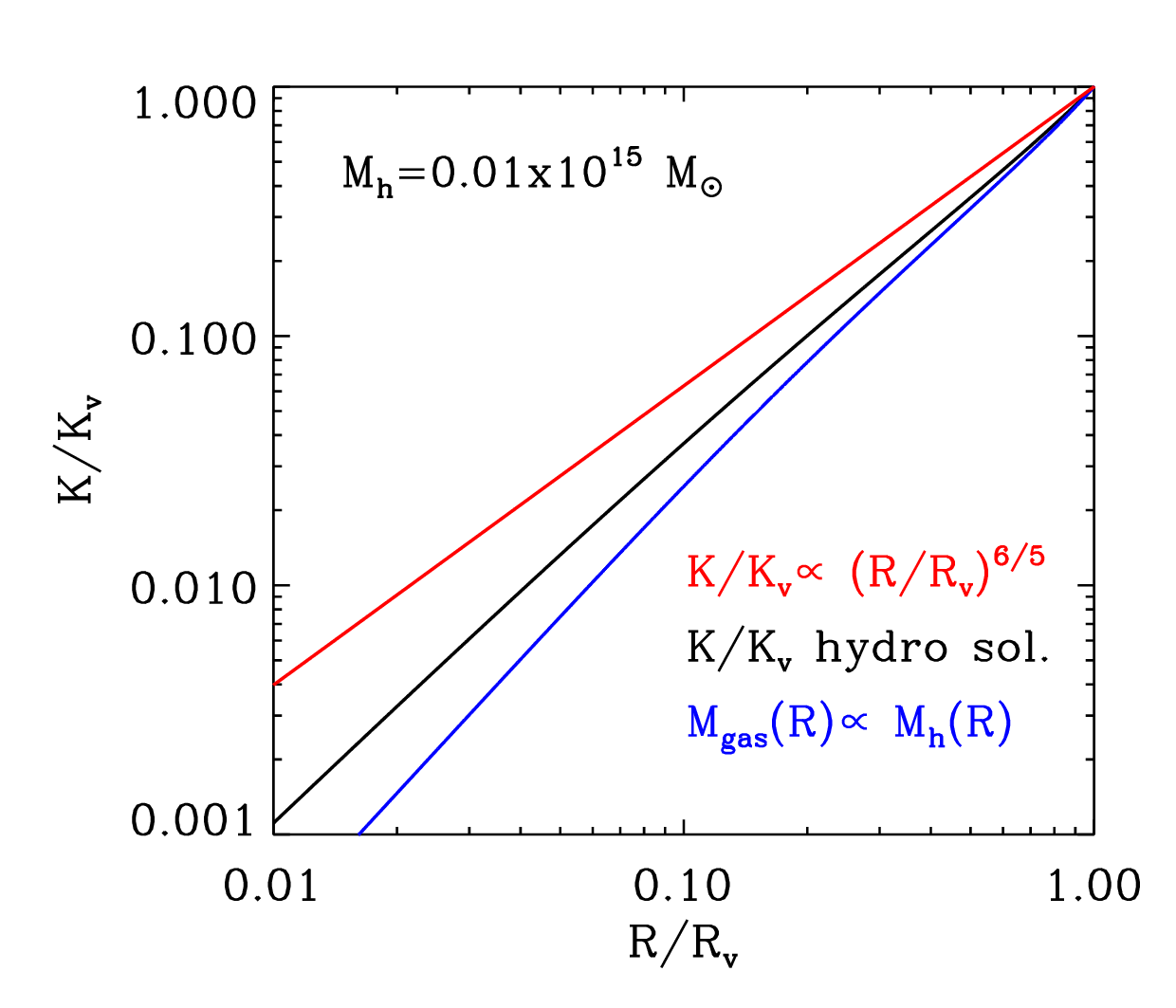}
    \hspace{-0.5cm}
	\includegraphics[width = 0.35\textwidth]{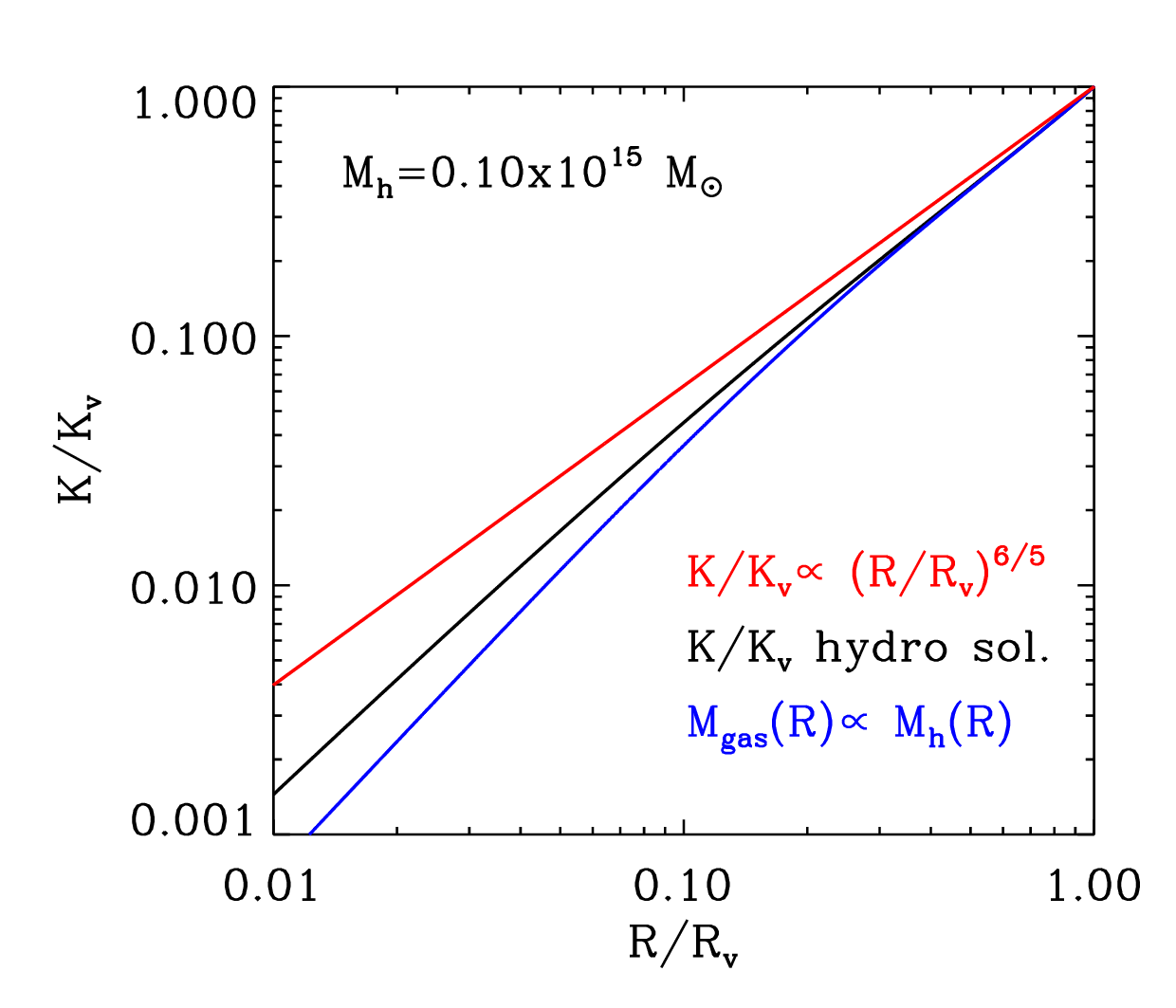}   
    \hspace{-0.5cm}
	\includegraphics[width = 0.35\textwidth]{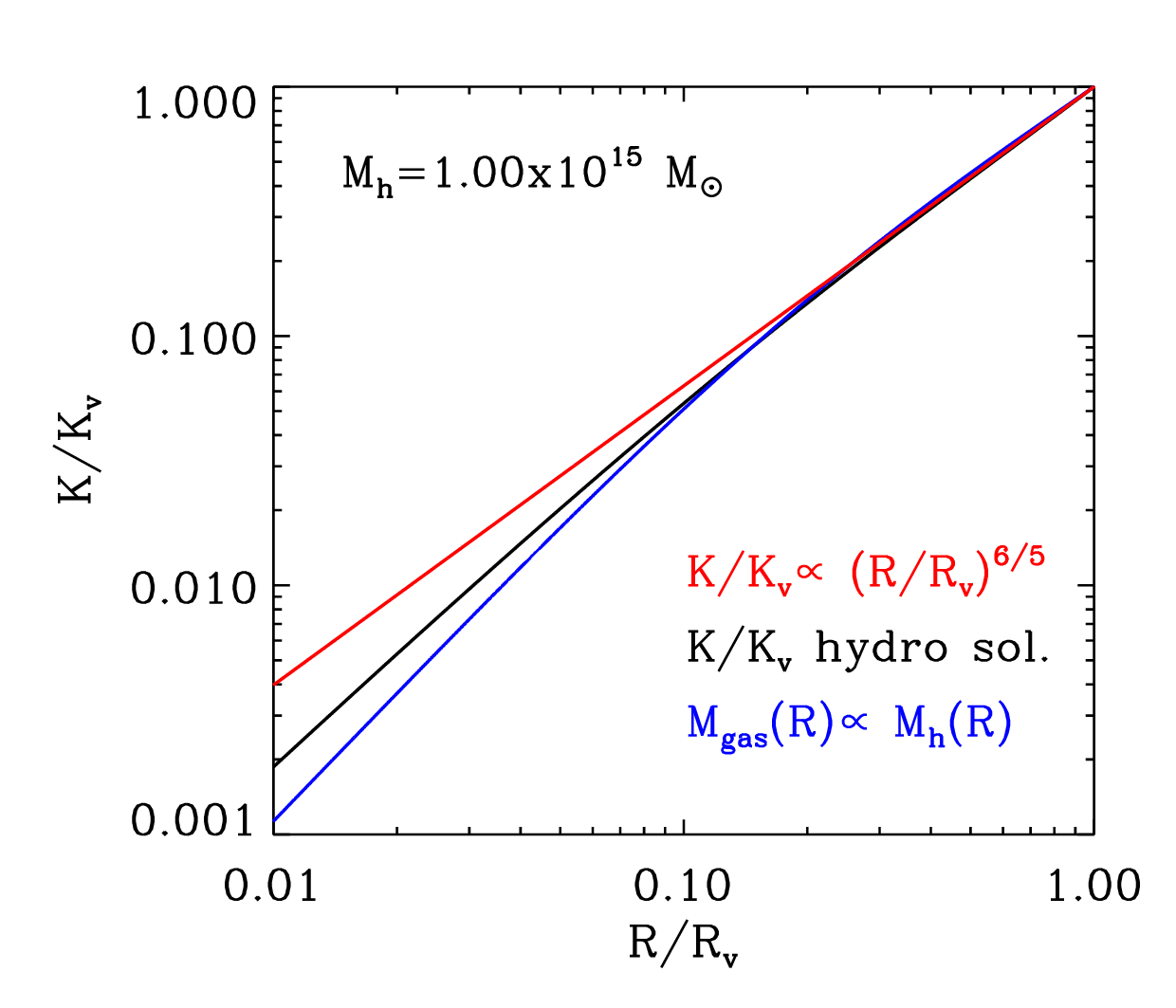}   
	\caption{Predicted entropy profile for a halo mass of  $10^{13}$ M$_\odot$ (left panel), $10^{14}$ M$_\odot$ (central panel) and $10^{15}$ M$_\odot$ (right panel) respectively. The radius is normalized to the shock (virial) radius. All profiles are normalized at the shock radius. In red we plot a reference powerlaw with slope 1.2, in blue the profile derived  from a combination of the functional forms: $K(R) = K_{\rm v}(M_{\rm h}(R))$ and in black the one derived by solving the hydrostatic equilibrium equation.}
	\label{fig:k_vs_r}
\end{figure*}

As can be appreciated, the slopes of the three entropy profiles  differ considerably. Since the term in Eq. \ref{eq:dlnk_dlnr} coming from Eq. \ref{eq:dlnm_dlnr} does not depend on mass (the NFW profile is self-similar) the difference between the three curves is entirely due to the term coming from Eq. \ref{eq:slope} and more specifically to $\alpha$ and $ {\rm d}\ln \alpha / {\rm d}\ln M_{\rm h}$.

\subsection{Radial Entropy profiles}\label{sec:sub:radprof}

As discussed at the beginning of this section, we compute entropy and other thermodynamic profiles in two different ways. In the first, we perform a straightforward combination of the functional forms: $K(R) = K_{\rm v}(M_{\rm h}(r))$, in the second we impose hydrostatic equilibrium on the shocked gas.
Note that, in the first case, the gas fraction averaged within the shock radius remains unchanged with respect to the one we impose on the infalling  shells.  In the second, gas is allowed to reorganize itself. As in \cite{Tozzi_Norman:2001}, we solve for hydrostatic equilibrium from the outside. We take the values of thermodynamic values at the shock radius as boundary conditions and numerically solve a set of three equations: the hydrostatic equilibrium equation, the gas mass conservation equation and the entropy conservation equation. This allows us to derive radial  profiles for all thermodynamic variables. A detailed description of the procedure is provided in App. \ref{sec:app2}.

In Fig. \ref{fig:k_vs_r} we plot radial entropy profiles for three different halo masses: $10^{13}$ M$_\odot$ (left panel), $10^{14}$ M$_\odot$ (central panel) and $10^{15}$ M$_\odot$ (right panel), note that the radius has been normalized to the shock (virial) radius and the entropy has been divided by the entropy at the virial radius, $K_{\rm v}$.
As we may surmise from the figure, the two solutions, shown respectively in blue and black, do not differ much from one another, both become steeper as we move inward as highlighted by a comparison with a reference power-law profile (red line). Numerical computation of the ${\rm d}\ln K/{\rm d}\ln r $ curves, obtained  from  the solution derived through a straightforward combination of functional forms, shows them to be in excellent agreement with the ones computed analytically (see Eq. \ref{eq:dlnk_dlnr}) and shown in Fig. \ref{fig:dlnk_dlnr}.

It is worth noting that, for the cluster mass range, $10^{14}$M$_\odot$-$10^{15}$M$_\odot$, these results, 
are consistent with those derived from non-radiative cosmological simulations performed using both Lagrangian smoothed particle hydrodynamics \citep[SPH see][]{Borgani:2002,Rasia:2004} and Eulerian adaptive mesh refinement (AMR) codes \citep[see][]{Voit_entropy:2005}.
This tells us something quite important:  a simple one dimensional model, based on mass and energy conversion arguments, provides a description of the entropy generation process which is, at least to first order, comparable to that which can be achieved through a much more detailed and computationally intensive analysis.

\begin{figure}
    \hspace{-0.5cm}
	\centerline{\includegraphics[angle=0,width=9.2cm]{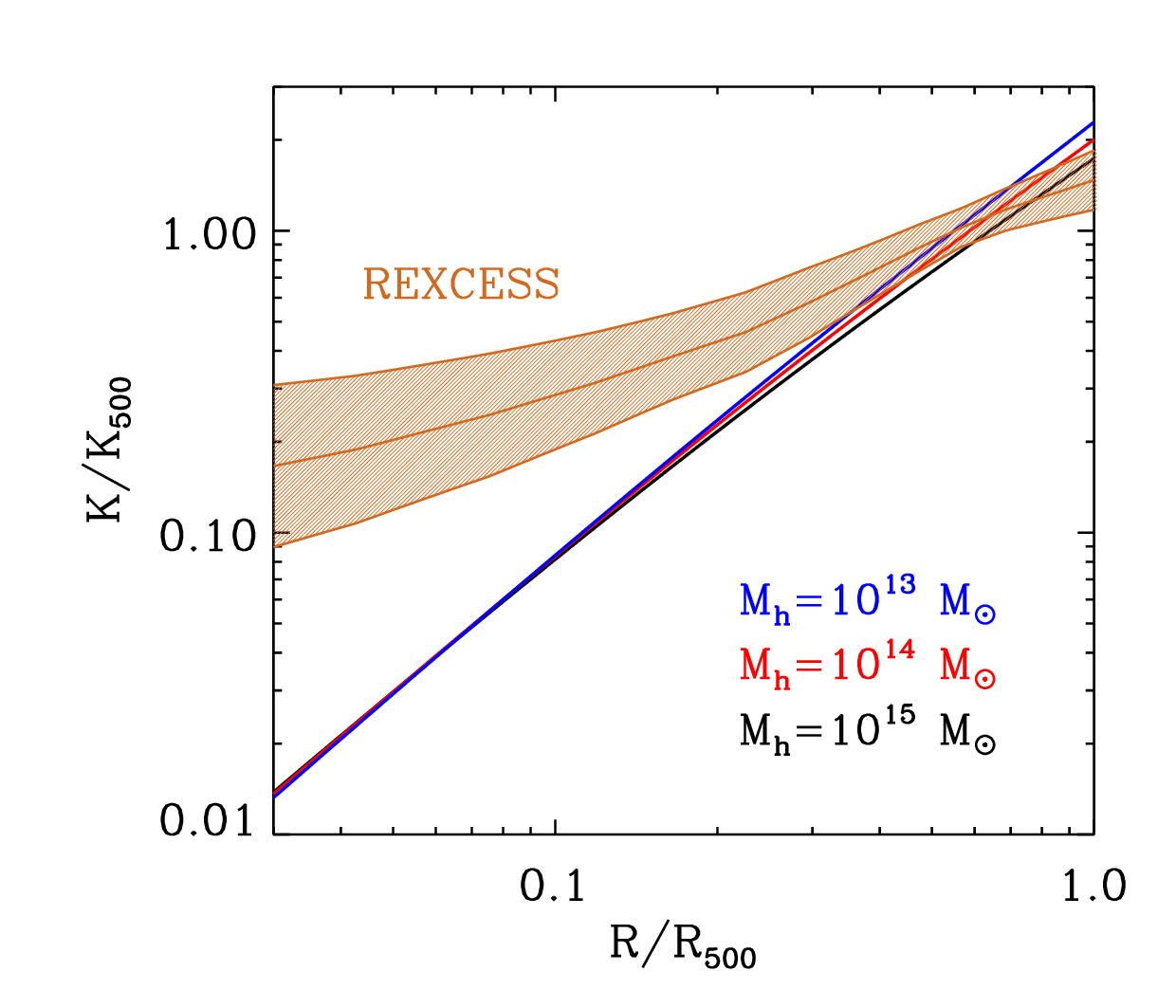}}
	\caption{Predicted entropy profiles for a halo mass of  $10^{13}$ M$_\odot$ (blue), $10^{14}$ M$_\odot$ (red) and $10^{15}$ M$_\odot$ (black) respectively, compared with the observed mean entropy profile and intrinsic scatter (orange shaded region) from the REXCESS sample of galaxy clusters, \citep{Pratt:2010}, which features a median mass of $M_{500} \sim 2.5 \times 10^{14}$M$_\odot$. To ease comparison with observational data, the predicted entropy has been rescaled by $K_{500}$ and radius by $R_{500}$. 
	}
	\label{fig:k_vs_r_obs}
\end{figure}

\subsection{A first comparison with observations}

A detailed comparison with observations will be presented in Sect. \ref{sec:obs}, here we wish to ascertaining if the simple model outlined above provides estimates that are at least broadly consistent with available data.
In Fig. \ref{fig:k_vs_r_obs} we compare the entropy profiles derived imposing hydrostatic equilibrium, and already presented in Fig. \ref{fig:k_vs_r}, with the median profile derived from the REXCESS sample of galaxy clusters \citep{Pratt:2010}. To ease comparison with observational data, entropy has been rescaled by $K_{500}$, which is defined as in \cite{Pratt:2010}, see their Eq. 3, and radius by $R_{500}$. 
At large radii, $R \sim R_{500}$, the predicted entropy is broadly consistent with the measured one, however, as we move inwards, the behavior of measured and predicted profiles diverge, the latter featuring a much steeper decline than the former. 
In more physical terms, the entropy generated  purely from  gravitational shocks is consistent with the observed one in the outskirts but severely under-predicts it at smaller radii. Both agreement and disagreement are rich in significance. The agreement suggests that, to first order, a description of the virialization process that is as crude  as the one we presented, can reproduce observed data. This implies that the entropy generation process is probably not too dissimilar from a total conversion of kinetic energy into thermal energy at a single shock radius\footnote{As the attentive reader will have undoubtedly realized, the slopes of observed and predicted profiles differ significantly. This point is discussed at length in Sect. \ref{sec:obs:out}}. It is worth pointing out that, although highly desirable, such an outcome is by no means guaranteed. Indeed, in light of the limited understanding of the physics of the accreted gas compounded by the dearth of observational data, the entropy generation process could just as easily have taken very different forms. 
Just as an example, \cite{Vazza:2009} present simulations of the accretion process on massive halos in which thermalization occurs over an extremely broad radial range, with kinetic energy exceeding thermal energy for radii larger than 0.75$ R_{\rm v}$ and remaining as high as 30\% of the thermal energy down to $\sim 0.3 R_{\rm v}$. 
%

Regarding the disagreement at smaller radii, it has been known since the time scaling relations were first established \citep{Kaiser:1986,Kaiser:1991}, that non-gravitational forms of heating, including AGN feeding/feedback \citep[e.g.][for a review]{Gaspari:2020}, may provide an important contribution. Some two decades ago, several authors  \cite[e.g.][]{Tozzi_Norman:2001,McCarthy:2004,Voit:2005} developed models in which the self-similar shape of entropy profiles was at least partially modified by assuming that the infalling gas was pre-heated to a fraction of the virial temperature. In the following section we revisit pre-heating.

\section{ Pre-heating}\label{sec:pre}
Pre-heating is the process by which gas is heated prior to being accreted.
As pointed out in \cite{Tozzi_Norman:2001}, the pre-heated gas  contracts in an approximately adiabatic fashion as long as its temperature is larger than the virial temperature of the halo onto which it is falling. Under such conditions no virial shock forms and the entropy of the accreted gas is essentially that of the pre-heated gas. As soon as the virial temperature grows beyond the temperature of the infalling pre-heated gas, an accretion shock forms. 

As discussed in a seminal paper  \citep{Kaiser:1991},  pre-heating is likely associated to the galaxy formation process.
Pre-heating introduces a flat region in the radial entropy profile. The gas residing in such an iso-entropic region has been accreted without being shocked and retains the entropy it had prior to its accretion. As shown in \cite{Voit:2003}, the entropy of the gas including the effects of preheating $K_{\rm WP}$, can be expressed reasonably well as the sum of the entropy of the gas without pre-heating, $K_{\rm v}$, and the entropy of the preheated gas, $K_{*}$ multiplied by a correction factor:  
\begin{equation}
	  K_{\rm WP} \simeq \, K_{\rm v}  + 0.84 K_{*} \, .
	\label{eq:ph}
\end{equation}
\begin{figure}
	\hspace{-0.5cm}
	\centerline{\includegraphics[angle=0,width=9.2cm]{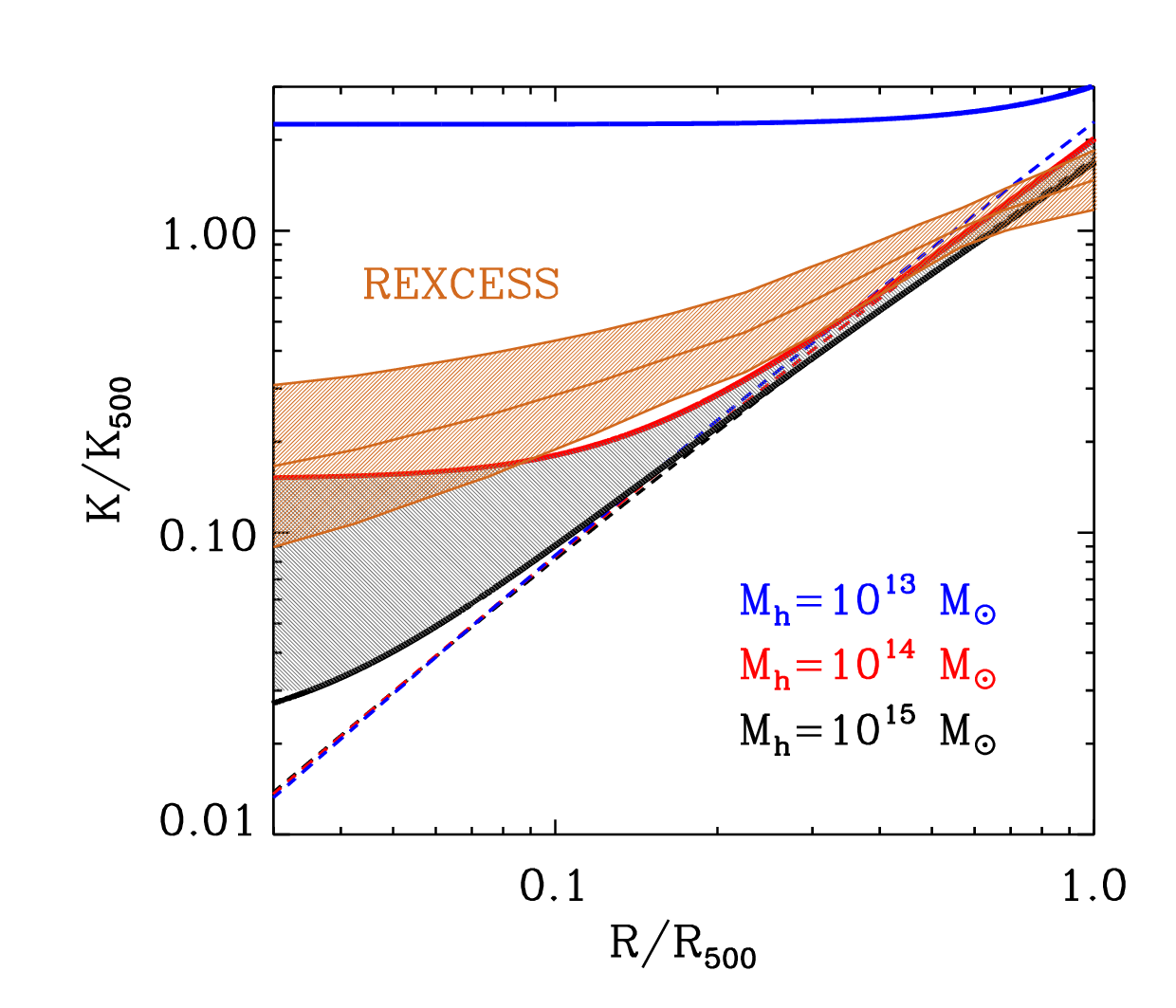}}
	\caption{Predicted entropy profiles for a halo mass of  $10^{13}$ M$_\odot$ (blue), $10^{14}$ M$_\odot$ (red) and $10^{15}$ M$_\odot$ (black) respectively. Dashed lines show entropy from gravitational processes only, solid lines include the effects of pre-heating.  The mean entropy profile and intrinsic scatter from the REXCESS sample of galaxy clusters, \citep{Pratt:2010}, which features a median mass of $M_{500} \sim 2.5 \times 10^{14}$M$_\odot$, is reported as an orange shaded region. The region between the $10^{14}$ M$_\odot$ and $10^{15}$ M$_\odot$ profiles is shown as a gray shaded region.}
	\label{fig:k_wp}
\end{figure}
A crucial point is the estimate of the pre-heating entropy. Several authors \cite[e.g.][]{Tozzi_Norman:2001,McCarthy:2002,Ponman_entropy_sc:2003,Voit:2003} have proposed values in the range  $1-5 \times 10^{33}$erg cm$^2$ g$^{-5/3}$. These estimates were arrived at by comparing observed scaling relations or entropy profiles with models including pre-heating. In Sect. \ref{sec:decoup} we discuss a different approach to estimating $K_*$.

In Fig. \ref{fig:k_wp} we show entropy profiles for a halo mass of  $10^{13}$ M$_\odot$, $10^{14}$ M$_\odot$ and $10^{15}$ M$_\odot$  respectively. The entropy of the pre-heated gas is set to the value generated through gravitational processes at the virial radius of a $10^{13}$ M$_\odot$ halo: $K_{*} =  K_{\rm v} (M_{\rm h}=10^{13}$ M$_\odot) \sim 3.5 \times 10^{33}$ erg cm$^2$ g$^{-5/3}$, for all three halo masses \citep[for a justification of this choice see][and Sect. \ref{sec:sub:profwdecph}]{Tozzi_Norman:2001}. As can be observed, the effect of pre-heating strongly depends on the halo mass. For the $10^{13}$ M$_\odot$ system,  it provides the dominant contribution at all radii. At the opposite mass end, $10^{15}$ M$_\odot$, its effects can be appreciated only for $R < 0.1 R_{500}$. In somewhat different terms, we can say that pre-heating produces a departure from self-similarity that is strongly halo mass dependent.  

As far as the comparison with observations is concerned, pre-heating does reduce significantly differences between modeled and  observed profiles but it does not allow for agreement between the two. With the pre-heating modification to the model, predicted entropy profiles of clusters fall between the thick red ($10^{14}$ M$_\odot$) and blue ( $10^{15}$ M$_\odot$) curves, observed ones do not. In particular, in observed clusters, the excess with respect to the gravity only model (dashed lines in Fig. \ref{fig:k_wp}) is  present up to intermediate radii $R \sim  0.5 R_{500}$,  while in modeled systems,  with halo masses $M_{\rm h } \gtrsim 10^{14}$ M$_\odot$, it is found only at smaller radii. 

It is worth pointing out that cooling has not been introduced  in the gravity only or the pre-heating modified model, this limits the applicability of the models to regions beyond the core, $R \gtrsim 0.1 R_{500}$, where the cooling time of the gas is sufficiently long, we will return to this point in Sect. \ref{sec:sub:clu}, where we perform a more detailed comparison with observations.

\section{Baryon decoupling}\label{sec:decoup}

The pre-heating model described in Sect. \ref{sec:pre} and in the literature \cite[see][and refs. therein]{Voit:2005} manifestly fails to reproduce the shape of entropy profiles in clusters.

In this section we investigate a different description of feedback processes in the hope it will  lead to a better agreement with observations, in doing so we take stock of the considerable advances that have taken place over the last two decades.

\subsection{Advancements}\label{sec:sub:adva}
One of the most important things we have learned is that the halo mass range between  a few $10^{12}$M$_\odot$ to $10^{13}$M$_\odot$ plays a critical role in the evolution of baryons. This is where star formation shuts down \citep[e.g.][]{Leauthaud:2012,Coupon:2012,Behroozi:2013,Moster:2013,Coupon:2015,Cowley:2018,Behroozi:2019,Legrand:2019,Girelli:2020,Shuntov:2022}.
Equally important is the fact that the position of the star formation peak  is largely insensitive to cosmic time, for $z \lesssim 4$ as shown in \citep{Behroozi:2013,Behroozi:2019,Legrand:2019}. The cold baryon reservoir from which stars form is suddenly and dramatically emptied, quite possibly through feedback from the central AGN that heats up and disperses the cold gas. The same process likely leads to a substantial reduction in the accretion on the central black hole. However, the significantly different scales on which star formation and AGN feeding operate likely lead to a delay between the shutting down of the latter with respect to the former. It is worth pointing out that the observed correlation between stellar mass and Black Hole mass,  \citep[see][and refs. therein]{Zhang:2024} as well as the one between hot halo properties and Super-Massive Black Hole (SMBH) mass \citep{Gaspari:2019} are consistent with this scenario. 

Relatively little is known of the heated gas, it is most likely in a hot phase either within  the halo, in which case it is referred to as the hot phase of the Circum Galactic Medium \citep[CGM, see][and refs. therein]{Tumlinson:2017} or outside it, in which case it takes the name of Warm Hot Intergalactic Medium \citep[see][and refs. therein]{Nicastro:2022}.
Another important piece of the puzzle that has emerged over the last decade is that the gas fraction, $f_{\rm g}$, in massive halos, correlates with halo mass \citep[see][and refs. therein]{Eckert:2021}, with lower mass systems featuring a significantly smaller baryon fraction than more massive ones.
Putting the two pieces together, we may surmise that somewhere around $\sim 10^{13}$ M$_\odot$, the feedback events responsible for shutting down the accretion also lead to a significant evacuation of baryons from the halo. The substantial scatter in gas fraction observed in groups most likely reflects the stochastic nature of the process. Feedback processes operate on a wide range of halo masses, to avoid confusion, we shall refer to feedback on $\sim 10^{13}$ M$_\odot$ halos as early-time feedback while those operating on more massive halos will be described as late-time feedback. 
Within this framework, the increase in gas fraction observed for halo masses larger than $\sim 10^{13}$ M$_\odot$, corresponds to a phase of slow re-accretion of gas previously ejected by less massive halos. It is worthwhile reminding our readers that we do have indirect evidence of the accretion of pre-enriched hot gas in massive halos through metals \citep[see][]{Molendi:2024}.

\subsection{A simple prescription}

Having taken stock of what has been learned over the last two decades, we modify the self-similar accretion model described in Sects. \ref{sec:engen} and \ref{sec:radprof} by decoupling the accretion of baryons from that of dark matter. 
Baryon decoupling provides a very simple modification, one might argue the simplest possible modification, to satisfy the constraints described in Sect. \ref{sec:sub:adva}. 
In this section we develop a model based on baryon decoupling and see how far
it takes us in interpreting the thermodynamic properties of
massive accretors.
Before proceeding with the actual modification it is instructive to perform a little bit of deductive reasoning to see how differences between measured and predicted entropy might be explained by baryon decoupling. 

Our approach is somewhat analogous to methods used by geologists when studying rock stratification. Similarly, in massive halos, gas stratifies by entropy, with gas accreted earlier positioned closer to the center, while gas accreted later resides in the outer regions.
Comparison with observations, see Figs. \ref{fig:k_vs_r_obs} and  \ref{fig:k_wp}, shows that the entropy generated when the halo was a fraction of its current mass is higher than expected. From Eq. \ref{eq:entro_def} we see that a higher entropy might result from a lower density. A lower density likely follows from a lower gas accretion rate with respect to the one assumed in Eq. \ref{eq:const_acc}, namely: 
\begin{equation}
	\dot{M}_{\rm g} < f_{\rm b}  \dot{M}_{\rm h} \, ,
	\label{eq:var_acc}
\end{equation}
or in other words, from baryon decoupling. 

At least qualitatively, baryon decoupling goes in the direction required to reconcile the self-similar model with observations. 
A more quantitative assessment requires the definition of a gas fraction versus halo mass relation.
To this end,  we assume decoupling to take place at a specific halo mass, roughly 10$^{13}$ M$_\odot$. At much smaller and larger masses we assume baryon accretion to be coupled to dark matter accretion. 
A simple way to describe mathematically this process is through the formula:

\begin{equation}
	\log(f_{{\rm g}}) =  \log(f_{{\rm b}}) + {\, l_{\rm n} \, {\exp\Bigl[-{1\over 2} \Bigl({ \log M_{\rm h} - l_{\rm m} \over l_{\rm \sigma} }\Bigr)^2 \Bigr]} } \, ,
	\label{eq:f_gas}
\end{equation}
where  $\log(f_{{\rm g}})$ is the base 10 logarithm of $f_{\rm g}$, $\log(f_{{\rm b}})$ is the base 10 logarithm of $f_{\rm b}$, $10^{l_{\rm m}}$M$_\odot$ is the halo mass at which baryon decoupling is strongest, i.e. where the deviation of the gas fraction, $f_{{\rm g}}$, from the cosmic baryon fraction reaches its maximum. The second parameter quantifies the decoupling: $10^{l_{\rm n}}$ is the ratio of the gas fraction over the cosmic baryon fraction at $M_{\rm h} = 10^{l_{\rm m}}$M$_\odot$, and finally $l_{\rm \sigma}$ determines the halo mass range over which the gas fraction is decoupled from the cosmic baryon fraction. 

\begin{figure}
    \hspace{-0.5cm}
	\centerline{\includegraphics[angle=0,width=9.2cm]{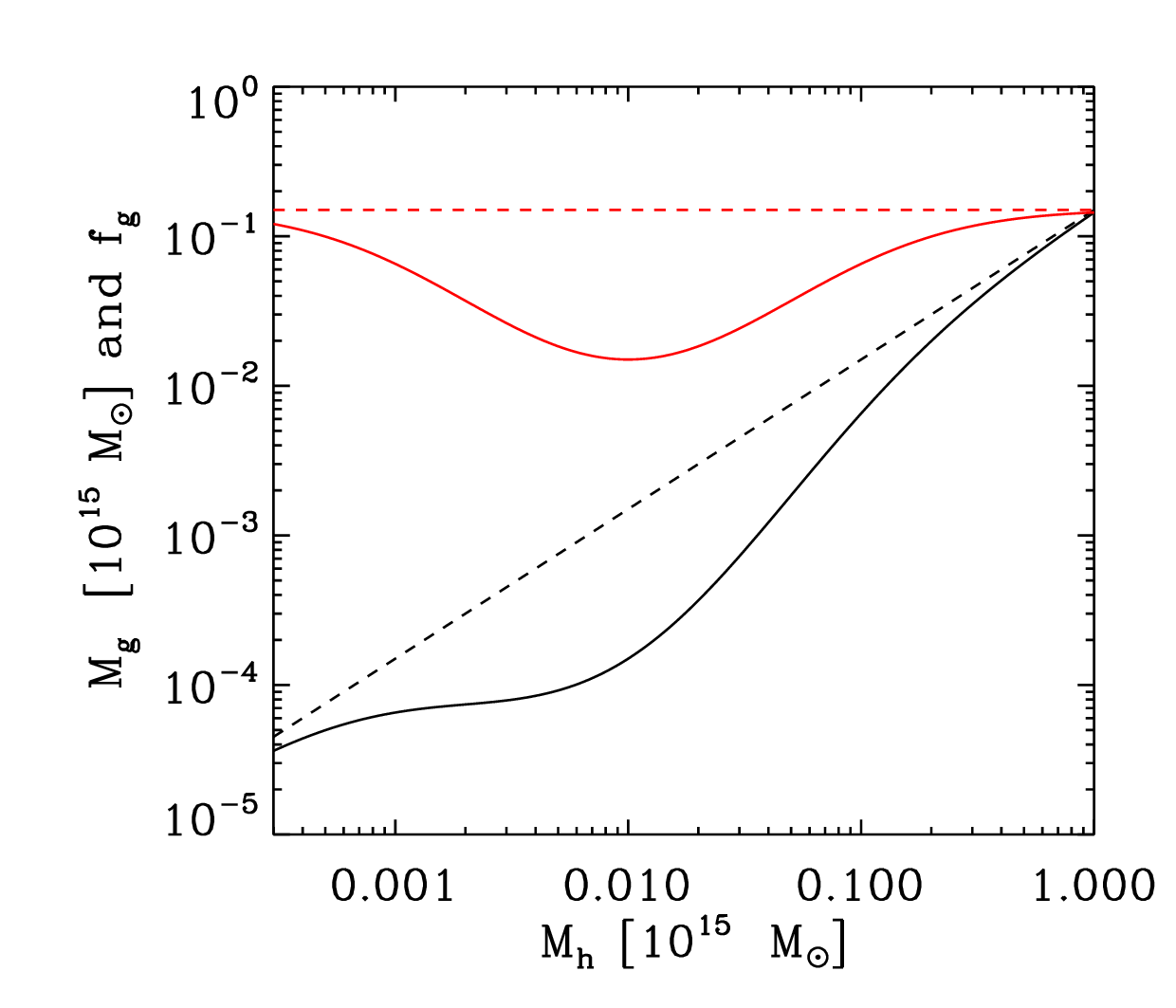}}
	\caption{Gas fraction, $f_g$, shown in red and gas mass, $M_g$, shown in black, as a function of halo mass. The dashed lines show the case where $f_g = f_b$ and $M_g = f_b M_h$ at all halo masses. The solid lines show the case modeled by Eq.\ref{eq:f_gas} with parameters chosen as follows: $l_m = 13$, $l_n = -1$ and $l_\sigma =0.6$. 		  
	}
	\label{fig:f_gas}
\end{figure}

In Fig. \ref{fig:f_gas} we show, as a red solid line, a realization of the gas fraction versus halo mass relation for a representative choice of parameters. Additionally, the figure shows the gas mass versus halo mass relation as a black solid line, highlighting another key relation.
Note how, despite the reduction in gas fraction with increasing halo mass for $M_{\rm h} < 10^{l_{\rm m}}$M$_\odot$, the gas mass always increases with halo mass. 


Our implementation of baryon decoupling provides a very simple modification of the gravitational entropy generation process. It identifies a critical mass at which baryon decoupling operates, $10^{l_{\rm m}}$M$_\odot$, and allows for variations in the strength and the range over which the modification operates, respectively through the $l_{\rm n}$ and $l_{\rm \sigma}$ parameters. In the next subsection we develop a model based on baryon decoupling and see how far it takes us in interpreting the thermodynamic properties of massive accretors.

\subsection{Entropy versus Mass relation with decoupling}
\label{sec:sub:k_vs_m}
\begin{figure}
    \hspace{-0.5cm}
	\centerline{\includegraphics[angle=0,width=9.2cm]{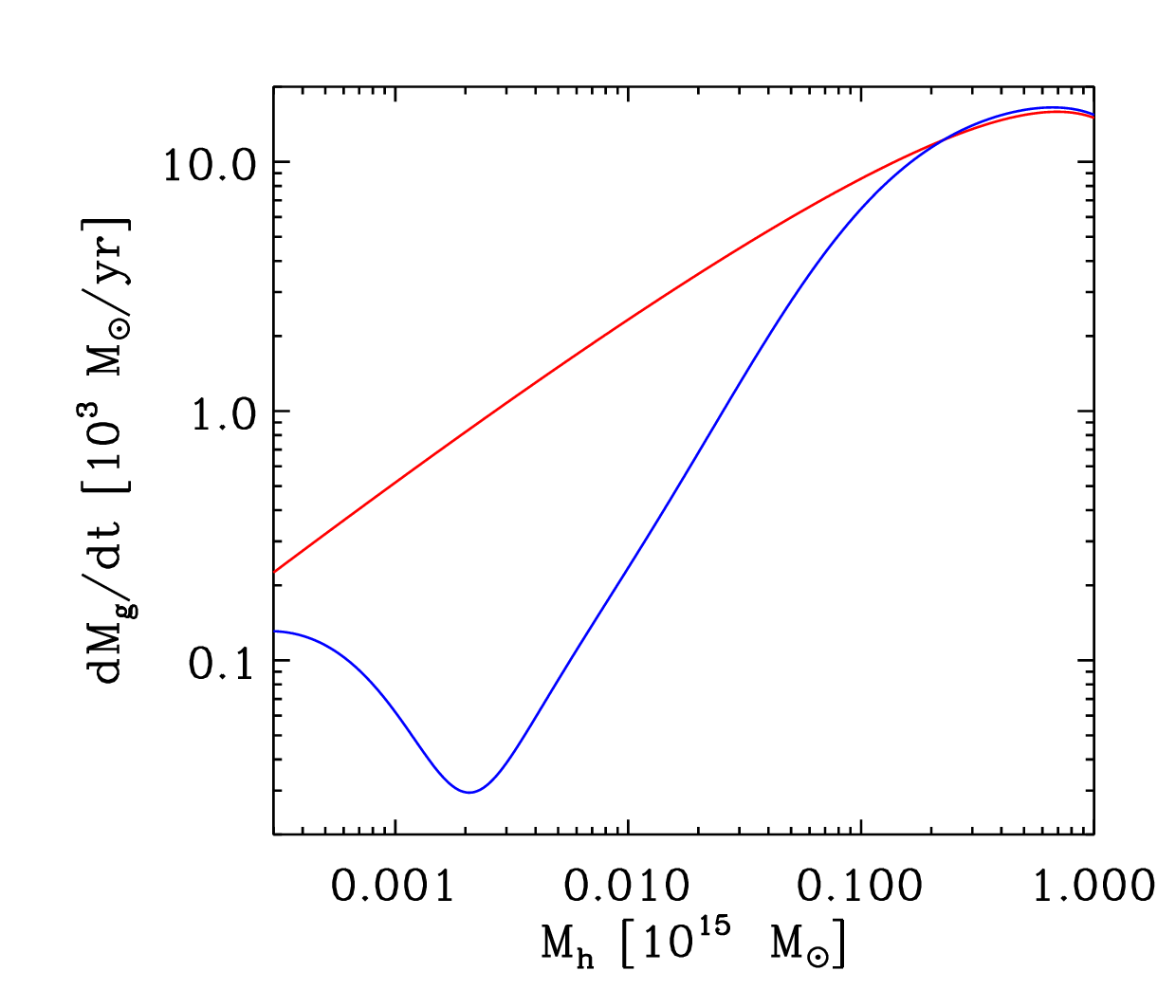}}
	\caption{Gas accretion rate versus halo mass relation for a halo of  $10^{15}$ M$_\odot$ at $z=0$. The blue and red lines have been computed assuming $f_{{\rm g}}$ follows respectively Eq. \ref{eq:f_gas} and $f_{{\rm g}}=f_{{\rm b}}$.   
	}
	\label{fig:mgasdotvsmh}
\end{figure}

In this subsection we  investigate the effects of
  baryon decoupling on the entropy versus halo mass relation. The first thing to note is that, since the gas fraction of the accreted shells is no longer constant,  Eq. \ref{eq:entro_sm2} is not valid and we have to fall back on Eq. \ref{eq:entro_sm}. 
  Furthermore, the $\dot{M_{\rm g}} $ term, which in the case $f_{{\rm g}}=f_{{\rm b}}$ is simply $\dot{M_g} = f_{{\rm g}} \dot{M} $, for a non constant $f_{{\rm g}}$ takes the form:
  
\begin{equation}
	\dot{M_{\rm g}} =  f_{\rm g}  \dot{M_{\rm h}} \, \biggl( 1 + {{\rm d}\ln f_{\rm g}  \over {\rm d}\ln M_{\rm h}  } \biggr) \, .
	\label{eq:mdotg}
\end{equation}
Since the logarithmic derivative ${\rm d}\ln f_{\rm g}  / {\rm d}\ln M_{\rm h}$ is negative for $M_{\rm h} < 10^{l_m}$M$_\odot$ and positive for $M_{\rm h} > 10^{l_m}$M$_\odot$, the term in parenthesis on the right hand side of Eq. \ref{eq:mdotg} reaches a minimum at  $M_{\rm h} = 10^{l_m}$M$_\odot$.  Multiplication by the $f_{\rm g}  \dot{M}_{\rm h} $ term shifts  the minimum to somewhat lower halo masses. This is clearly seen in Fig. \ref{fig:mgasdotvsmh}, where we plot $\dot{M_{\rm g}}$ as a function of halo mass for a halo of $10^{15}$ M$_\odot$ at $z=0$.

In more physical terms, baryon decoupling reduces the gas accretion rate with respect to the $f_{\rm g} = f_{\rm b}$ case. Most importantly, it introduces a critical mass $\sim 1/4 \, 10^{l_m}$M$_\odot$, below which  the gas accretion rate decreases and above which it increases, with increasing halo mass. For halos that grow well beyond the critical mass, the accretion rate slowly converges to $f_{{\rm g}} \dot{M_{\rm h}} $, see Fig. \ref{fig:f_gas}. In other words, baryon decoupling does not halt accretion, it delays it.

\begin{figure}
    \hspace{-0.5cm}
	\centerline{\includegraphics[angle=0,width=9.2cm]{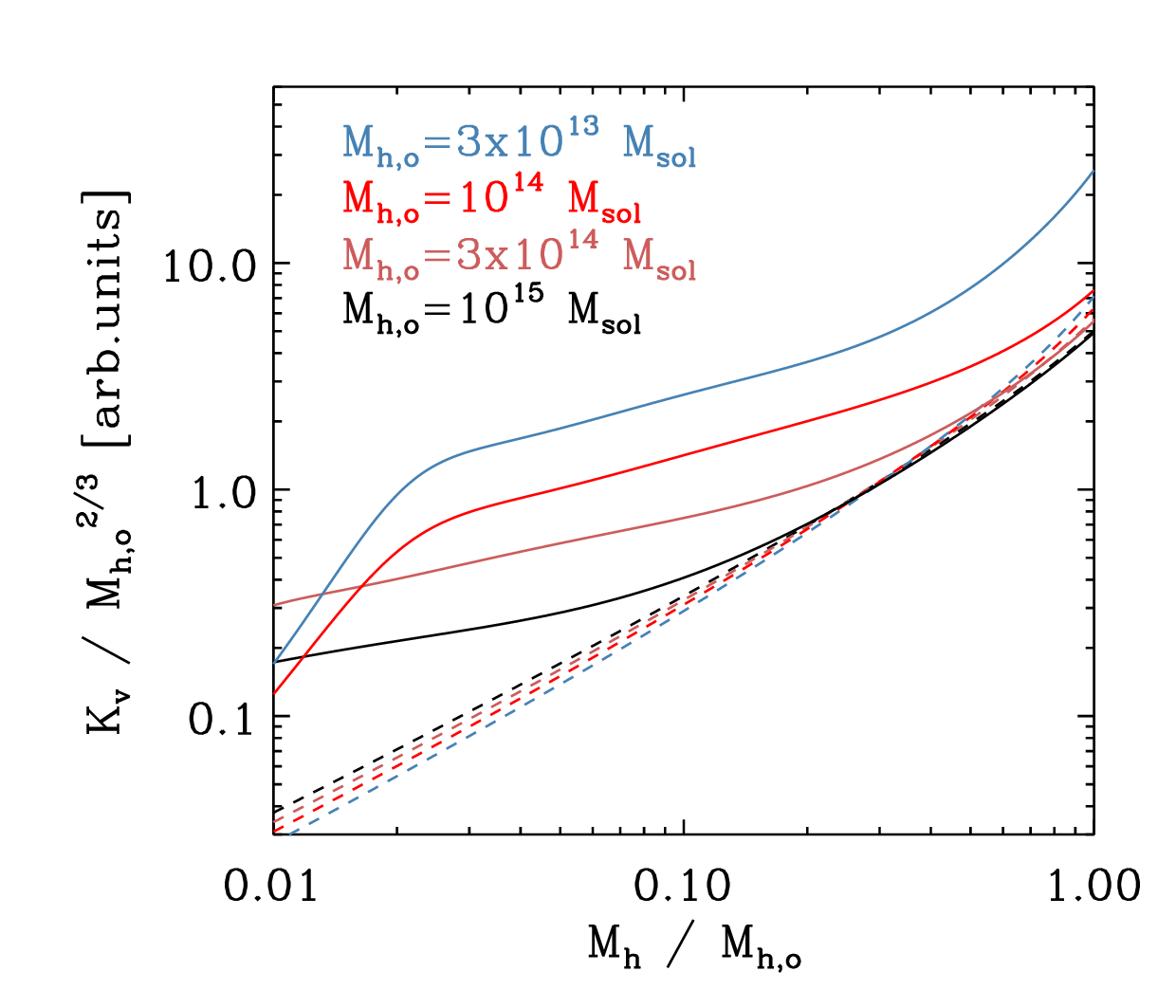}}
	\caption{Rescaled entropy versus halo mass relation for four halos of $3\cdot 10^{13}$ M$_\odot$ , $10^{14}$ M$_\odot$,  $3 \cdot 10^{14}$ M$_\odot$ and $10^{15}$ M$_\odot$ at $z=0$ respectively. Filled lines have been computed assuming $f_{{\rm g}}$ follows Eq.\ref{eq:f_gas}, dashed ones assuming $f_{\rm g}=f_{\rm b}$. Note how baryon decoupling produces a break in self-similarity. The change in slope observed at small  $M_{\rm h}/M_{\rm h,o}$ values for the $3\cdot 10^{13}$ M$_\odot$ and $10^{14}$ M$_\odot$ at  $z=0$ halos is associated to the minimum in  $\dot{M_{\rm g}} $, see Fig. \ref{fig:mgasdotvsmh}. In the case of $3 \cdot 10^{14}$ M$_\odot$ and $10^{15}$ M$_\odot$ at $z=0$ profiles, the minimum occurs at smaller $M_{\rm h}/M_{\rm h,o}$ values and falls outside the plotted range.
	}
	\label{fig:kvsmh2}
\end{figure}

As shown in Fig. \ref{fig:kvsmh2},  baryon decoupling induced deviations propagate from the gas accretion rate  to the entropy.  
Comparison of the solid curves derived assuming  $f_{{\rm g}}$ follows Eq. \ref{eq:f_gas} with dashed ones for which $f_{{\rm g}}=f_{{\rm b}}$, shows that baryon decoupling has a strong impact on the rescaled entropy versus halo mass relation. The reduction in accretion rate leads to an increase in entropy.  As expected, the effect is largest for the least massive halos ($3\cdot 10^{13}$ M$_\odot$, $10^{14}$ M$_\odot$ at $z=0$), where the reduction in accretion rate is strongest. Another important point that is clearly observed in  Fig. \ref{fig:kvsmh2} is that baryon decoupling leads to a break in self-similarity; once again it is the less massive systems that deviate most. Finally, the change in slope observed for $M_{\rm h} / M_{\rm h,o} \sim 0.2$, for the cases $M_{\rm h,o} = 3\cdot 10^{13}$ M$_\odot$ and $M_{\rm h,o} = 10^{14}$ M$_\odot$, is related to the minimum in  $\dot{M}_{\rm g}$ observed in Fig. \ref{fig:mgasdotvsmh}. 
 
\begin{figure}
	\hspace{-0.5cm}
	\centerline{\includegraphics[angle=0,width=9.2cm]{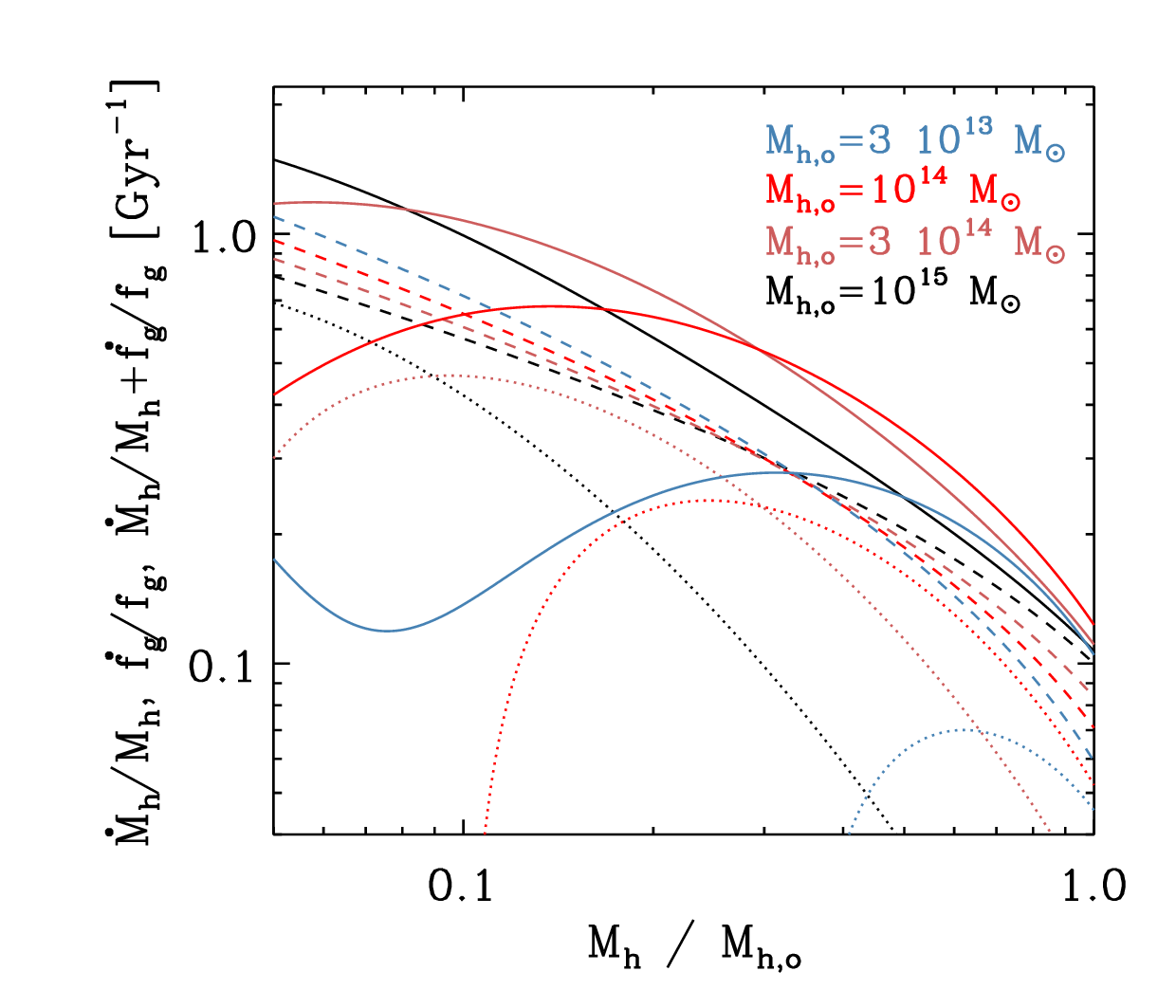}}
	\caption{$\dot{M}_{\rm h} / M_{\rm h}$ (dashed lines),  ${\dot{f}_{\rm g} / f_{\rm g}}$ (dotted lines) and ${\dot{f}_{\rm g}/f_{\rm g}} + {\dot{M}_{\rm h}/M_{\rm h}}$ (filled lines) as a function of halo mass for 4 halos of $3\cdot 10^{13}$ M$_\odot$ , $10^{14}$ M$_\odot$,  $3 \cdot 10^{14}$ M$_\odot$ and $10^{15}$ M$_\odot$ at $z=0$ respectively.
	}
	\label{fig:spdotvsmh}
\end{figure}

\subsection{Gas fraction rescaling} \label{sec:sub:gas_fr}
With a little algebra Eq. \ref{eq:entro_sm} can be rewritten in the following fashion:
\begin{equation}
	K_{\rm v} =  {1 \over 3} \, (\pi G^2)^{2/3} \, \left( { M_{\rm h} \over f_{\rm g} } \right)^{2/3}  
	\left[ {\dot{f}_{\rm g} \over f_{\rm g}} + {\dot{M}_{\rm h} \over M_{\rm h}} \right]^{-2/3} \, .
	\label{eq:ksmfgmh}
\end{equation}
This form highlights the similarity between the role played by $M_{\rm h}$ and $f_{\rm g}$. We recall that the self-similar scaling with respect to mass follows from the fact that $\dot{M}_{\rm h} / M_{\rm h}$ depends only weakly on mass, see Fig. \ref{fig:mdotvsm}. Something similar may happen with $f_{\rm g}$ if ${\dot{f}_{\rm g} / f_{\rm g}}$  is independent or at least weakly dependent on $f_{\rm g}$. Having identified a specific prescription for $f_{\rm g}$, see Eq. \ref{eq:f_gas}, we can verify to what extent  $f_{\rm g}$ scaling might work.
In Fig. \ref{fig:spdotvsmh} we plot  $\dot{M}_{\rm h} / M_{\rm h}$,  ${\dot{f}_{\rm g} / f_{\rm g}}$ and ${\dot{f}_{\rm g}/f_{\rm g}} + {\dot{M}_{\rm h}/M_{\rm h}}$ as a function of halo mass for four halos of $3\cdot 10^{13}$ M$_\odot$, $10^{14}$ M$_\odot$, $3 \cdot 10^{14}$ M$_\odot$ and $10^{15}$ M$_\odot$ at $z=0$ respectively.
The $\dot{M}_{\rm h} / M_{\rm h}$ (dashed lines)  are very similar to one another, this has already been shown in Fig. \ref{fig:mdotvsm}. It is this behavior that is responsible for self-similarity in the $f_{\rm g}=f_{\rm b}$ case. The ${\dot{f}_{\rm g} / f_{\rm g}}$ curves (dotted lines) show somewhat larger variations. However, since $\dot{M}_{\rm h} / M_{\rm h}$ and ${\dot{f}_{\rm g} / f_{\rm g}}$ curves have opposite dependencies on halo mass, the ${\dot{f}_{\rm g}/f_{\rm g}} + {\dot{M}_{\rm h}/M_{\rm h}}$ (filled lines) show smaller variations on halo mass, with typical  deviations from one another of less than $\sim 30\%$  for $M_{\rm h} / M_{\rm h,o} \gtrsim 0.3$, suggesting that self-similarity can be recovered, at least in part, by normalizing on the gas fraction.

\begin{figure}
	\hspace{-0.5cm}
	\centerline{\includegraphics[angle=0,width=9.2cm]{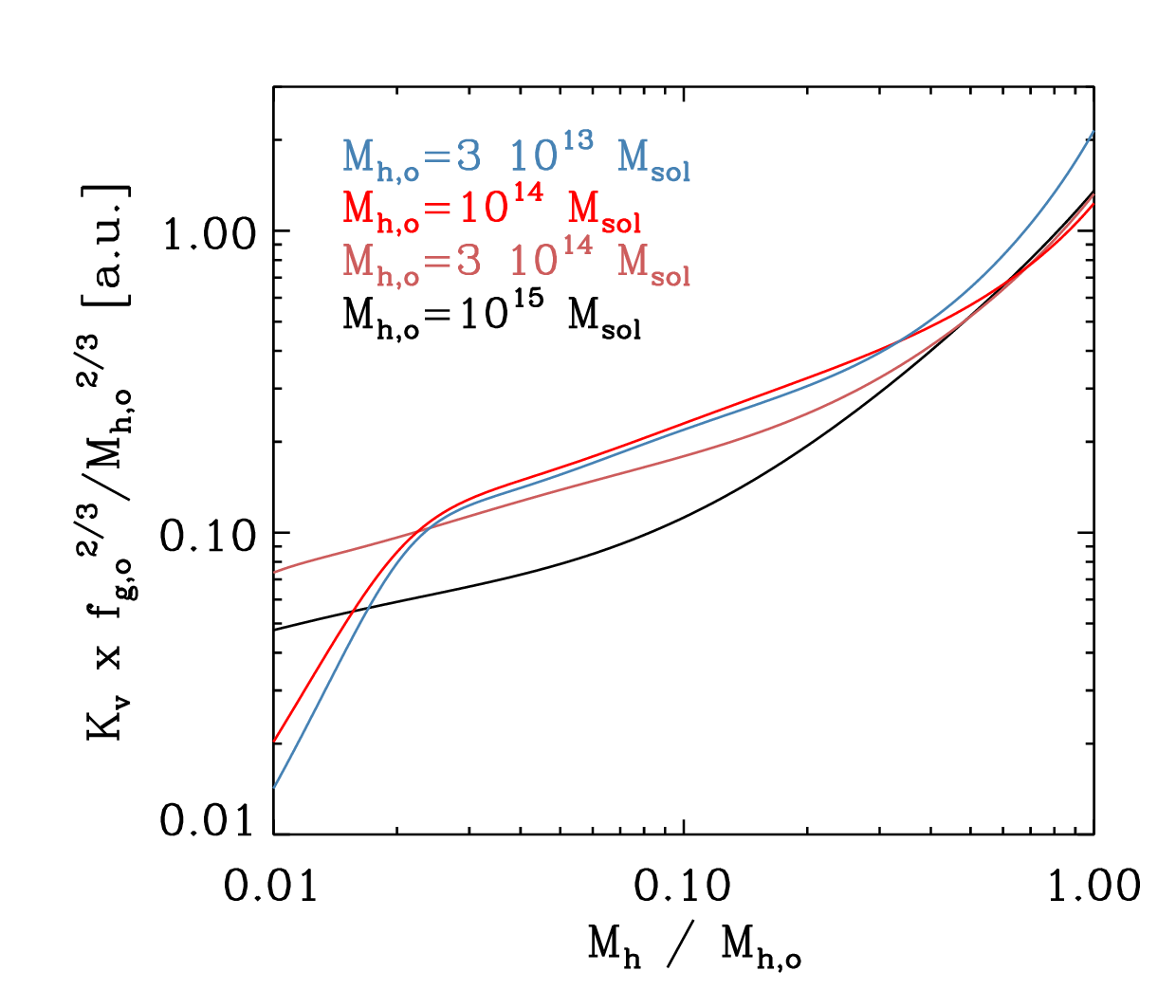}}
	\caption{Halo mass and gas fraction scaled entropy as a function of halo mass for 4 halos of $3\cdot 10^{13}$ M$_\odot$ , $10^{14}$ M$_\odot$,  $3 \cdot 10^{14}$ M$_\odot$ and $10^{15}$ M$_\odot$ at $z=0$ respectively.
	}
	\label{fig:ksmfgvsmh}
\end{figure}


\begin{figure}
	\hspace{-0.5cm}
	\centerline{\includegraphics[angle=0,width=9.2cm]{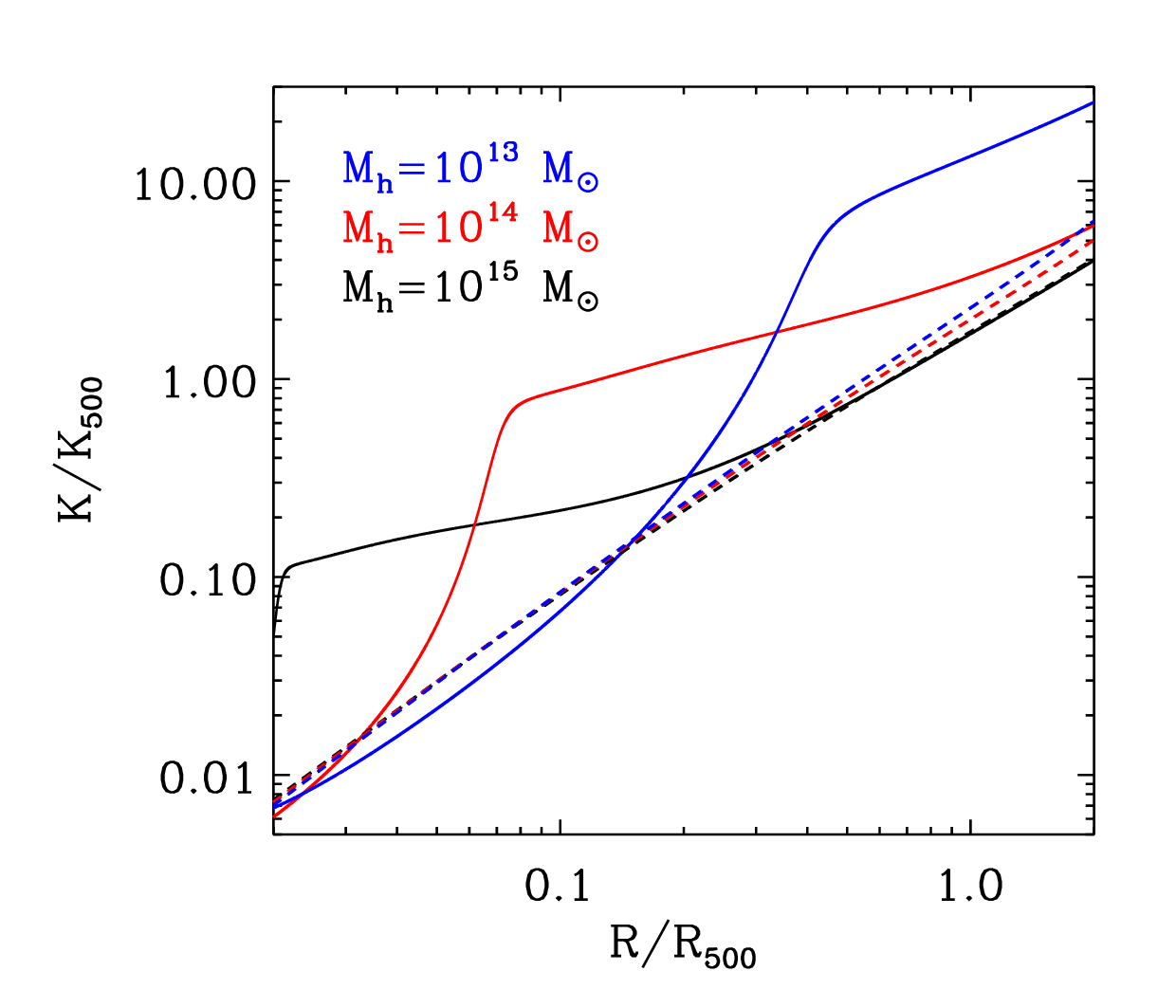}}
	\caption{Radial entropy profiles derived by imposing hydrostatic equilibrium for a halo mass of  $10^{13}$ M$_\odot$ (blue) , $10^{14}$ M$_\odot$ (red) and $10^{15}$ M$_\odot$ (black) respectively. The radius is normalized to $R_{500}$ and the entropy to $K_{500}$. Filled lines have been computed with baryon decoupling, dashed ones without.}
	\label{fig:k_vs_r_bd}
\end{figure}

A comparison of Fig. \ref{fig:ksmfgvsmh} with  Fig. \ref{fig:kvsmh2}  shows that rescaling by gas fraction does reduce variability between the different halos in a substantial way. This implies that the break in self-similarity associated to the decoupling between gas and dark matter accretion can be in large part recovered by rescaling the entropy by the gas fraction. Note however that, while at large fractional masses, all profiles are broadly similar, as we move to smaller fractional masses, the lower mass systems deviate before the higher mass ones. 

\subsection{Entropy profiles with baryon decoupling}\label{sec:sub:profwdec}
Following the procedure described in Sect. \ref{sec:sub:radprof} we derive radial entropy profiles with baryon decoupling.  As expected, we observe substantial deviations with respect to the self-similar profiles, see  Fig.\ref{fig:k_vs_r_bd}. More specifically, baryon decoupling augments entropy for all halo masses but it does so in different radial ranges for different halo masses. 
In broad terms we may say that, for $M_{\rm h} = 10^{13}$ M$_\odot$ (blue line) the increase is concentrated at large radii, for $M_{\rm h} = 10^{14}$ M$_\odot$ (red line) at intermediate radii and at small radii for $M_{\rm h} = 10^{15}$ M$_\odot$ (black line). Intriguingly, the  black solid line, which includes baryon decoupling,  coincides with the dashed one which is derived assuming the baryon fraction of the accreting gas is constant, for radii larger than $\sim 0.4 R_{500}$. This is expected, as shown in Fig. \ref{fig:f_gas}, matter accreted on massive halos is close to having a cosmological baryon fraction. As pointed out in Sect. \ref{sec:sub:k_vs_m}, baryon decoupling delays accretion, it does not stop it.  An important consequence is that, in the outer regions of the most massive systems  the entropy generation process is dominated by gravity, or in more explicit terms by the conversion of kinetic into thermal energy at the virial shock. We will return to this important point in Sect. \ref{sec:obs:out},
The careful reader may note that also for the inner radii of low mass systems deviations from the gravitational heating model are modest. However, as discussed in Sect. \ref{sec:sub:clu}, for $R < 0.1R_{500}$ radiative cooling, which is not included in our model,  significantly modifies  entropy profiles.

\begin{figure}
	\hspace{-0.3cm}
	\centerline{\includegraphics[angle=0,width=9.2cm]{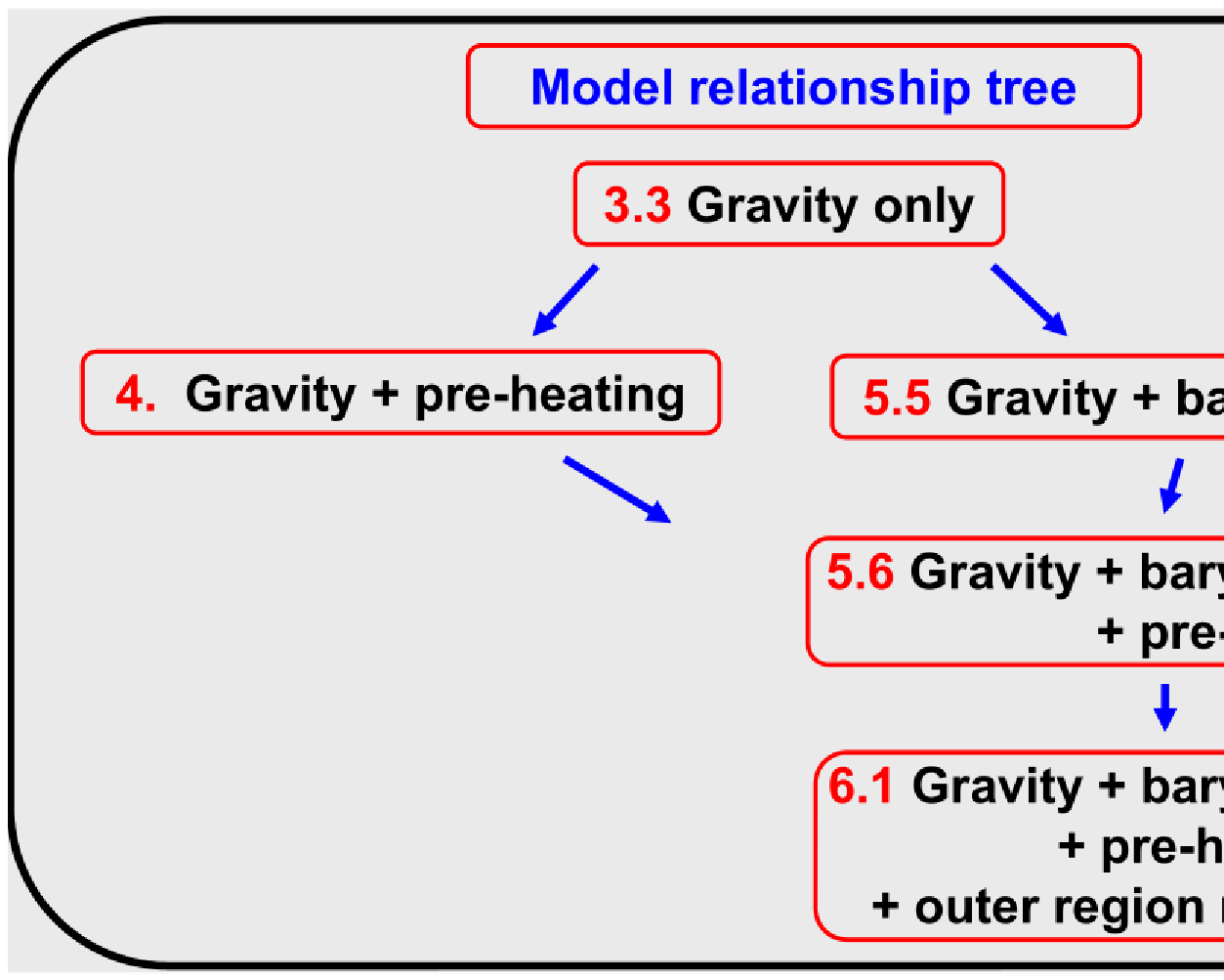}}
	\caption{Relationship tree connecting the models explored in this paper. The numbers in red on the left of the model name identify the section where the model is discussed.}
	\label{fig:mod_rel_tree}
\end{figure}

To help readers keep track of the modifications we have made and will make to our model (see Sects. \ref{sec:sub:profwdecph} and \ref{sec:obs:out}), in Fig. \ref{fig:mod_rel_tree}, we provide a model relationship tree.


\subsection{Entropy profiles with decoupling and preheating}\label{sec:sub:profwdecph}
Decoupling and pre-heating are not mutually exclusive, quite the contrary, early-time feedback processes that remove gas  from shallow potential wells may also heat it.
In this subsection we modify entropy profiles derived in Sect. \ref{sec:sub:profwdec} to include the effects of preheating.

As already discussed in Sect. \ref{sec:pre}, to provide a realistic entropy model, it is critical to estimate $K_*$, the pre-heating contribution. To this end, we make use of what has been learned about feedback mechanisms over the last two decades.
We proceed by estimating the temperature and the density of the preheated gas. Expulsion from low mass halos occurs when the temperature of the hotter phase of the gas exceeds the virial temperature. We set  the temperature of the expelled gas to be: $T_{\rm exp} = \Phi T_{\rm v}(M_{\rm h}=0.3 -1\times 10^{13}$ M$_\odot)$, where $\Phi$ is some yet to be determined factor.
Regarding density, we start by noting that the mass of gas that is accreted onto the halo is much larger than that which is involved in the star formation or central black hole feeding process. 
For halos of $\sim 0.3-1.0 \times 10^{13}$ M$_\odot$ roughly 1/7 of the baryon mass goes into stars \citep[see][and refs. therein]{Behroozi:2019}.
Assuming that roughly 1/3 of the gas that goes into stars is fed back \cite[see discussion on the $r_o$ factor in][and refs. therein]{Ghizzardi:2021}, and that the contribution from the AGN feedback is negligible, as suggested by the fact that the black hole mass is less that $\sim 0.01$ of the stellar mass \citep[e.g.][]{Reines:2015}, we have that roughly $1/7 \times 1/3 \sim 5$\% of the baryons are involved in the feed-back process. 
This implies that while the energy of the outflow will contribute significantly to that of the expelled gas, the mass contribution will be modest and for the present estimate negligible. Since the gas that requires least energy  to be expelled is the one that has just been accreted, we set the density of the expelled gas to that of the gas that has recently sustained  a virial shock,  $\rho_{\rm exp} \sim \rho_{\rm v}(M_{\rm h}= 0.3 -1\times 10^{13}$ M$_\odot)$, coupling this with $T_{\rm exp} = \Phi T_{\rm v}(M_{\rm h}=0.3 -1\times 10^{13}$ M$_\odot)$, which we derived earlier we get: $K_{\rm exp} = \Phi K_{\rm v}(M_{\rm h}=0.3 -1\times 10^{13}$ M$_\odot)$. 
Finally, we estimate that $\Phi$  cannot be much larger than a few, if it were $\gtrsim 10^{2/3}$ then the thermal energy of the gas would have been sufficient to inhibit accretion on objects that are 10 times more massive (clusters), which would in turn require clusters to have gas fractions comparable to those of less massive systems (groups). 
On the basis of this discussion, we include a preheating entropy of $K_{*} =  K_{\rm v} (M_{\rm h} = M_*) \sim 3.5 \times 10^{33}$erg cm$^2$ g$^{-5/3}$, where $M_*=10^{13}$ M$_\odot$. Since preheating only affects gas that is accreted on halos that are larger than the critical mass, $M_*$, 
we implement the preheating correction on entropy by multiplying $K_*$ by the factor $\exp(-M_*/M_{\rm h})$ which goes respectively to 1 for  $M_{\rm h} \gg M_*$ and to 0 for  $M_{\rm h} \ll M_*$. This accounts for the fact that gas accreting on halos with mass comparable to $M_*$ is likely a mixture of cold and preheated, with cold becoming dominant as we move downwards in halo mass and preheated if we move in the opposite direction. 

\begin{figure}
	\hspace{-0.5cm}
	\centerline{\includegraphics[angle=0,width=9.2cm]{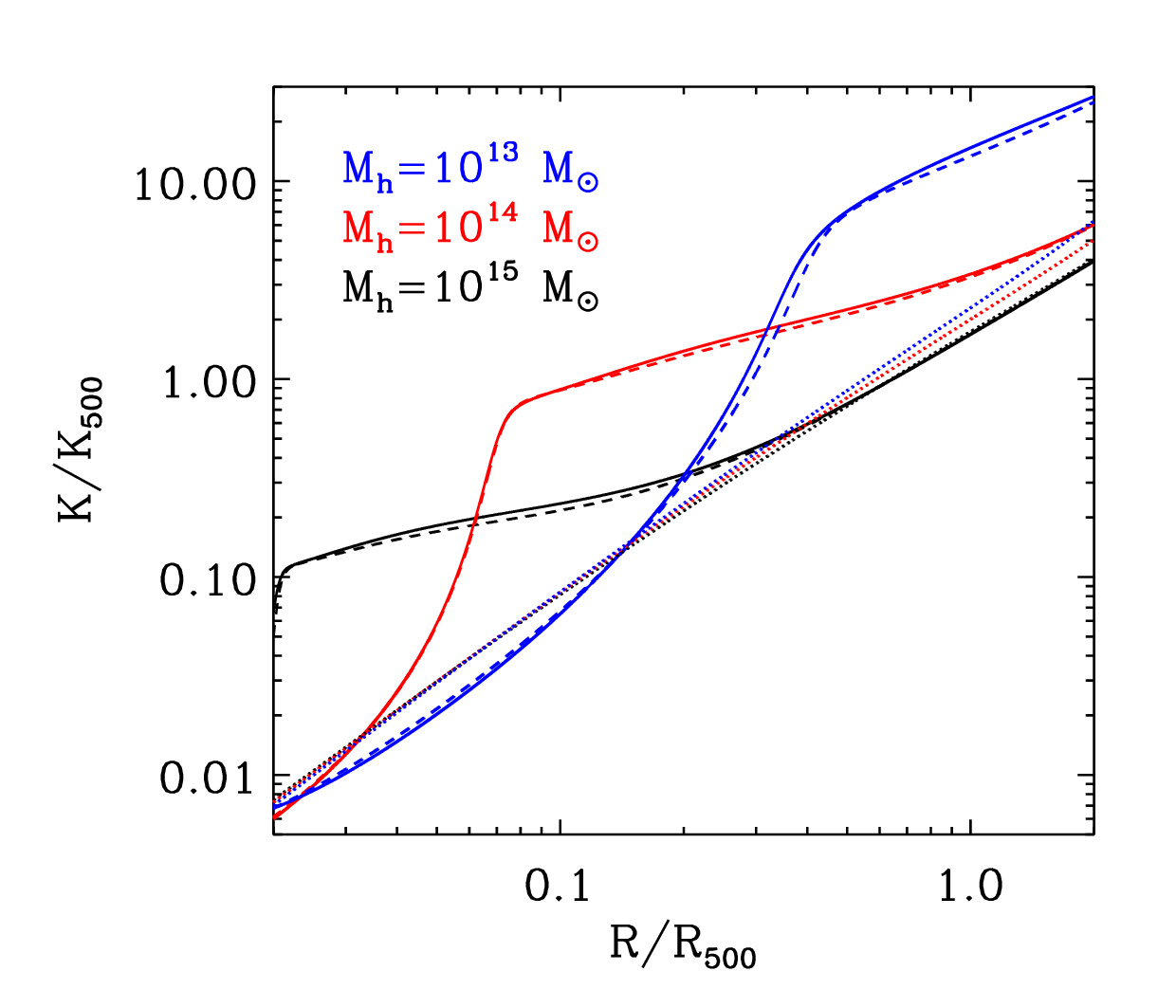}}
	\caption{Entropy profiles derived by imposing hydrostatic equilibrium for a halo mass of  $10^{13}$ M$_\odot$ (blue) , $10^{14}$ M$_\odot$ (red) and $10^{15}$ M$_\odot$ (black) respectively. The radius is normalized to $R_{500}$ and the entropy to $K_{500}$. Filled lines have been computed with baryon decoupling and pre-heating, dashed lines with baryon decoupling and no preheating, dotted lines without either.}
	\label{fig:k_vs_r_bd_wp}
\end{figure}
 
In Fig. \ref{fig:k_vs_r_bd_wp} we show entropy profiles for a halo mass of  $10^{13}$ M$_\odot$ (blue), $10^{14}$ M$_\odot$ (red) and $10^{15}$ M$_\odot$ (black) respectively with baryon decoupling and pre-heating (filled lines), with baryon decoupling and no preheating (dashed lines) , without either (dotted lines). As already shown in Fig. \ref{fig:k_vs_r_bd}, baryon decoupling introduces significant modification to the entropy profiles. Conversely, including pre-heating on top of baryon decoupling introduces negligible modifications. This is fairly easy to understand, for small halo masses, $M_{\rm h} < 10^{l_{\rm m}}$M$_\odot$, this is achieved by construction through the exponential cutoff term, $\exp(-M_*/M_{\rm h})$ described above. For  large halo masses, the accreted gas is all pre-heated, $\exp(-M_*/M_{\rm h}) \sim 1 $, but its contribution to entropy is negligible as in the preheating only model (see Fig. \ref{fig:k_vs_r_obs}). We point out that changing the value of $K_{*} $ by a factor of a few results in very minor modifications, this justifies, in retrospect, the order of magnitude estimate we have made of this quantity. 

In summary, 
within massive halos, $M_{\rm h} \sim 10^{15}$ M$_\odot$, almost all the heat conferred to the gas is of gravitational nature. The role of feedback processes, while vital, requires relatively little energy, all that needs to be done is to take gas out of shallow potential wells and let gravity do the rest. The situation is radically different at the low mass end, $M_{\rm h} \sim 10^{13}$ M$_\odot$, here the energy of the expelled gas is sufficient to prevent, at least in part, its re-accretion. 

\section{Comparison with observations}\label{sec:obs}
This section is dedicated to a detailed comparison of the model we have developed with state of the art observations. We begin by focusing on the outer regions of massive clusters.

\subsection{The outer regions of massive halos}\label{sec:obs:out}
On the base of the model described in the previous Section, it seems that the only place where entropy profiles depend only upon cosmological accretion is  large radii in massive systems. Indeed, this is where both baryon decoupling and pre-heating play a negligible role, see Fig. \ref{fig:k_vs_r_bd_wp}.
As shown in Figs. \ref{fig:k_vs_r_obs} and \ref{fig:k_wp} the entropy predicted by the model is broadly consistent with the measured one, perhaps a little on the high side, however, if we look a little more in detail, we see that the slopes are quite different, with observed profiles typically ${\rm d}\ln K/{\rm d}\ln R \sim 0.8$ \citep{Pratt_entropy:2006,Ghirardini:2019} and our model ${\rm d}\ln K/{\rm d}\ln R \sim 1.2$. 

As pointed out in Sect. \ref{sec:engen}, our modeling of the entropy generation at the accretion shock is relatively simple. Here we take a closer look at the process to gauge how a more realistic description might solve this issue. In very general terms there are two classes of corrections that may be applied. In the first, for some reason,  the temperature and therefore the entropy of the shock heated gas does not increase with halo mass as $ M_{\rm h}^{2/3}$ but at a slightly milder rate.  In the second, the kinetic to thermal conversion occurs over a radial range, rather than just at one radius. In this case the entropy, rather than jumping from zero to the value in Eq. \ref{eq:entro_sm2} at a specific radius, grows over a yet to be determined  radial range.

\subsubsection{From smooth to mixed accretion}\label{sec:sub:mix}
 In this subsection we discuss a minor modification to the smooth accretion framework that can have  an effect on the  entropy profiles in the outer regions of massive halos. Part of the gas undergoing shock heating is unbound, part is bound to infalling sub-halos. For the former part, all the kinetic energy associated to the infall is transformed into thermal energy, for the latter a fraction is required to strip the gas from the sub-halo. Thus, the mean temperature of the shocked gas will be lower than that estimated in the model described in Sect. \ref{sec:engen}, because part of the kinetic energy has been used up to free gas from sub-halos. As can be seen in  Fig. 7 of \cite{Eckert:2021}, the fraction of gas bound to the halo, goes up with increasing halo mass.
 This implies that the mean temperature of the shock heated gas is progressively reduced as we go up in mass, with respect to what is predicted on the basis of Eq. \ref{eq:kt2}. 
 Through some relatively straightforward algebra we find that the ratio between the "mixed accretion" temperature, $T_{\rm ma}$ and the temperature computed from Eq. \ref{eq:kt2}, can be expressed as:
 \begin{multline}
	{T_{\rm ma}(M_{\rm h}) \over T(M_{\rm h})} =  1 \; - \\ \left( {M_{\rm h}\over M_{\rm h,o}} \right)^{-2/3} \, {\bigintss_{M_{\rm min}/M_{\rm h,o}}^{M_{\rm h}/M_{\rm h,o}} f_{\rm g}(m_{\rm h}) \; m_{\rm h}^{5/3} \; {\rm d}n/{\rm d}m_{\rm h} \; {\rm d}m_{\rm h} \over \bigintss_{M_{\rm min}/M_{\rm h,o}}^{M_{\rm h}/M_{\rm h,o}}  m_{\rm h} \; {{\rm d}n / {\rm d}m_{\rm h}}  \; {\rm d}m_{\rm h}} \; ,
	\label{eq:tred}
 \end{multline}
where ${{\rm d}n / {\rm d}m_{\rm h}}$ is the halo mass function, $m_{\rm h}$ is a renormalized halo mass, $M_{\rm h,o}$ is the mass over which renormalization is carried out, $M_{\rm min}$ is the minimum mass over which integration is extended, for which we adopt a value $5 \times 10^{12}$ M$_\odot$. We note that integration is fairly insensitive to the specific value of $M_{\rm min}$ as long as it is smaller than $\sim 10^{13}$ M$_\odot$. For the gas fraction we adopt the expression presented in Eq. \ref{eq:f_gas} with parameters used in Fig. \ref{fig:f_gas}. We note that the computation is insensitive to small variations on the parameters.  A derivation of Eq. \ref{eq:tred} is provided in App. \ref{sec:app3}.
 Integration of Eq. \ref{eq:tred} leads to variation of ${T_{\rm ma}(M_{\rm h}) / T(M_{\rm h})}$ of a few percent in the range $10^{13}$ M$_\odot$ to $ 10^{15}$ M$_\odot$. More specifically, when going from $10^{14}$ M$_\odot$ to $ 10^{15}$ M$_\odot$ we find a fractional temperature reduction of about 4\%, see Fig. \ref{fig:ap3:tlos}, this is because more massive halos feature a larger fraction of bound gas.  
 The net effect is a slightly lower growth of temperature and therefore entropy with increasing mass, with respect to the case where evacuation of gas from sub-halos is not considered (see Sect. \ref{sec:sub:profwdecph} and Fig. \ref{fig:mod_rel_tree}), leading in turn  to a flattening of the entropy profile in the outer regions of massive clusters.
 The modification to the slope of the entropy profile, $\Delta({ {\rm d}\ln K / {\rm d}\ln r}) \sim 0.04 $, is however very modest and insufficient to reach agreement with observations.
 It is worth pointing out that ours is not the only approach that can be adopted to estimate a correction for mixed accretion. Furthermore, given the very gradual increase in the fraction of bound (or unbound) gas with increasing halo mass, the correction will likely be a modest one, regardless of the approach used.
  
   \begin{figure}
    \hspace{-0.5cm}
 	\centerline{\includegraphics[angle=0,width=9.2cm]{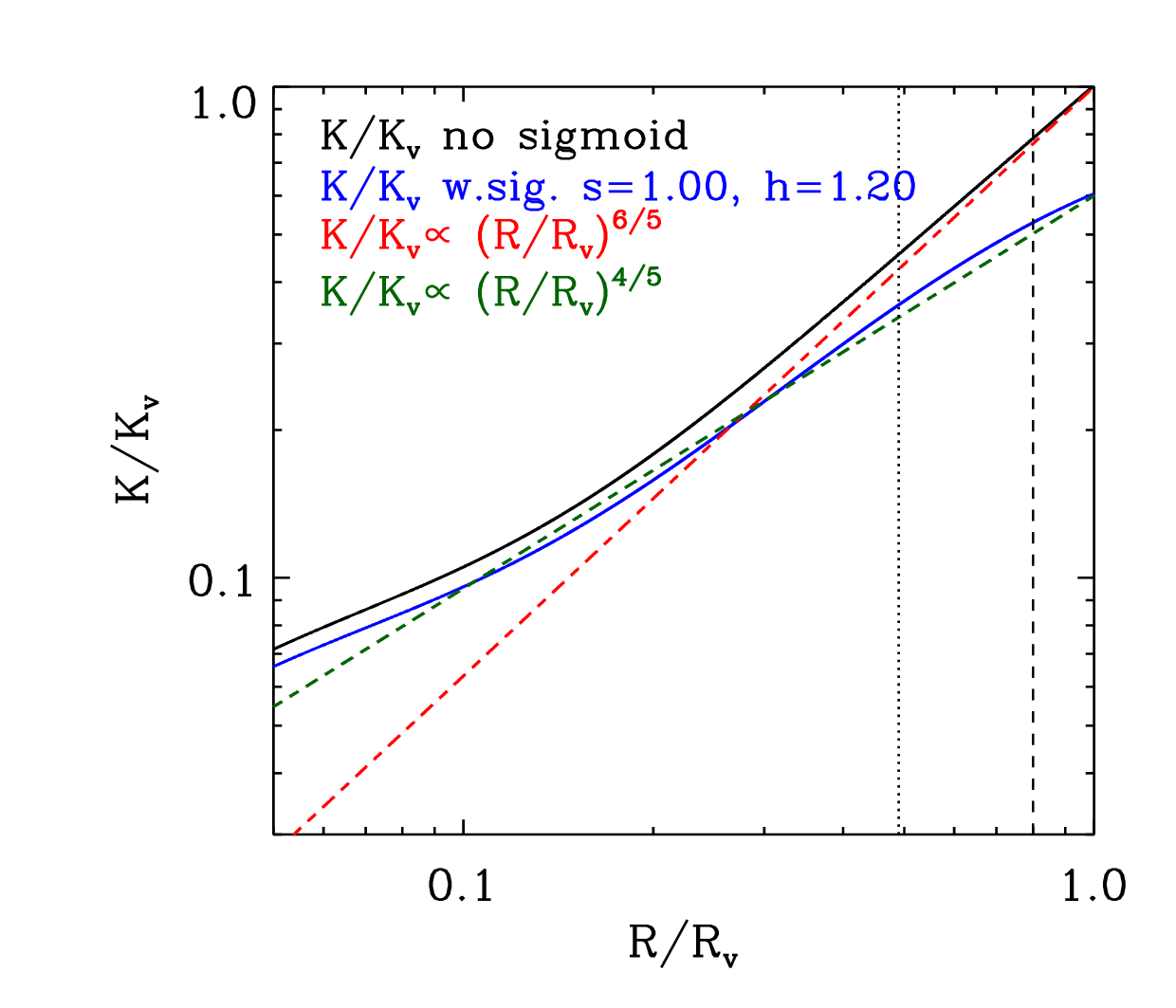}}
 	\caption{Radial entropy profiles for a halo of  $10^{15}$ M$_\odot$ at $z=0$. The black filled line shows the profile in the case of single shock at the virial radius. The blue filled line shows a profile modified with a sigmoid of the from reported in Eq. \ref{eq:sigm} and parameters shown in the figure. The vertical dotted and dashed lines mark the position of $R_{500}$ and $R_{200}$ respectively. The slope of the profiles can be gauged by comparison with the power laws plotted as dashed lines.}
 	\label{fig:k_vs_r_noc_nor}
 \end{figure}

 \subsubsection{From shock radius to shock region}\label{sec:sub:rad_reg} 
 The model described in Sect. \ref{sec:engen} is based on the assumption that all the kinetic energy generated by gas motions within the gravitation field of the forming halo is converted into thermal energy through one shock occurring at one specific radius. 
 There are some good reasons to think this cannot be the case.
  \begin{figure}[t]
 	\hspace{-0.5cm}
 	\centerline{\includegraphics[angle=0,width=9.2cm]{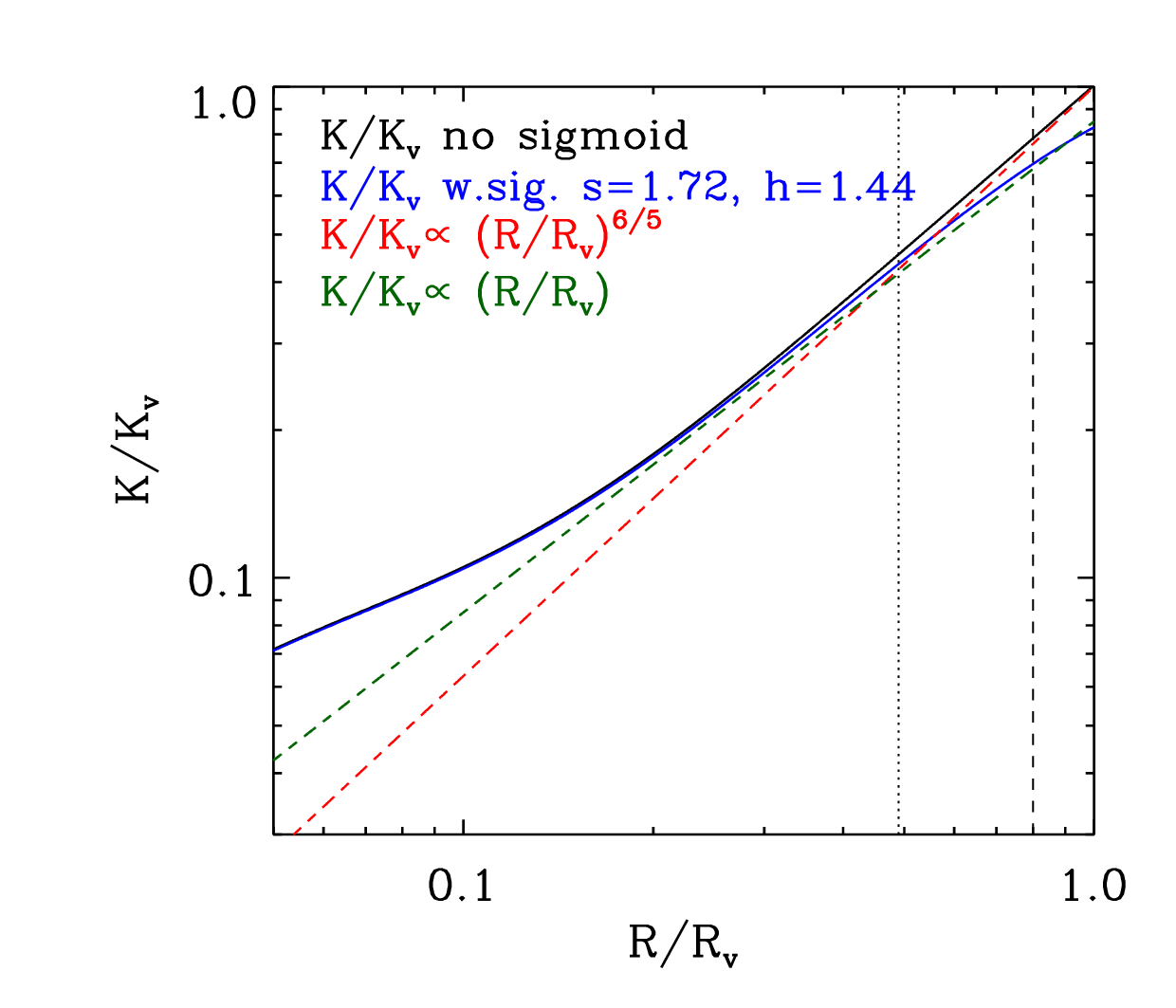}}
 	\caption{Radial entropy profiles for a halo of  $10^{15}$ M$_\odot$ at $z=0$. The black filled line shows the profile in the case of single shock at the virial radius. The blue filled line shows a profile modified with a sigmoid of the from reported in Eq. \ref{eq:sigm} and parameters shown in the figure. The vertical dotted and dashed lines mark the position of $R_{500}$ and $R_{200}$ respectively. The slope of the profiles can be gauged by comparison with the power laws  plotted  as dashed lines.}
 	\label{fig:k_vs_r_noc_nor2}
 \end{figure}
 
\begin{figure*}[!t]
	\centering
	\includegraphics[width = 0.49\textwidth]{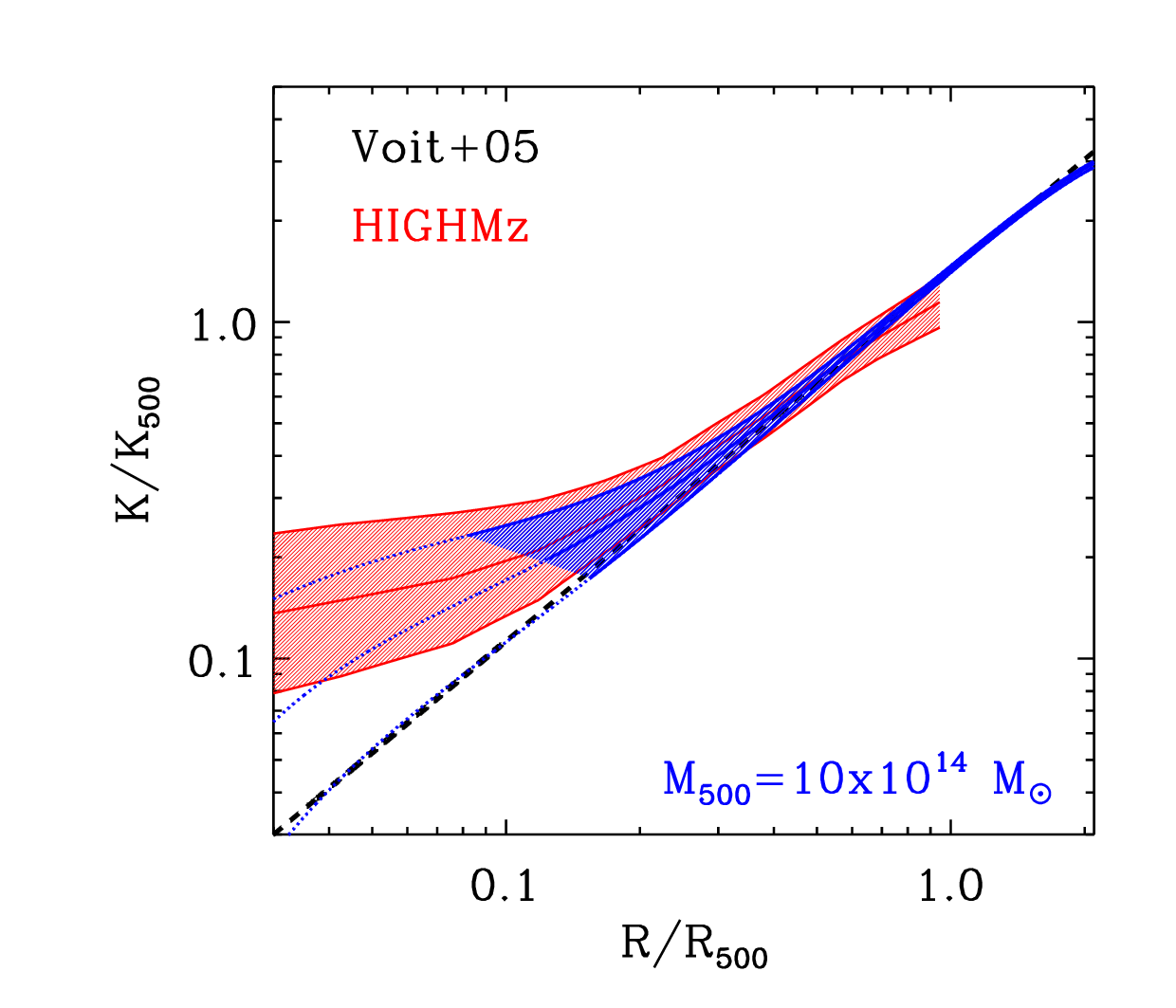}
	\includegraphics[width = 0.49\textwidth]{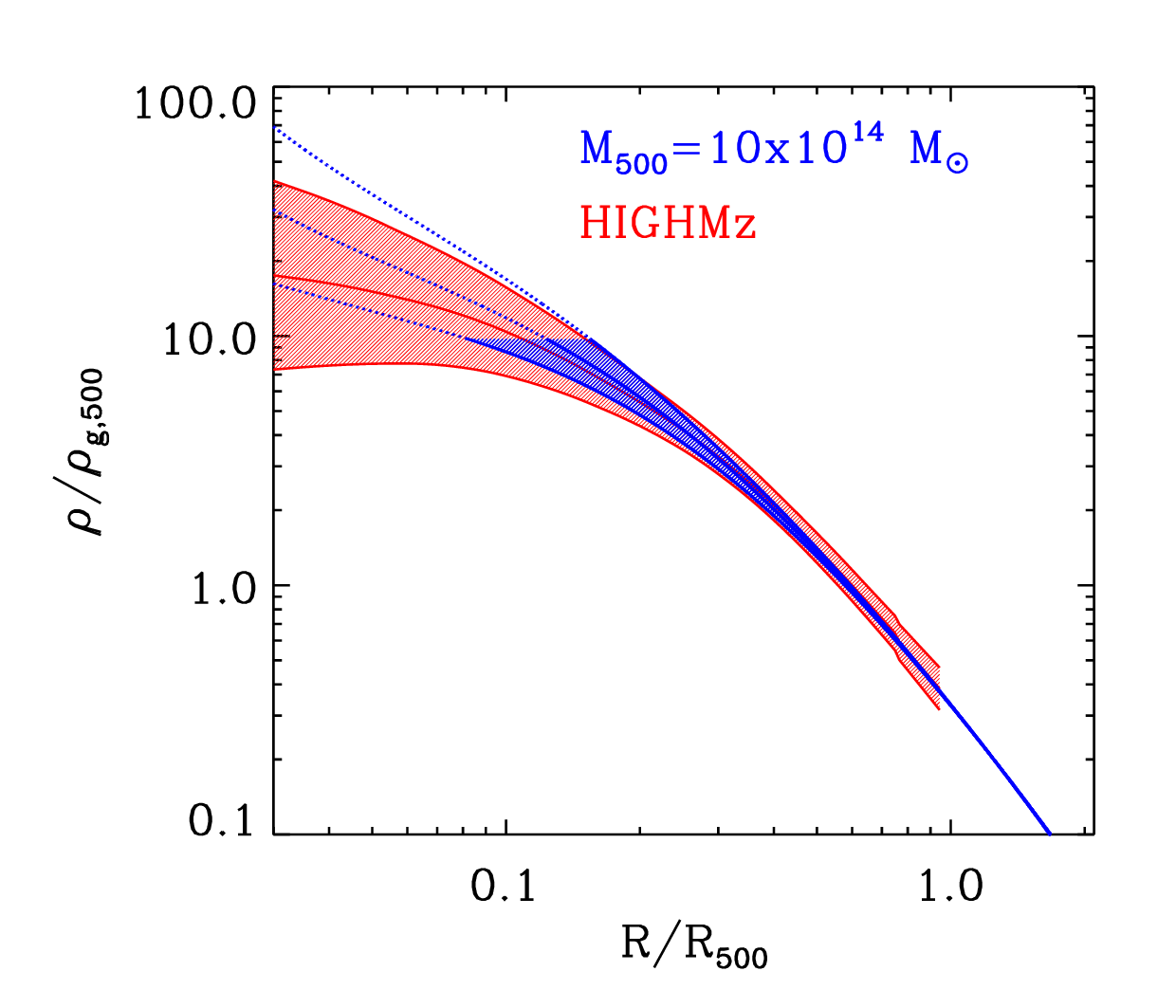}  
	\hfill
	\caption{{\bf Left panel:} Entropy in units of $K_{500}$ versus radius in units of $R_{500}$. The three blue lines are derived from the model presented in Sect. \ref{sec:decoup} with corrections described  in Sects. \ref{sec:sub:mix} and \ref{sec:sub:rad_reg}. All 3 profiles are computed assuming  $ M_{500} = 10^{15}$ M$_\odot$ and for the gas fraction parameters: $l_{\rm m} = 13.2$ and $l_{\rm s} = -0.75$. The $l_{\rm n}$ parameter is set to -1.1, -0.7 and -0.3 respectively for the top, medium and bottom curves. The region between the top and bottom curves is shaded for radii where the cooling time is longer than the time since the gas was shock heated. The red shaded region is  the median entropy and intrinsic scatter from the HIGHMz sample presented in \cite{Riva:2024}. The black dashed line shows the self-similar approximation introduced in \cite{Voit_entropy:2005}. {\bf Right panel:} Density in units of $\rho_{\rm g,500}$ versus radius in units of $R_{500}$. The three blue lines are derived from the model presented in Sect. \ref{sec:decoup} with corrections described  in Sects. \ref{sec:sub:mix} and \ref{sec:sub:rad_reg} using the  same parameters for the gas fraction  adopted for the entropy profile presented in the left panel. The region between the top and bottom curves is shaded for radii where the cooling time is longer than the time since the gas was shock heated. The red shaded region is the median and intrinsic scatter derived from the HIGHMz sample.}
	\label{fig:k_rho_10}
\end{figure*}

 As discussed in Sect. \ref{sec:decoup}, a sizable fraction of the gas falling onto a massive cluster has been previously accreted and ejected by smaller halos. It will be characterized by a distribution in entropy and other thermodynamic variables. Within the infall region, conversion of kinetic into thermal energy through adiabatic compression will operate more effectively in hotter subregions \citep[see Eq. 3 and 3.5 respectively in][]{Tozzi_Norman:2001,Pringle:2007}, this will lead to velocity gradient with cooler subregions falling in more rapidly than hotter ones. As soon as velocity differences become of the order of the sound speed, shocks (i.e. non adiabatic thermalization) will kick in and further amplify velocity gradients, likely leading to a runaway process. Furthermore, the process by which collisionless shocks convert kinetic into thermal energy is far from being well understood \citep[see][and refs. therein]{Goodrich:2023}. 
 For all these reasons, the radial extent of the virialization region is not easy to estimate, so we take an heuristic approach. We substitute the step function operating on entropy in the model described in Sect. \ref{sec:engen} with a sigmoid which we implement through an hyperbolic tangent of the form: 
 \begin{equation}
 	\mathcal{S}(R) =  {1 \over 2}\tanh\left( -s  \; {R - h R_{\rm v}  \over R_{\rm v} }\right) \; + \; {1 \over 2} \; ,
 	\label{eq:sigm}
 \end{equation}
 where $R_{\rm v}$ is the virial radius and $s$ and $h$ are two parameters that control the shape of the sigmoid. Note that $\mathcal{S}(R)$ goes to 1 for $R \ll R_{\rm v}$ and 0 for $R \gg R_{\rm v}$. Note also that the modified entropy profile,  $K^\prime(R)$, is related to the non-modified one through the relation:   $K^\prime(R) = \mathcal{S}(R) \; K(R)$.
 
 We constrain the parameters characterizing the sigmoid through two measurements: 1)
 the  entropy profiles in massive clusters, which, as pointed out above, have flatter slope than those predicted from our model; 2) the fraction of pressure in non-thermal form. The latter has been measured by several authors using different methods.
 \cite{Eckert_non_th_XCOP:2019}, by comparing the hydrostatic mass with the total mass derived from the gas mass under the assumption that the gas fraction of massive clusters is close to the baryon fraction,  find a non-thermal pressure support of $\sim 5\%$  at $R_{500}$ and $\sim 10\%$ at $R_{200}$. \cite{Wicker:2023}, applying a similar approach, come up with an hydrostatic mass bias factor\footnote{The factor $b$ is defined through the relationship $b = 1 - M_{\rm HSE}/M_{\rm T}$, where $M_{\rm HSE}$ is the mass derived from the hydrostatic equilibrium equation and  $M_{\rm T}$ is the total mass of the halo; to first approximation, the non-thermal pressure fraction, defined as the ratio of non-thermal pressure to total pressure is equal to the $b$ parameter \citep{Eckert_non_th_XCOP:2019}.} $b$ of $ \sim 20\%$. \cite{Dupourque:2024} analyzing surface brightness fluctuations on the CHEX-MATE sample, estimate a non-thermal pressure support of $9\pm 6\%$ at $R_{500}$. \cite{Lovisari:2024}, applying a similar approach to temperature maps from the CHEX-MATE sample, estimate a non-thermal pressure support of $10^{+20}_{-10}\%$. Finally, \cite{Munoz:2024}  derive an hydrostatic to lensing mass bias factor of $\sim 25\pm 7 \%$. 
  Given the relatively large uncertainty in the non-thermal pressure contribution suggested by these measurements, we consider two scenarios: one assuming a larger contribution and the other a smaller one.
 
 In Fig.\ref{fig:k_vs_r_noc_nor} we plot the radial entropy profile  for a halo of  $10^{15}$ M$_\odot$ at $z=0$.  The black filled line shows the profile in the case of a single shock at the virial radius, the blue filled line shows the case of a profile modified with a sigmoid  to account for an extended shock region. Comparison with the dashed lines shows that the change introduced by the sigmoid flattens the slope to a value consistent with the one measured in observations. 
 
 The relationship between entropy modification and non-thermal pressure support depends on the details of the virialization process which, as pointed out above, are unknown. Assuming a simplified model in which the gas undergoes gradual heating within a given radial range accompanied by full instantaneous compression, $n^\prime(R) = n(R) $,  we have that the relationship connecting the modified temperature with the unmodified one is: $T^\prime(R) = \mathcal{S}(R) \; T(R) $. Similarly the modified pressure can be expressed as: $p^\prime(R) = \mathcal{S}(R) \; p(R) $ and the fraction of pressure support of non-thermal origin is given by $ (p(R) - p^\prime(R))/p(R) = 1 - \mathcal{S}(R)$. Since $\mathcal{S}(R_{500}) \sim 0.8$ and $\mathcal{S}(R_{200}) \sim 0.7$, the modification shown in Fig. \ref{fig:k_vs_r_noc_nor} implies a relatively high fractional non-thermal pressure support of 20\% at $R_{500}$ and 30\% at $R_{200}$,  comparable to measurements such as  \cite{Munoz:2024} and \cite{Wicker:2023}. Note also that if we relax the condition of full instantaneous compression and write: $n^\prime(R) = \mathcal{R}(R) \; n(R) $, with $0 < \mathcal{R}(R)< 1$, the relation between modified and unmodified pressure becomes: 
 $p^\prime(R) = \mathcal{S}(R) \; \mathcal{R}(R)^{5/3} \; p(R) $ and the fraction of pressure support of non-thermal origin:  $1 - \mathcal{S}(R) \; \mathcal{R}(R)^{5/3} $,  leading to an even larger non-thermal pressure support. 
  
 In Fig. \ref{fig:k_vs_r_noc_nor2} we show that by applying modifications that satisfy tighter constraints on the non-thermal pressure, roughly 5\% at $R_{500}$ and 10\% at $R_{200}$, changes in the entropy profile are more modest, with the slope of the modified entropy profile in the outskirts reaching about 1.0. 
 
We note that the flattening in the entropy profile could have a different physical origin from the one we have proposed above. In such a case our heuristic modeling would need to be reinterpreted within a different framework.
 
By combining modifications discussed in Sect. \ref{sec:sub:mix} with the ones that assume a larger contribution from non-thermal pressure, we come up with an entropy profile that is in good agreement with the observed one. Conversely, if we assume a smaller contribution from non-thermal pressure, we derive an entropy profile that is still steeper than the observed one but only slightly so. 
The difference is sufficiently small to allow us to question if the remaining deviations might be associated with issues on the observational side. We will return to this point in Sect. \ref{sec:sub:clu}.

\begin{figure*}[t]
	{	\centering
	{%
		\label{subfig:a}%
		\includegraphics[width=.38\linewidth]{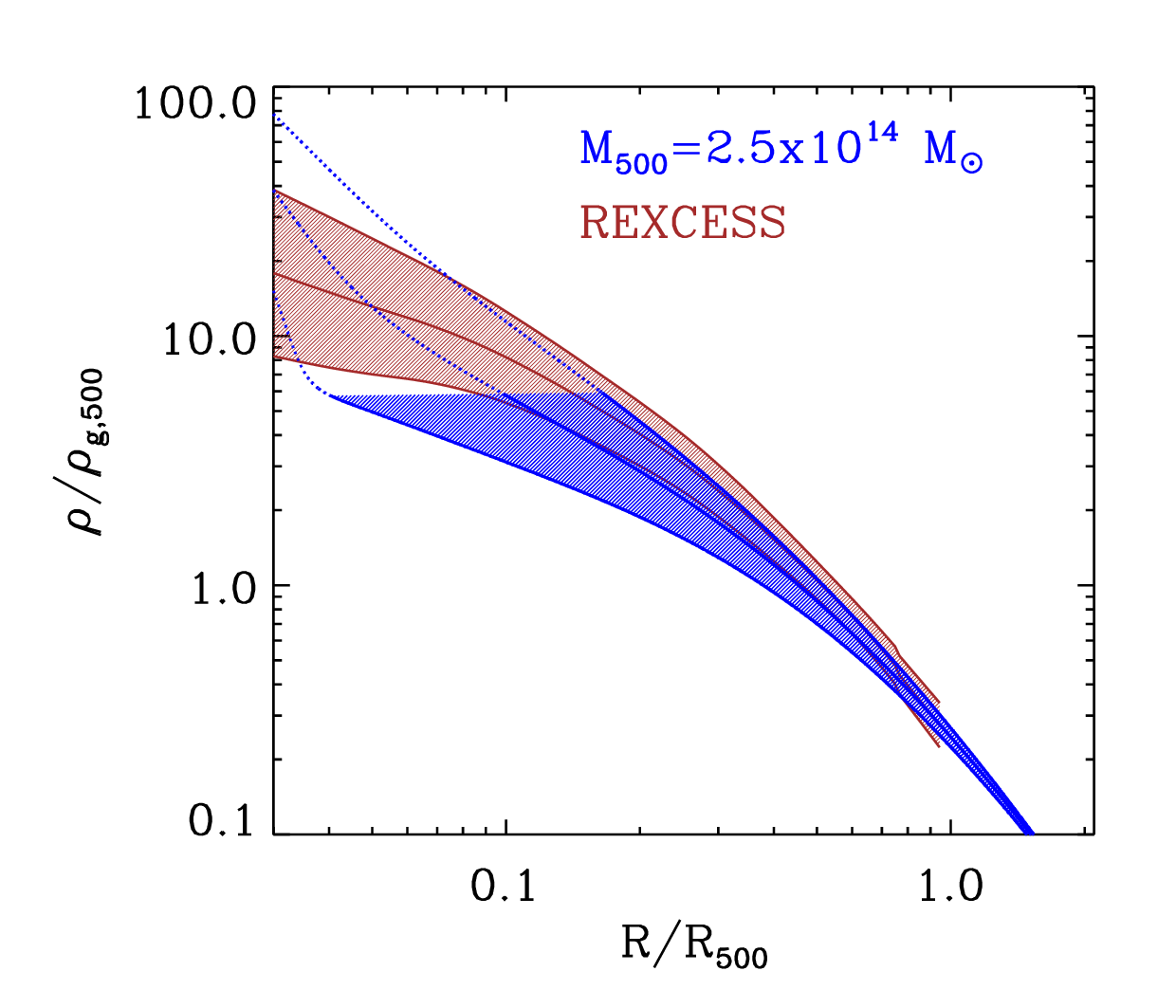}%
	}\hspace{-0.8cm}
	{%
		\includegraphics[width=.38\linewidth]{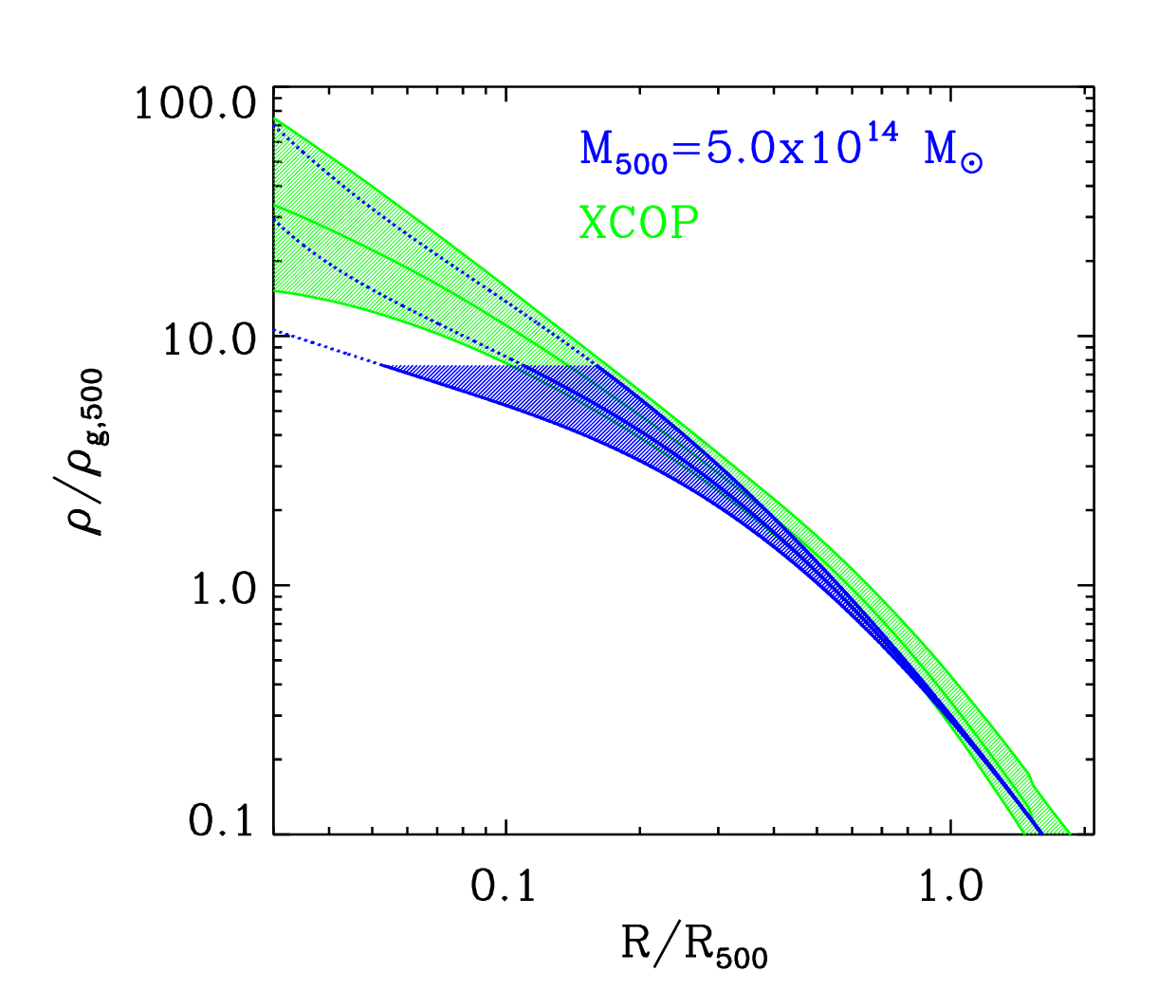}%
		\label{subfig:b}%
	}\hspace{-0.8cm}
	{%
		\includegraphics[width=.38\linewidth]{rho_500_vs_r_wp_noc_riva_10.00_fgas_-1.10_1.20_0.75_fgas_-0.70_1.20_0.75_fgas_-0.30_1.20_0.75_cor3_.ps}%
		\label{subfig:c}%
    }
    }
{	\centering
	{{%
       \includegraphics[width=.38\linewidth]{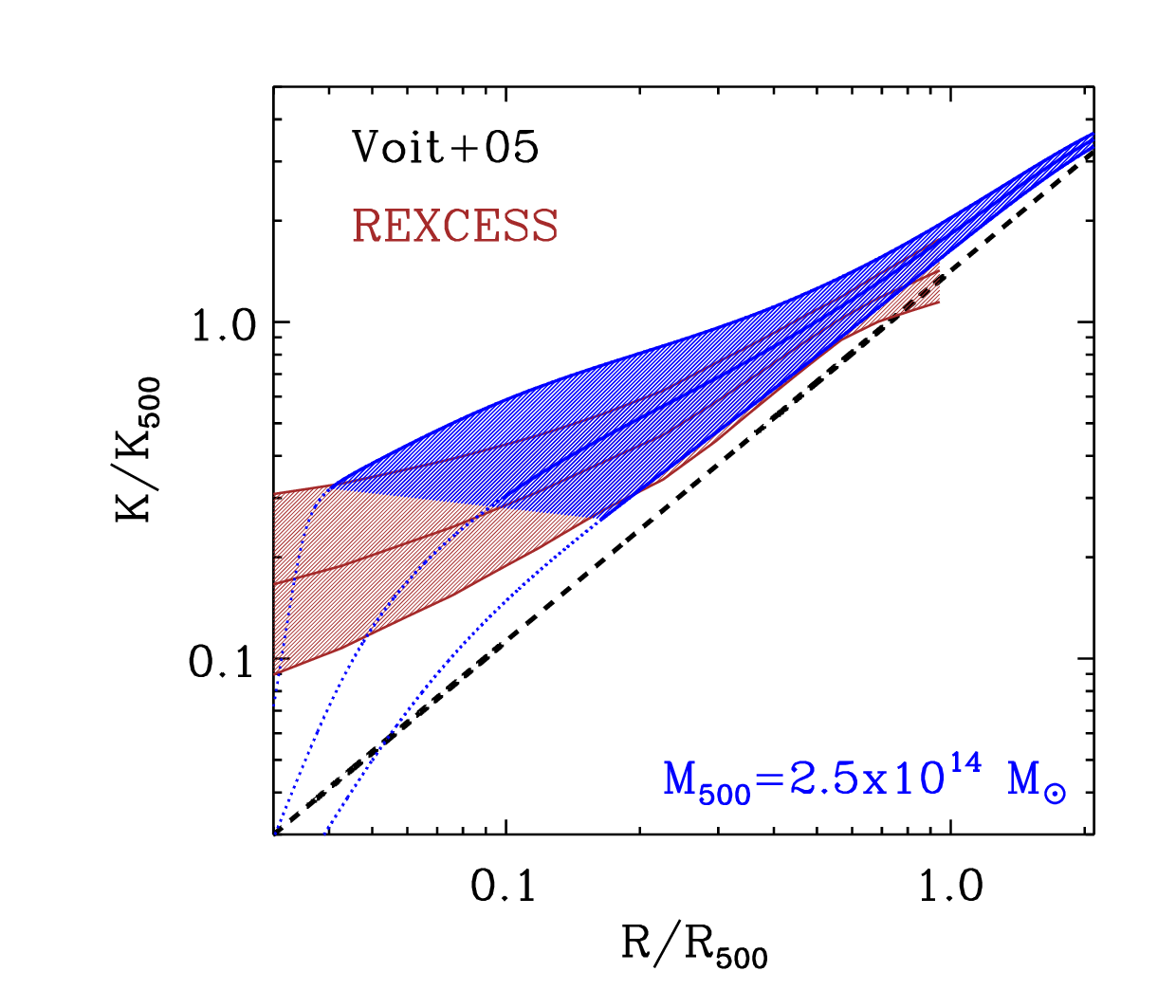}%
		\label{subfig:d}%
	}\hspace{-0.8cm}
	{%
		\includegraphics[width=.38\linewidth]{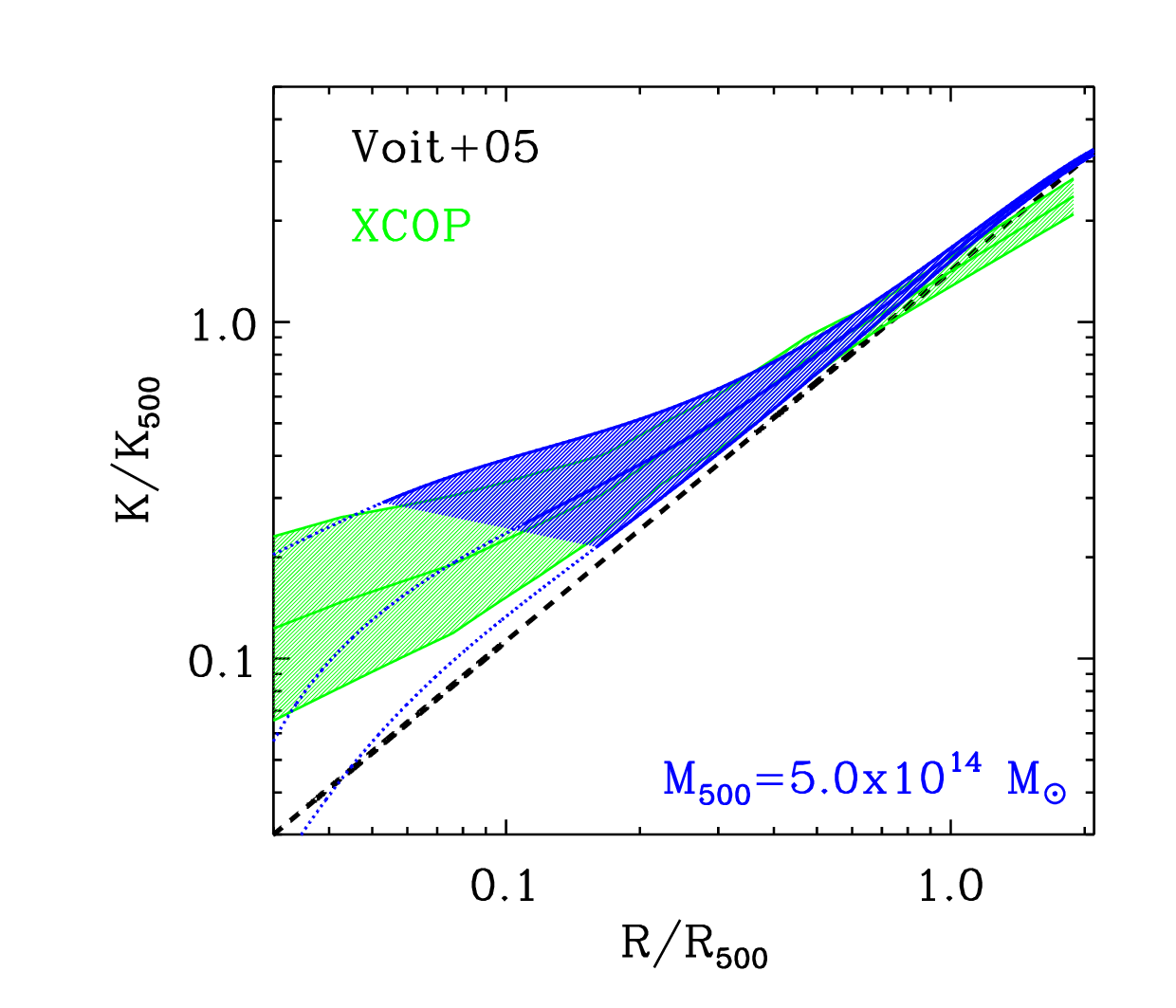}%
		\label{subfig:e}%
	}\hspace{-0.8cm}
	{%
		\includegraphics[width=.38\linewidth]{k_hy_sm_500_vs_r_wp_noc_riva_10.00_fgas_-1.10_1.20_0.75_fgas_-0.70_1.20_0.75_fgas_-0.30_1.20_0.75_cor3_.ps}%
		\label{subfig:f}%
	}
}
}
	\caption{{\bf Top panels:} Gas density profiles from the model presented in Sect. \ref{sec:decoup} with corrections described  in Sects. \ref{sec:sub:mix} and \ref{sec:sub:rad_reg} for: $M_{500} = 2.5\times10^{14}$M$_\odot$ (left), $M_{500} = 5.0\times10^{14}$M$_\odot$ (center) and $M_{500} = 10^{15}$M$_\odot$ (right), compared with observations from the REXCESS, XCOP and HIGHMz samples respectively. For further details see the label of Fig. \ref{fig:k_rho_10}. {\bf Bottom panels:} Same as top panels for the entropy profiles.}
	\label{fig:k_rho_clu}
\end{figure*}

\subsection{Comparison with Clusters}\label{sec:sub:clu}
We compare predictions based on the model presented in Sect. \ref{sec:decoup} with modifications described  in Sects. \ref{sec:sub:mix} and \ref{sec:sub:rad_reg} (see also Fig. \ref{fig:mod_rel_tree})  with 3 different cluster samples: HIGHMz \citep{Riva:2024}, XCOP \citep{Ghirardini:2019} and REXCESS \citep{Pratt:2010} with median halo masses roughly scaling by a factor of 2, see \cite{Riva:2024}\footnote{All entropy profiles have been extracted from  \cite{Riva:2024}, density profiles are extracted from the papers where they were first presented.}.  
Since the choice between larger and smaller non-thermal pressure support discussed in Sect. \ref{sec:sub:rad_reg} has an impact only in the outermost regions we consider only one namely the latter. 

Before proceeding with the comparison, we corrected the observed profiles for a hydro-static scaling bias of 15\%, b=0.15 \footnote{Moderate ($\pm 0.1$) variation on this quantity do not change our results in a substantial way.}.
We do not limit our comparison to entropy, we also include another thermo-dynamic variable, we choose density, for reason that will become apparent later. Since the remaining two variables can be derived from the former, the comparison is informative of thermo-dynamic variables in general.

In Fig. \ref{fig:k_rho_10} we show the comparison between entropy and density profiles predicted from our model and those measured from the HIGHMz sample \citep{Riva:2024}, the density profiles are renormalized through the characteristic density $\rho_{\rm g,500}$ defined by the relation: $\rho_{\rm g,500} = 500 f_{\rm b} \rho_{\rm c}$, where $f_{\rm b}$ is the cosmic baryon fraction and $\rho_{\rm c}$ is the critical density of the Universe at redshift zero.  The model was computed for a mass enclosed within an overdensity of 500 of $10^{15}$M$_\odot$. 
The agreement between observed data and model has been achieved by operating on the gas fraction parameters (see Eq. \ref{eq:f_gas}). This was not done through a formal fitting procedure but by adopting a trial and error approach. 
At the end of this procedure, two of the three parameters have been fixed to the following values: $l_{\rm m} = 13.2$ and $l_{\rm \sigma} = 0.75$.
The shaded blue region is obtained by allowing the third parameter, $l_{\rm n}$, to vary in the range [-1.1,-0.3], with a central value of -0.7. This is an attempt to explain the scatter within the observed sample as a variation in $l_{\rm n}$.  Note that shading stops at $R \sim 0.1 R_{500}$, for smaller radii the cooling time is shorter than the time since the gas was shock heated and  entropy is reduced via radiation, something we do not account for in our model. For the radial region where the model can be meaningfully compared to observation, $ 0.1 R_{500} \lesssim R \lesssim R_{500}$, the agreement is quite good both for the entropy and the density (and therefore for all thermo-dynamic profiles).

It is worth pointing out that there is a certain degree of degeneracy between the 3 parameters describing the gas fraction versus halo mass relation.  For example, a more gradual increase of the gas fraction can be obtained by increasing the value of $l_{\rm \sigma}$ or by reducing that  of $l_{\rm m}$.
As already alluded to, we have chosen to reproduce the observed scatter by varying just one parameter,  $l_{\rm n}$. This allows us to associate a physical interpretation to the intrinsic scatter measured in the observed sample.  
More specifically, objects that feature an entropy larger than average are interpreted  as those for which deviation from the self-similar prediction is strongest, $l_{\rm n} \sim -1.1$, conversely, systems that feature an entropy smaller than average, $l_{\rm n} \sim -0.3$, are understood has having the smallest deviations from self-similarity.


Having derived a set of $f_{\rm g}$ parameters that reproduces reasonably well the observed profiles of the HIGHMz sample, we perform a comparison with the lower mass samples XCOP  and REXCESS . Note that in performing these comparisons we change the halo mass, which we fix to a value close to that of the median of the sample, while the parameters describing the gas fraction are fixed to the values determined for the HIGHMz sample. In Fig. \ref{fig:k_rho_clu} we show a comparison of the model with density and entropy profiles for the three datasets. 
For the most part we find good agreement, we will return in the next paragraphs on minor differences.
However, the most important point is that the model reproduces two key features that are found in the data. First: for each entropy profile deviations from the self similar prediction increase with decreasing radius. Second: when comparing entropy profiles, we see that, as we go down in halo mass, observations and model feature an increasing offset with respect to the self-similar prediction.
Within the model that we propose, these two features are closely connected, 
indeed in both instances, the growing departure from self-similarity as we move down in mass or radius stems from the increasing decoupling of baryon accretion from dark matter accretion.  It is this result, much more than the quality of the individual fits, that suggests to us that baryon decoupling captures a key feature of entropy generation.



The discrepancy between the modeled and observed entropy at large radii could be resolved by assuming a greater contribution from non-thermal pressure support (see Sect. \ref{sec:sub:rad_reg}). This would suggest a substantial non-thermal pressure fraction, $\sim 20\%$ at $R_{500}$. However, we are cautious about adopting this interpretation because the deviation between the model and the data (see Fig. \ref{fig:k_rho_clu}, bottom panel) is at the threshold of statistical significance and background related systematics may play a role \citep[see][]{Rossetti:2024}. Therefore, we find it more prudent to conclude that, within the scope of our model, current entropy measurements are consistent with a  non-thermal pressure fraction within a range of 5\% to 20\% at $R_{500}$.



\subsection{Comparison with Groups}\label{sec:sub:gro}

We  move on to comparing our model with groups, $M_{500} \lesssim 10^{14}$M$_\odot$. 
Much less is known about these systems. While clusters thermodynamic profiles are reasonably well constrained, the same is not true for groups.
More specifically there is no general consensus on the mean, median and scatter of thermodynamic profiles \citep[see][and refs. therein]{Popesso:2024}.
One point recognized by all workers \citep[see][]{Eckert:2021,Oppenheimer:2021} is that group thermodynamic profiles feature a significantly larger scatter than those of clusters. We start off by making  a comparison  with the \cite{Sun:2009} sample \citep[see also][]{Sun:2012}, which was drawn form archival data and, although not X-ray selected in a strict sense, is X-ray bright.

\begin{figure*}[t!]
	{\includegraphics[width=.48\linewidth]{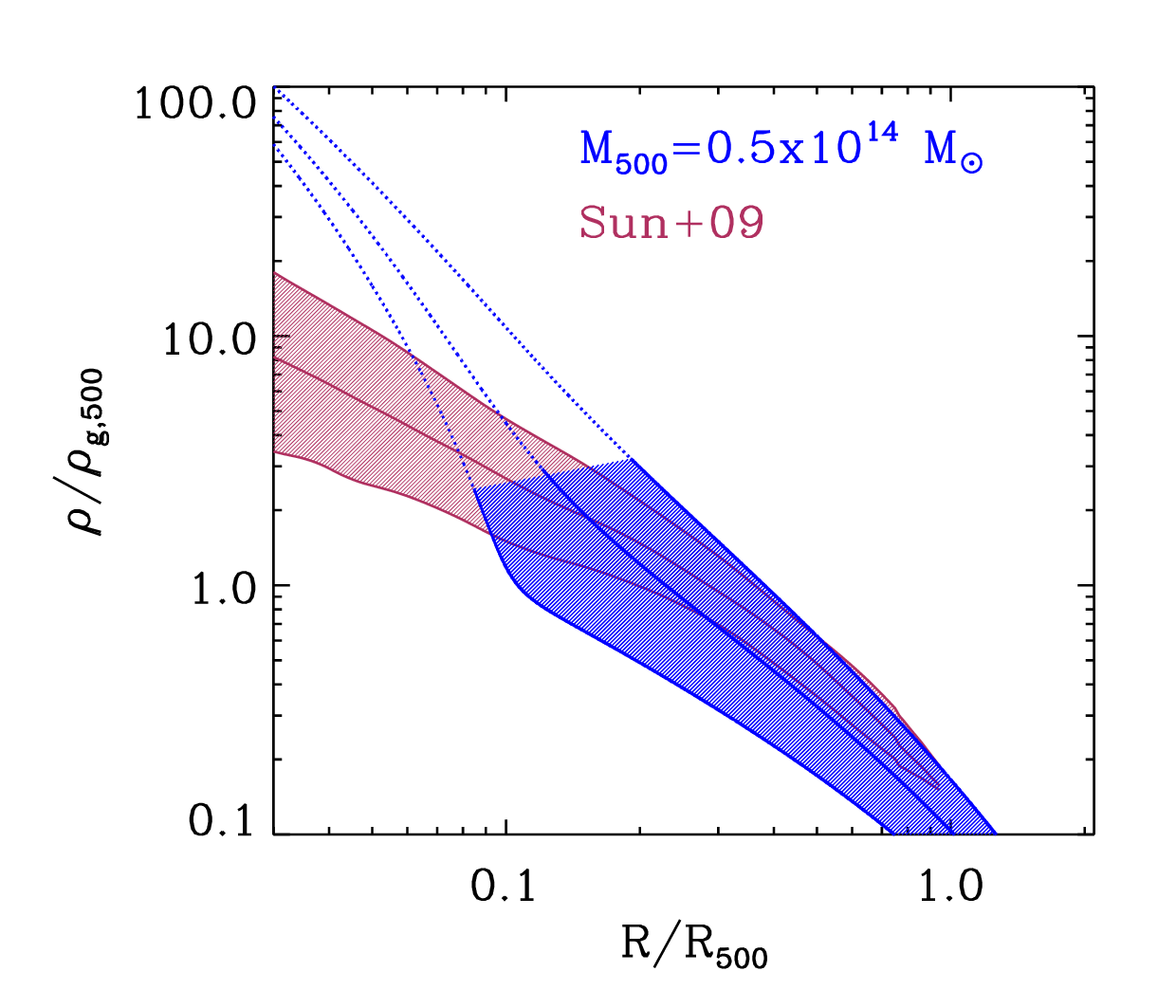}%
	}\hfill
	{\includegraphics[width=.48\linewidth]{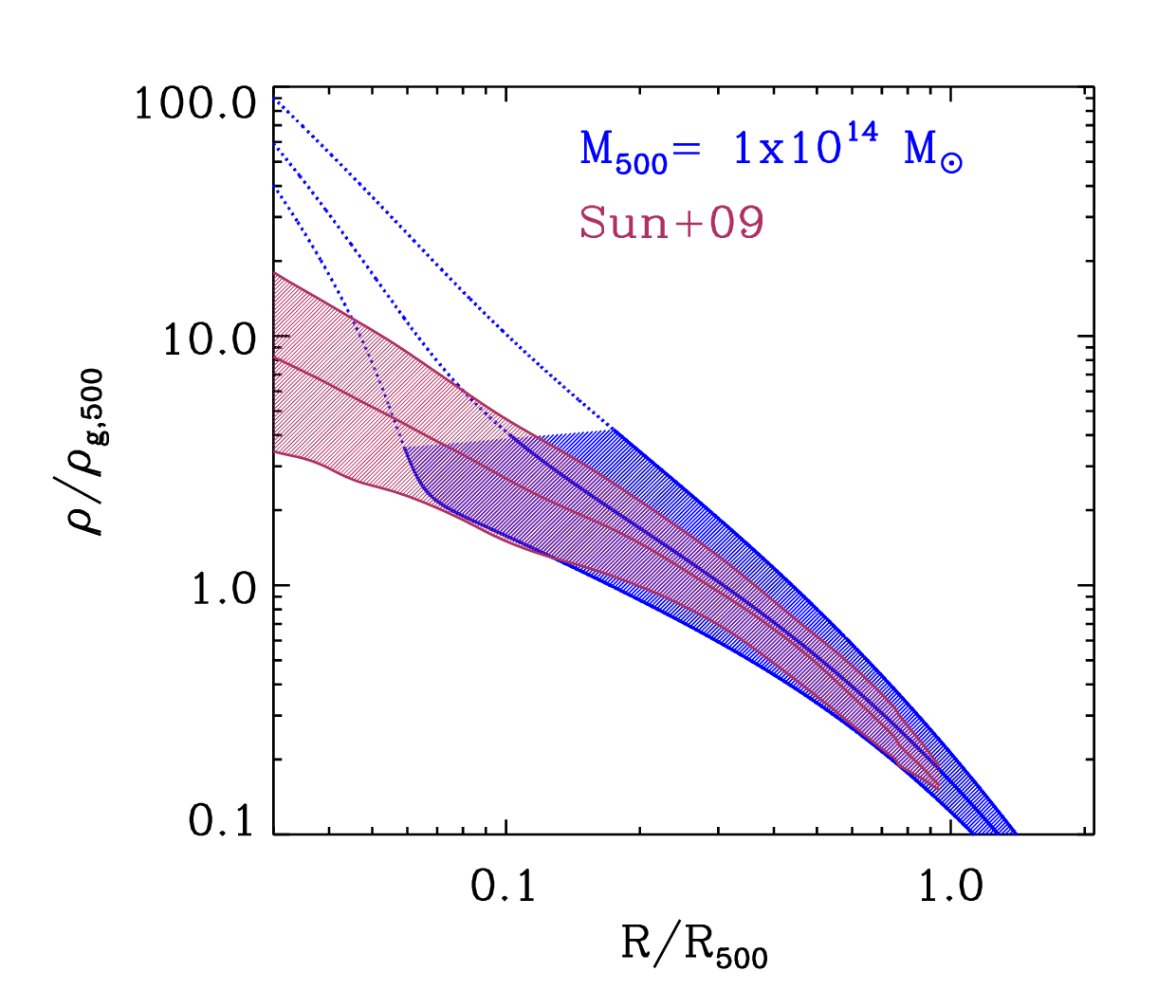}%
	}\hfill
	\\
	{\includegraphics[width=.48\linewidth]{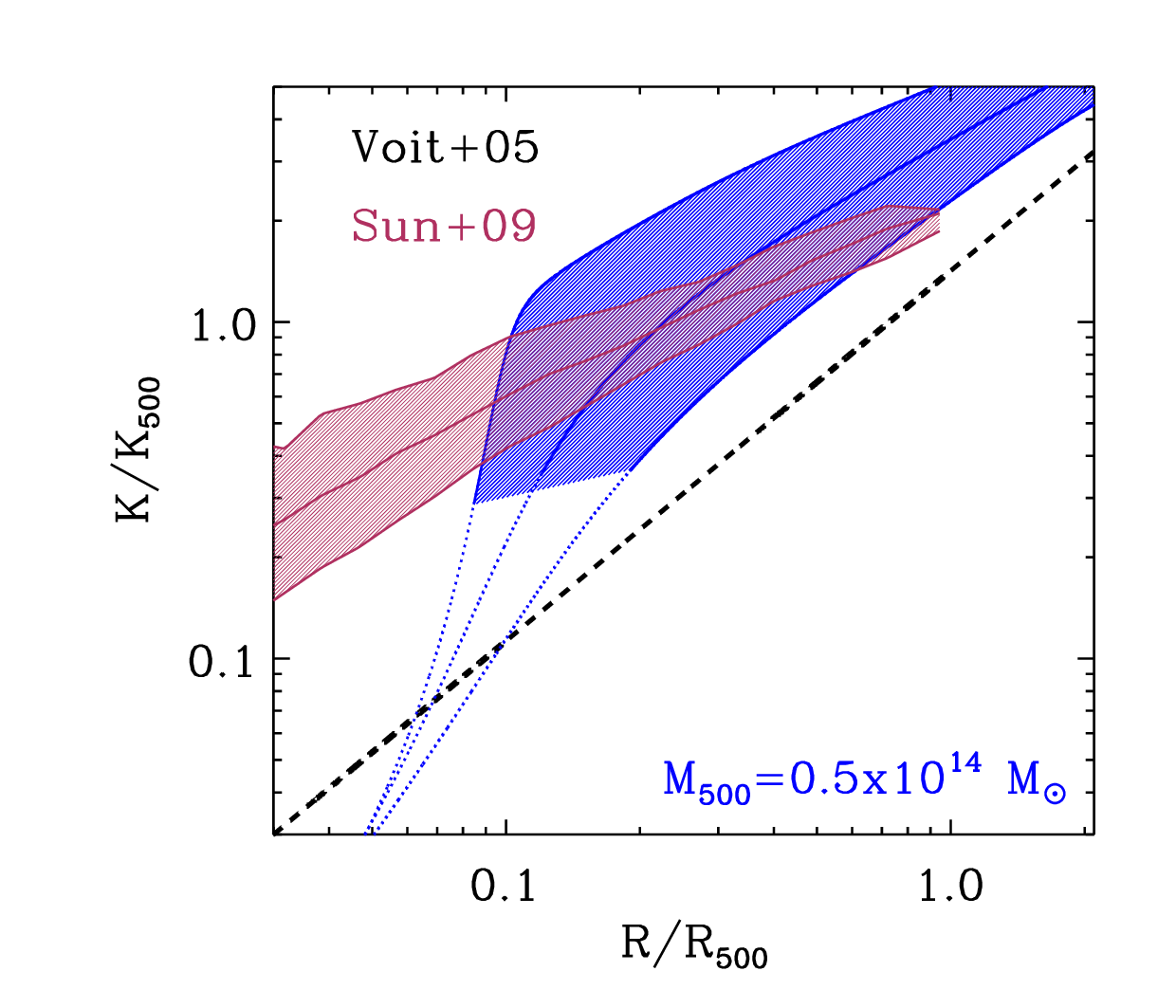}%
	}\hfill
	{\includegraphics[width=.48\linewidth]{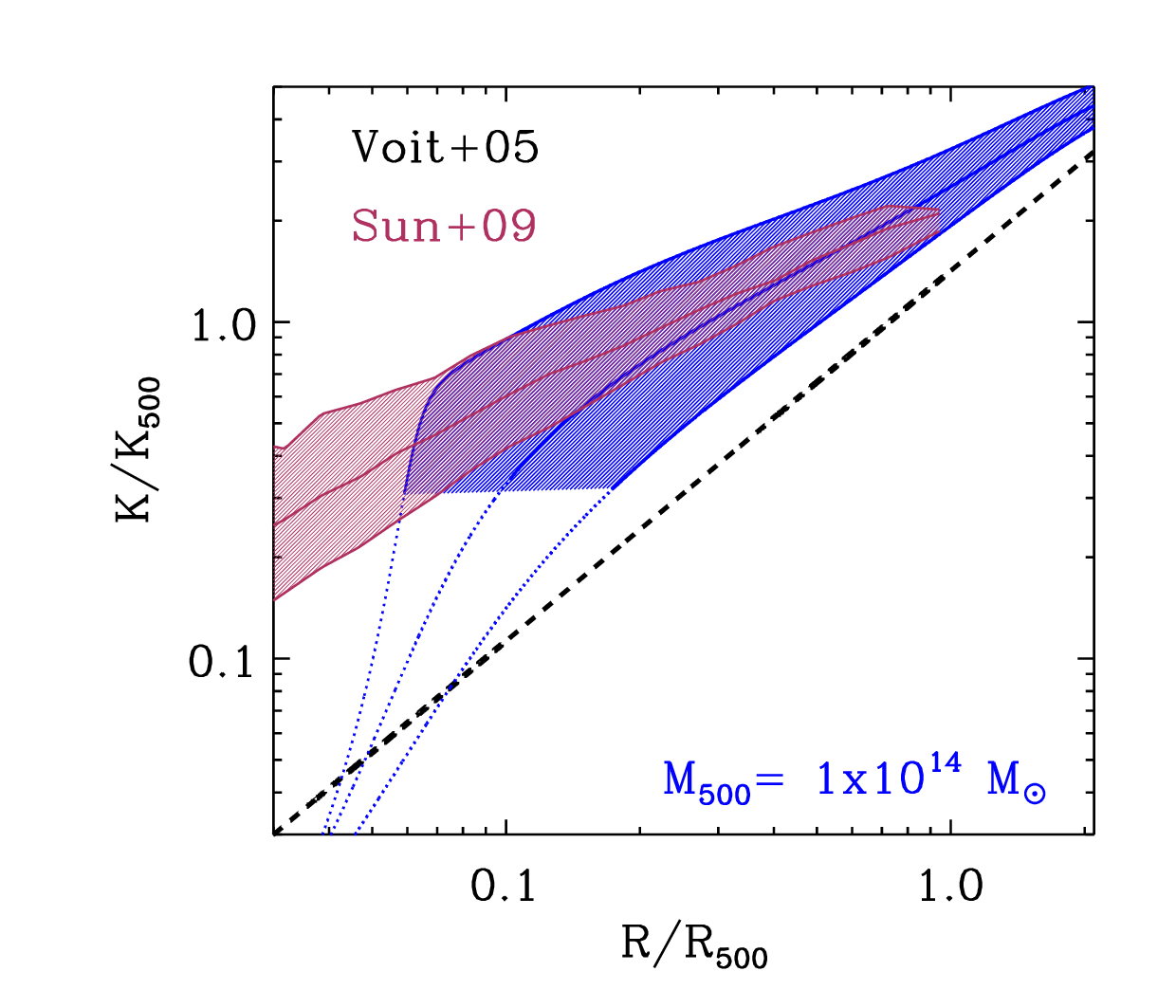}%
	}\hfill
	\caption{{\bf Top panels:} Density profiles predicted from our model for halo masses of  $5.0\times10^{13}$M$_\odot$ (left) and $10^{14}$M$_\odot$ (right), shown as blue shaded regions, compared with observations from the \cite{Sun:2009} sample shown as a red shaded region.  For further details see the label of Fig. \ref{fig:k_rho_10}. {\bf Bottom panels:} Same as top panels for entropy profiles.}
	\label{fig:k_rho_gro}
\end{figure*}

In Fig. \ref{fig:k_rho_gro} we compare our model with observations from the \cite{Sun:2009} sample, the only one for which density and entropy profiles are currently available.
As for the comparison with the XCOP and REXCESS samples no tuning of the $f_{\rm g}$ parameters is made, they are fixed to the values determined by comparison with the HIGHMz sample. Since the sample comprises mostly objects in the $5.0\times10^{13}$M$_\odot$  to $10^{14}$M$_\odot$ range, we compare it with two different realizations of the model: one for  $M_{500} = 5.0\times10^{13}$M$_\odot$ and the other for $M_{500} = 10^{14}$M$_\odot$. The model shows reasonable agreement with the data. As might be expected, since the median mass of the sample is  $ \sim 7\times10^{13}$M$_\odot$, the measured density profile is on the high side of the region predicted by the model for  $M_{500} = 5.0\times10^{13}$M$_\odot$ and on the low side for a halo mass of  $M_{500} = 10^{14}$M$_\odot$. A similar variation, albeit in the opposite direction, is observed for the entropy profiles.
All in all, given that the parameters describing the model where tuned by comparison with data of systems that are 10 times more massive than those considered here, the agreement is remarkable. However, before drawing more general conclusions, we must point out that there are groups which feature very different radial profiles.
 
It has been known for some time that certain systems feature entropy profiles that display a rapid rise in the core followed by a quasi-iso-entropic range extending all the way to the outskirts.
Two such systems are reported in \cite{Baldi:2009}, a few more can be found in \cite{Humphrey:2012}. More recently, several more have been identified in the XGAP sample \citep{Eckert:2024}. 
\begin{figure*}
	\centering
	\includegraphics[width = 0.48\textwidth]{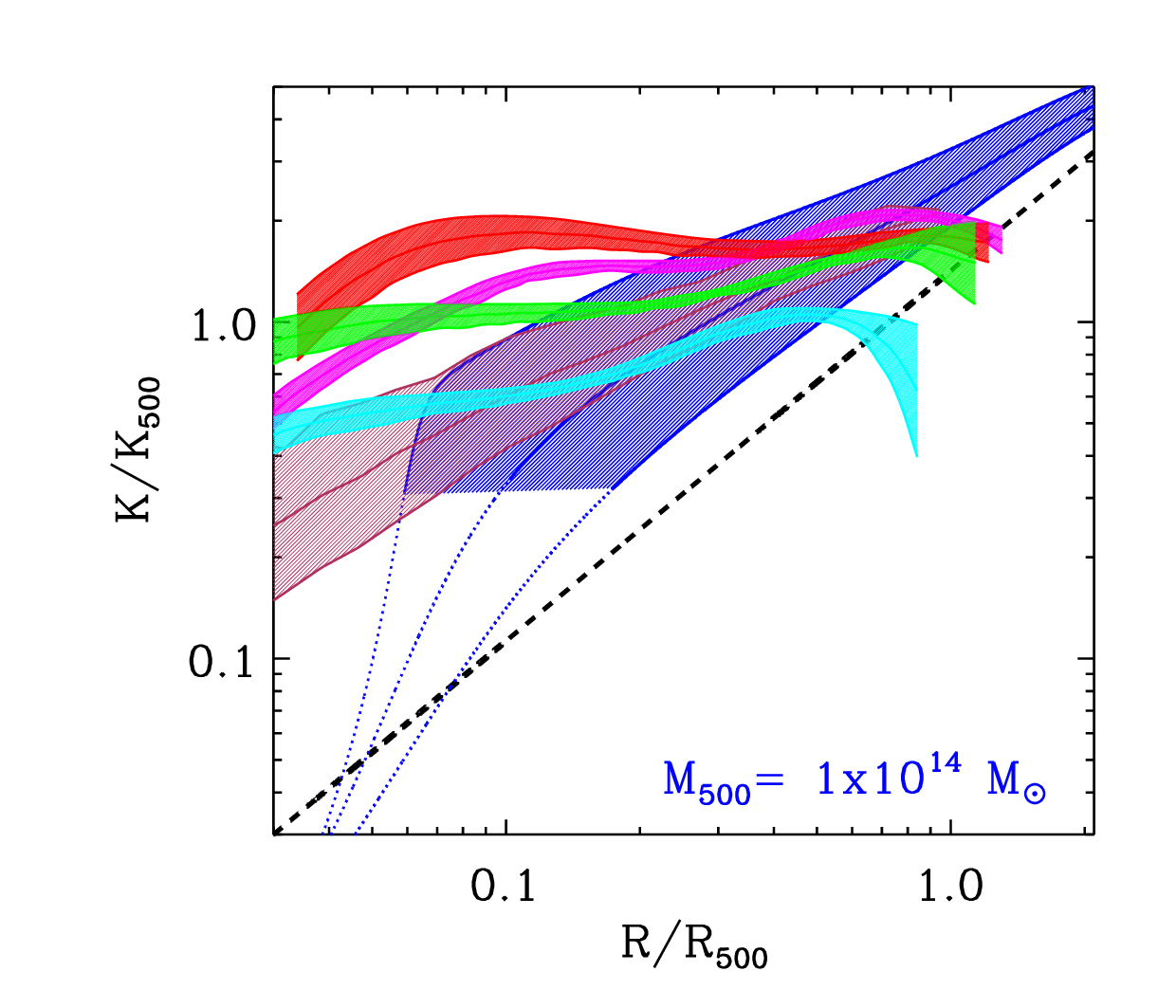}
	\hfill
	\includegraphics[width = 0.48\textwidth]{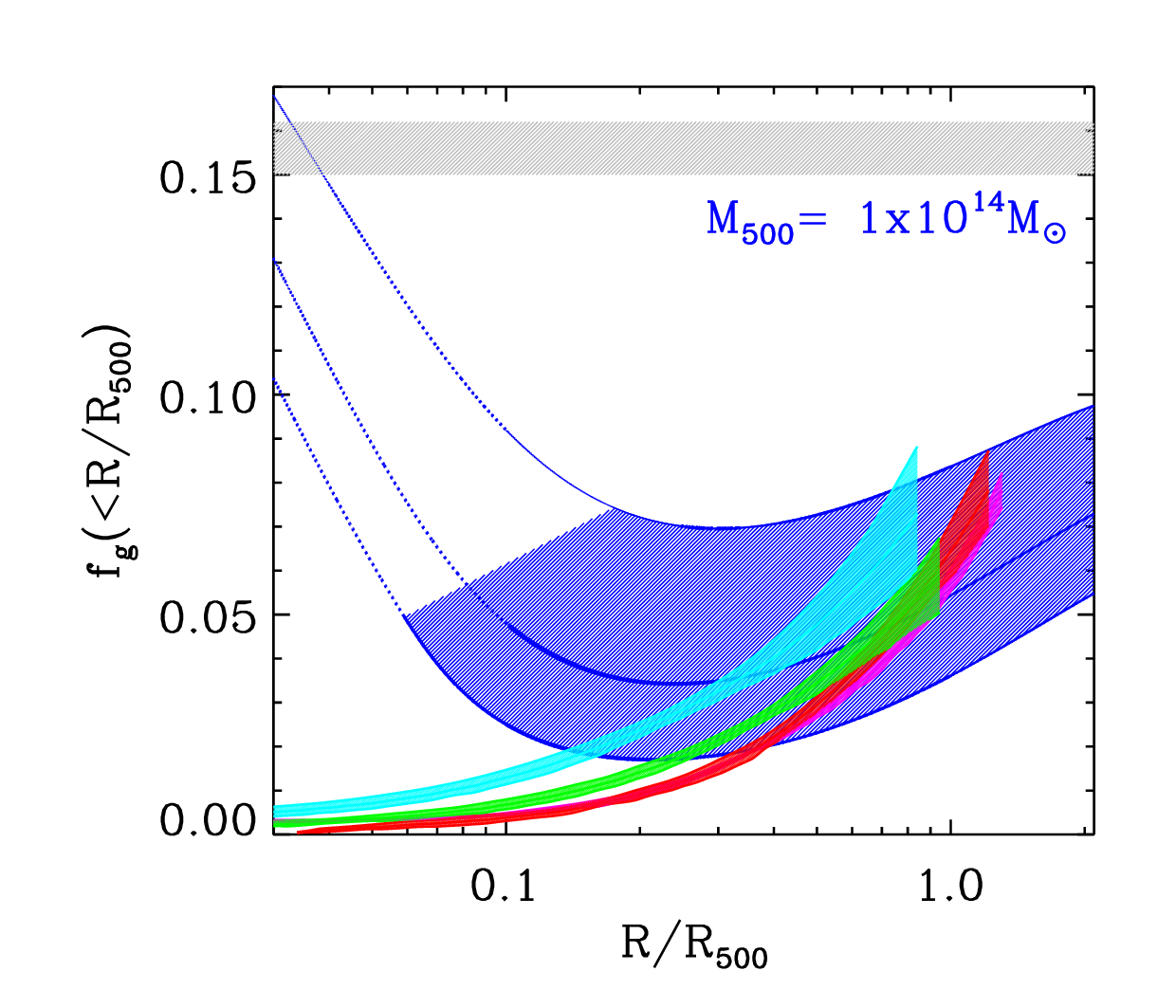}  
	\hfill
	\caption{{\bf Left panel:} Radial entropy profiles.  The blue  and brown shaded regions show respectively, the model, for $M_{500} = 10^{14}$ M$_\odot$, and the \cite{Sun:2009} profiles, already reported in Fig. \ref{fig:k_rho_gro}. The red, green, violet and cyan shaded regions represent profiles reported in \cite{Eckert:2024}. {\bf Right panel:} same as the left panel, for the gas fraction profiles. The gray shaded region shows the cosmic baryon fraction.}
    \label{fig:k_gro_eck}
\end{figure*}

In the left panel of Fig. \ref{fig:k_gro_eck}, we can appreciate the difference between modeled and X-ray bright profiles on one side and those reported in \cite{Eckert:2024}. The latter are all characterized by a rapid rise followed by a flat (quasi-iso-entropic) region. The major difference between the four XGAP profiles is the radius at which the transition occurs and the value of the entropy in the flat region. For radii smaller than 0.1-0.2 $R_{500}$, the entropy is in excess of both model predictions and observed data from the \cite{Sun:2009} sample. The simplest possible explanation is that these systems have recently undergone significant heating events in their cores, the likeliest culprit being the central AGN. Given that our model does not include late-time AGN heating\footnote{As discussed in Sect. \ref{sec:decoup}, our model assumes that early-time AGN feedback is responsible for expelling large amounts of gas in halos with masses around $10^{13}$M$_\odot$. It does not include a treatment for late-time AGN feedback at the group or cluster scales.}, radiative cooling or self regulated Chaotic Cold Accretion \citep[e.g.][]{Gaspari:2020}, it is not surprising that it fails to reproduce the \cite{Eckert:2024} profiles in the central regions. By looking at the right panel of Fig. \ref{fig:k_gro_eck} we see that for radii larger than 0.2-0.3 $R_{500}$, the gas fraction of these systems is consistent with the one derived from our model, suggesting that the late-time AGN heating rearranges gas only within the inner regions of the halos \citep[see][]{Gaspari:2012}.


Although our model does not include a central heating source, it is relatively straightforward to estimate the amount of heat required to account for the entropy excess and gas defect characterizing the core of the \cite{Eckert:2024} systems when compared to our model or the \cite{Sun:2009} sample. To first approximation, the required heat has to be comparable to the thermal energy contained within
0.1-0.3 $R_{500}$. A larger heat input would lead to an entropy excess extending to larger radii.
The total thermal energy is the enthalpy integrated over the region of interest, it can be expressed as:
\begin{equation}
	E_{\rm th}(x) = {15 \over 2} \; p_{500} V_{500} \; \int_0^x   \, \text{\cursive{p}} (\mathrm{x})  \, \mathrm{x}^2 {\rm d}\mathrm{x} \; ,
	\label{eq:therm_ene}
\end{equation}
where: $p_{500}$ is the characteristic pressure \citep[see][for a definition]{Arnaud:2010}, $V_{500} = 4/3 \pi R_{500}^3$, $x = R/R_{500}$ and $\text{\cursive{p}}(x) = p(x)/p_{500}$. Note also that within the integral we used the symbol $\mathrm{x}$ rather than $x$ to avoid confusion with the upper integration limit.
A derivation of the above formula is provided in App. \ref{sec:app4}.
By carrying out the calculation we find the thermal energy within 0.1-0.3 $R_{500}$ to be somewhere between 10\% and 30\% of the thermal energy within $R_{500}$. Furthermore, since the energy required to go from a \cite{Voit:2005} profile to an XGAP one is of the order of the thermal energy content within $R_{500}$, we conclude that: if, prior to the heating event, the XGAP systems had thermodynamic profiles similar to  \cite{Sun:2009} ones, the required energy to make the transition is about 10\% to 30\% of what would be needed if the starting profiles featured a self-similar behavior down to $\sim 0.1 R_{500}$.

\section{Comparison with theoretical work}  \label{sec:theo}

Over the last two decades, there have been several attempts to reproduce observed entropy profiles, sometimes through models, more often via simulations. 
\begin{figure*}[!t]
	\centering
	\includegraphics[width = 0.48\textwidth]{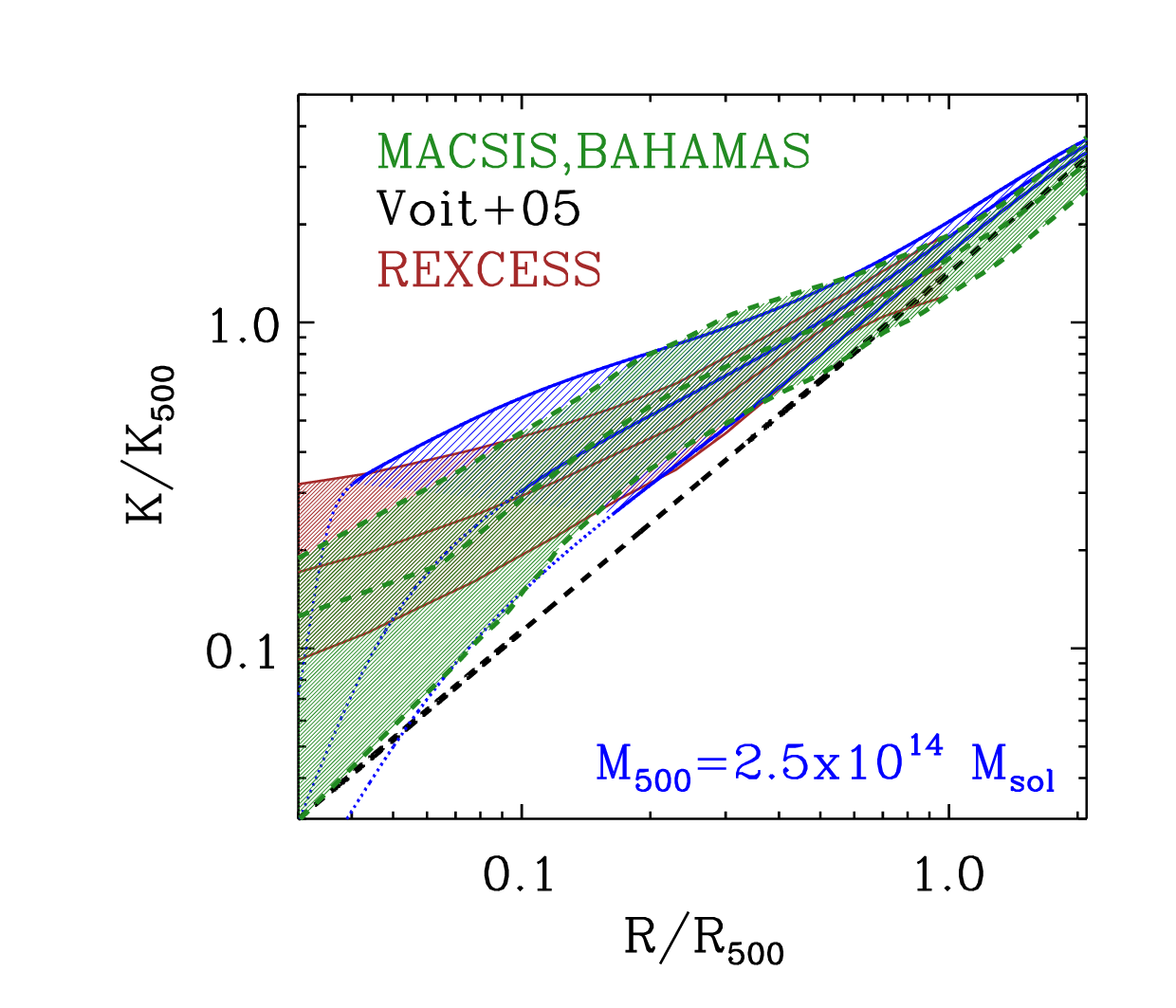} 
	\hfill 
	\includegraphics[width = 0.48\textwidth]{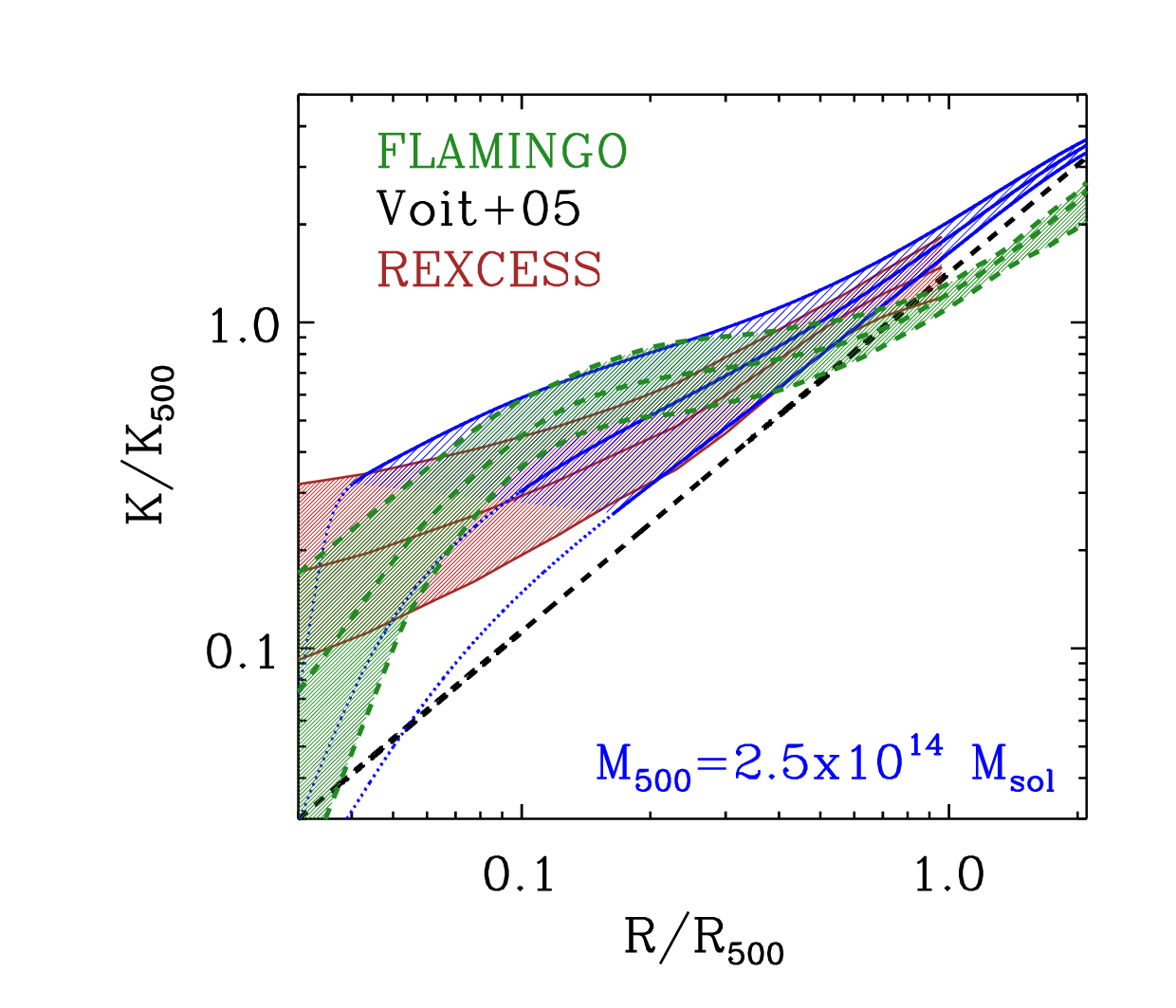}
	\hfill
	\caption{{\bf Left panel:} Radial entropy profiles for the combined sample extracted from the  MACSIS and BAHAMAS simulations \citep{Barnes:2017} compared with the one derived from our model for  $M_{500}= 2.5 \times 10^{14}$M$_\odot$ and with the observed REXCESS profile. {\bf Right panel:} Radial entropy profiles from the FLAMINGO simulation for objects in the $M_{500} = 10^{14}-10^{14.5}$M$_\odot$ mass bin compared with the one derived from our model for  $M_{500}= 2.5 \times 10^{14}$M$_\odot$ and with the observed REXCESS profile.}
	\label{fig:k_vs_r_sim}
\end{figure*}
\subsection{Analytical modeling}\label{sec:sub:ana_mod}
Very recently \cite{Sullivan:2024} presented an analytical model for cluster entropy incorporating non-thermal pressure contributions. They find, as we do (see our Fig. \ref{fig:k_vs_r_bd_wp} and their Fig. 4, left panel) that, at large radii, the predicted entropy profile is steeper than the observed one. As we have done (see Sect.\ref{sec:obs:out}) they interpret this as evidence for non thermal pressure support.  More specifically they  show that the predicted entropy profile can be reconciled with the observed one  by allowing roughly  20\% of the energy content of the gas at $R_{500}$ to be in a non-thermal form (see their Fig. 4, right panel). This is quite similar to what we achieve, through a somewhat more heuristic approach, see our Fig. \ref{fig:k_vs_r_noc_nor}. 

\subsection{Simulations}
In \cite{Barnes:2017}, the authors produce thermodynamic profiles from a large suite of simulations. Their "hot"  subsample, constituted by  massive systems, features a median entropy profile which, beyond the core region, is in good agreement with the self-similar prediction \citep{Voit_entropy:2005}, our own model for the $M_{500}= 10^{15}$M$_\odot$ case and the observed HIGHMz profile \citep{Riva:2024}. Furthermore, the median entropy profile derived from their combined sample, with a median mass $M_{500} \sim 3 \times 10^{14}$M$_\odot$, is in agreement with our model for the $M_{500}= 2.5 \times 10^{14}$M$_\odot$ case and the observed REXCESS profile (see Fig. \ref{fig:k_vs_r_sim}, left panel). According to the authors, the ability of their simulations to reproduce deviations from self-similarity observed in the data has to do with the inclusion of feedback from AGN. Unfortunately, from their paper it is not possible for us to determine how feedback actually achieves that objective.

Recently \cite{Braspenning:2024} have published thermodynamic profiles from the FLAMINGO suite of simulations \citep{Schaye:2023}. Their  entropy profile for the $M_{500} = 10^{14}-10^{14.5}$M$_\odot$ mass bin is compared to our prediction for $M_{500} = 2.5 \times 10^{14}$M$_\odot$ and the REXCESS observed profile in Fig. \ref{fig:k_vs_r_sim}, right panel. As we can see, the FLAMINGO profile appears to be flatter than our own and, more importantly, of the observed profile. The tendency of simulated entropy profiles to be flatter than observed ones has been noted before \citep[see][]{Altamura:2023}. 
As discussed by these authors, there are several possible explanation. One possibility is that numerical viscosity and conduction, present in SPH simulations, but absent from our model, favor mixing and heat transfer in the gas which in turn lead to a reduction of the entropy gradient \citep[see Fig. 6 of][]{Altamura:2023}. As discussed in detail in \cite{Molendi:2023} \citep[see also][]{Zhuravleva:2019} there is significant evidence that both conduction and viscosity are heavily suppressed in the ICM.

\subsection{Scaling relations}
\begin{figure}
	\hspace{-0.5cm}
	\centerline{\includegraphics[angle=0,width=9.2cm]{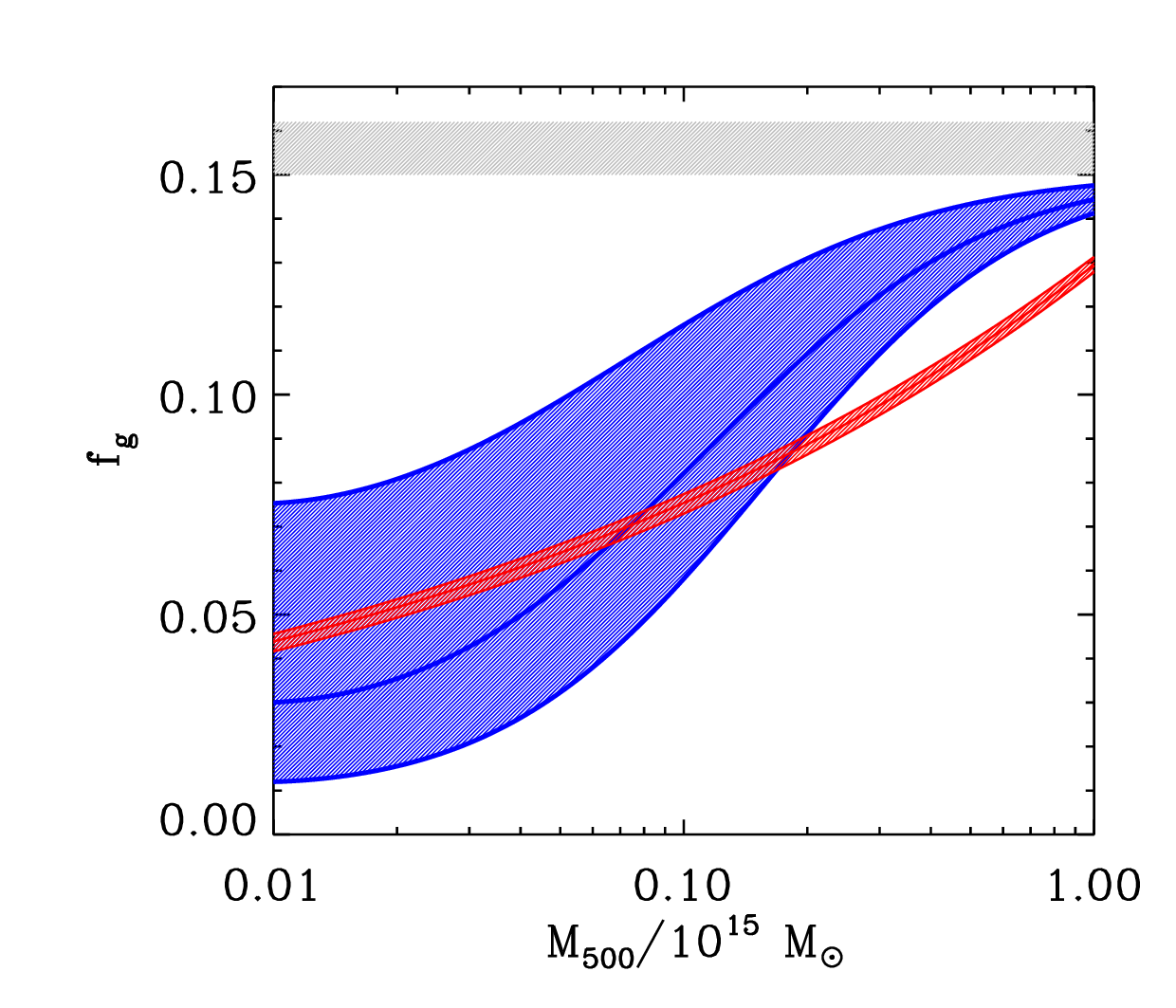}}
	\caption{Gas fraction as a function of halo mass. The solid blue lines show the profiles for the three choices of parameters adopted to fit cluster and group thermodynamic profiles. The blue shaded region spans the region between the different profiles. The red shaded region shows the gas fraction derived in \cite{Ettori:2023} to minimize scatter in cluster scaling relations. The gray shaded region shows the cosmic baryon fraction. 
	}
	\label{fig:fgas_fit}
\end{figure}

A different kind of connection can be made between our model and a phenomenological approach that has been pursued over the last two decades.  Several authors have attempted to reduce the scatter in scaling relations and thermodynamic profiles that is left after applying self-similar scaling, by introducing modifications to the scaling laws \citep[see][for a review]{Lovisari:2022}.
\cite{Pratt:2010} have shown that the scatter on the entropy profile of REXCESS clusters can be substantially reduced by  multiplying it by the gas mass fraction. 
 Recently,  \cite{Ettori:2023}, have identified a correction scheme that modifies the self-similar predictions through two quantities: the gas mass fraction and the ratio between the global spectroscopic temperature and the temperature at $R_{500}$, with the former providing the most significant contribution. 
Interestingly, the authors derive a gas fraction versus halo mass relation that is broadly consistent with the one we derive by fitting the thermodynamic profiles of massive halos, see Sect. \ref{sec:sub:clu} and Fig. \ref{fig:fgas_fit}. It would therefore seem that a model based on baryon decoupling can provide a physical basis for the phenomenological findings reported in \cite{Pratt:2010} and \cite{Ettori:2023}.

Another connection can be made between our model and scaling relations. In \cite{Bower:2008}, the authors employ a semi-analytical model to investigate the luminosity temperature relation. To bring predictions into agreement with observations, they require the gas fraction in less massive halos to be smaller than in more massive ones. Intriguingly, they point out that the major problem for their model is the high level of heating required from the AGN, which follows from assuming that the heating occurs after the system has collapsed. As discussed in Sect. \ref{sec:sub:gro}, baryon decoupling offers a simple and viable solution to this problem: the low gas fraction in less massive systems is achieved by delaying baryon infall rather than by ejecting the gas. 
 
\section{Discussion}\label{sec:discussion}
We have shown that a simple model based on baryon decoupling is consistent with thermodynamic profiles at cluster scale and provides valuable insight on processes operating at the group scale.
In this section we explore the implications of our results. Before doing so we take care of an important  preliminary issues.

\subsection{Is there evidence against a scale of massive gas ejection?}

Inspection of Fig. \ref{fig:fgas_fit} reveals that the deviations from self-similarity in entropy at the cluster scale (see Figs. \ref{fig:k_rho_10} and \ref{fig:k_rho_clu}) stem from a phase of extremely low gas content centered around $10^{13}$M$_\odot$. According to our model, halos at this mass scale are missing between 50\% and 90\% of their baryons, corresponding to a gas fraction ranging between 2\% and 8\%.
As discussed in Sect. \ref{sec:sub:adva}, we interpret this severe depletion as the result of a phase of massive baryon evacuation. This finding is quite significant, and before proceeding, we must critically evaluate whether it aligns or conflicts with current observational data.

The gas content of halos around $10^{13}$ M$_\odot$ remains largely unknown, primarily because the relevant measurements are extremely challenging to secure \citep[see][and refs. therein]{Tumlinson:2017}. 
In a  recent and brave attempt, \cite{ZhangY:2024}, stacking X-ray emission around $\sim 10^5$ SDD galaxies, found that the gas fraction within $R_{500}$ of halos of mass $ \sim 10^{13}$M$_\odot$ is in the order of 0.02-0.03, which is  well within the range of our estimates. A detailed analysis of the systematics associated to the stacking procedure \citep{Popesso:2024b}, while highlighting several criticalities inherent to the technique, confirm the \cite{ZhangY:2024} result, albeit with significantly larger uncertainties: $f_{\rm g}(R<R_{500}) \sim 0.00-0.07$ \citep[][see their Fig. 4]{Popesso:2024c}.   
It is also worth pointing out that, in the few cases where measurements are available on individual systems, they are likely not representative of the general population \citep[see][and refs. therein]{Popesso:2024}. Furthermore, we do have indirect evidence that large amounts of enriched gas are expelled at these scales, \citep[see][]{Molendi:2024}.  We conclude that, at least for the moment, the little  observational evidence that is available does not challenge our inference of substantial gas mass ejection at the $10^{13}$M$_\odot$ scale.

\subsection{Understanding entropy in massive halos}
The success of our model in reproducing thermodynamic profiles of massive halos allows us make a few inferences about the underlying physics.
The rough agreement between observed and modeled entropy and density in the outskirts of massive clusters (see Fig. \ref{fig:k_vs_r_obs}), combined with evidence for moderate non-thermal pressure support in the $R_{500}$-$R_{200}$, see discussion at the end of Sect. \ref{sec:sub:clu}, tells us that, to first order: 1) entropy is generated close to the virial radius, through an accretion shock; 2) the processes operating  in such a shock, whatever they may be, must convert the bulk of the kinetic energy into thermal energy in a relatively narrow radial range.

The formation of an accretion shock at the virial radius can be understood in relatively simple terms. We know that within our Universe halos virialize at a specific overdensity \citep[see][and refs, therein]{Voit:2005}. In a collisionless fluid, such as dark matter, virialization is the process by which matter  reorganizes itself by converting its velocity from radial to isotropic.
This  leads to a sharp increase in matter density around the virial radius. Since baryons are as affected by gravity as dark matter is, they also redistribute themselves in phase space like dark matter. However, unlike dark matter, baryons are collisional, which means that the interaction between low density baryons that are radially infalling and higher density baryons on isotropic orbits results in the formation of a shock front. It is for this reason that, as the halo mass grows by orders of magnitude, the shock tracks  the virial radius as it moves outwards. Simulations \citep[e.g.][]{Vazza:2009} suggest the gas undergoes multiple shocks spread out over a large radial range. However, from the point of view of entropy generation, the bulk is likely produced in just one shock, roughly located at the virial radius. 

On the second point there is much less we can say. More than 60 years after their discovery, collisionless shocks in space are still very much the mystery they have always been \citep[see][and refs. therein]{Goodrich:2023}. While we can infer that accretion shocks must be quite efficient in converting kinetic energy into thermal energy, we have little to say about how this is achieved. 

The good agreement between observed thermodynamic profiles  and predictions from our model, over a broad range of halo masses (see Figs. \ref{fig:k_rho_10}, \ref{fig:k_rho_clu} and \ref{fig:k_rho_gro}), tells us that the modifications to the thermodynamics of  massive halos, with respect to the self-similar model, can be explained through baryon decoupling.  More specifically, for halo masses larger than $\sim 10^{13}$M$_\odot$, accreted matter contains a baryon fraction that is smaller than the cosmic one, this results in an increase in entropy with respect to the self-similar model. Since the baryon deficiency is larger for accretion on  smaller halos: 1)  for any given halo, the deviation in the radial profiles with respect to the self-similar prediction is larger at smaller radii and 2) the thermodynamic profiles of less massive halos feature a general offset, with respect to the self-similar model, that is larger than in more massive ones. It is this result, much more than the quality of the individual fits, that suggests to us that baryon decoupling captures a fundamental feature of entropy generation.


\subsection{Non-thermal pressure support}
As discussed in Sects. \ref{sec:sub:rad_reg}, \ref{sec:sub:clu} and \ref{sec:sub:ana_mod}, within our and similar models \citep[e.g.][]{Sullivan:2024}, there is a tight connection between non-thermal pressure support and the slope of the entropy profile in the outer regions of clusters, with steeper profiles requiring a smaller non-thermal pressure fraction. Current estimates place the non-thermal pressure fraction at $R_{500}$ between 5\% and 20\%, while the entropy profile slopes in the $R_{500}$-$R_{200}$ range are estimated  to be between 0.7 and 0.9.
The limitations of current measurements for both parameters hinder our ability to use one to precisely constrain the other, leaving the radial extent of the shock-heated region uncertain.

Progress will require new and improved measurements of the entropy or the non-thermal pressure in cluster outskirts. 
Entropy profiles could be secured through a combination of high sensitivity surface brightness measurements in X-rays and Sunyaev-Zeldovich (SZ) observation in the sub-millimeter range. The former 
with missions such as AXIS \citep{Reynolds:2024}, recently selected for a phase A study by NASA and through the WFI  experiment on board the L-class mission NewAthena currently under study by ESA; the latter  with future facilities such as AtLAST \citep{DiMascolo:2024}. The most effective way of estimating non-thermal pressure is through high resolution spectroscopic observations. The currently operational XRISM observatory \citep{Tashiro_XRISM:2018} might provide some first results, much more will become possible with the XIFU experiment \citep{Peille:2025} on board NewAthena.


\subsection{Baryon decoupling versus pre-heating}
The baryon decoupling mechanism we propose exhibits substantial differences from the pre-heating models of the early 2000s , see Sect. \ref{sec:pre}. In our model, in massive halos, $M_{\rm h} \sim 10^{14} -10^{15}$ M$_\odot$, almost all the heat conferred to the gas is of gravitational nature. 
Although feedback processes remain crucial, they require comparatively little energy; the key is to remove gas from shallow potential wells and allow gravity to do the rest.
 The situation is radically different at the low mass end, $M_{\rm h} \sim 10^{13}$ M$_\odot$, here the energy of the expelled gas is sufficient to prevent, at least in part, its re-accretion.

\subsection{Gas fraction parametrization}
The broad range of minimum baryon content suggested by our fits,  $f_{\rm g}/f_{\rm b} = 0.1-0.5$ for $M_{\rm h} \sim 10^{13}$M$_\odot$, see Fig. \ref{fig:fgas_fit}, likely follows from the stochastic nature of accretion onto the halo and of AGN feedback processes. 
As far as the mass range over which this minimum is reached, it cannot be estimated from our fits alone because they are essentially insensitive to accretion on halos less massive than about $M_{\rm h} \sim 10^{13}$M$_\odot$. As discussed in Sect. \ref{sec:sub:gas_fr}, see also Figs. \ref{fig:mgasdotvsmh} and   \ref{fig:kvsmh2}, the presence of a minimum in the gas fraction versus halo mass relation leads to a marked change in slope in the radial entropy profile. As it happens, for more massive halos (see Fig. \ref{fig:k_rho_clu}), this variation occurs in the core of these systems where gas cooling becomes important and, as indicated by the comparison with data, washes it away. 
For less massive systems (see Fig. \ref{fig:k_rho_clu}), although the variation occurs beyond the region where cooling becomes important, it is still within the reach of feedback from the SMBH \citep{Gaspari:2014b} operating at late-times, i.e. when the halo mass has grown well beyond $10^{13}$M$_\odot$. 

The minimum on $10^{l_{\rm m}}$ can be set at $\sim 5 \cdot 10^{12} M_\odot$ by recalling that the stellar mass fraction versus halo mass relation reaches a maximum around a few $10^{12}$M$_\odot$, for halos at $z \sim 2-3$, which are the progenitors of current groups and clusters, suggesting that up to this scale feedback effects do not prevent star formation by heating up the gas and therefore are unlikely to heat it up to virial temperatures.
The maximum can be estimated at $\sim 3 \cdot 10^{13} M_\odot$ by noting that observed entropy profiles at the cluster mass scale show no evidence of the rapid change in slope discussed earlier in this subsection.

\subsection{Scatter and bias in group samples}\label{sec:sub:bias}

Our analysis indicates that reproducing the scatter in thermodynamic profiles at the cluster scale requires the gas fraction to exhibit an increasing scatter with decreasing halo mass, as illustrated in Fig. \ref{fig:fgas_fit}. 
We propose that this trend arises from a shift in accretion modes.
At the "ejection" scale, around $10^{13}$M$_\odot$, outbursts from the central galaxy's supermassive black hole (SMBH) expel gas from the halo. Due to the stochastic nature of these outbursts, different systems experience significantly varying levels of baryon loss. Over cosmic time, as the halo mass increases, the source of re-accreted baryons changes. Initially, the re-accreted material  originates primarily from the halo's own ejected baryons. However, as the halo mass continues to grow, re-accreted baryons increasingly come from a large number of smaller halos.
Since individual halos expel varying amounts of gas, this process progressively reduces the scatter in the total amount of accreted gas, leading to the trend observed in Fig. \ref{fig:fgas_fit}.

Whatever the origin of the increase in scatter with decreasing halo mass, it implies that for $ M_{\rm h} \lessapprox 5 \cdot 10^{13}$M$_\odot$, a substantial portion of the population must have a very low baryon fraction. Consequently, any X-ray selected group sample is likely to be significantly biased. Notably, a similar conclusion has been reached by other studies using entirely independent methods \citep[see][and references therein]{Popesso:2024}. It is also important to note that this bias may be even more pronounced than our model predicts, as it does not account for late-time AGN heating effects, which are clearly observed \citep[see][]{Eckert:2024}.

 
\subsection{Transitions}
From an observational point of view, the "ejection" scale around $10^{13}$M$_\odot$ provides a physically meaningful divide between lower mass halos dominated by individual galaxies and higher mass multi galaxy systems. 
Below this threshold, stellar mass builds up rapidly from the accretion of cold gas, above it, the accreted gas is predominantly hot, leading to a sharp decline in the central galaxy's star formation rate, often to the point of near cessation. As halo mass increases beyond the ejection scale, the stellar mass associated with sub-halos becomes more prominent, making the system more easily identifiable as a multi-galaxy system \citep[see Fig. 9 of][]{Coupon:2015}.
 
Baryons at the  group and cluster scales clearly show significant differences.
Groups feature a larger scatter in their properties, their baryon fraction is smaller and late-time heating from the central AGN contributes more significantly to their overall thermodynamic structure.
That said, it is important to recognize that there is no clear separation  between these two regimes. As far as we can tell, all these properties slowly vary with halo mass. In this sense the group/cluster scale separation appears to be less fundamental than the galaxy/group one, which is characterized by a change of the primary accretion mode from cold to hot.  

\subsection{AGN Heating}
In the framework we have sketched, the group and cluster scales  ($M_{\rm h} \gtrsim 10^{13}$M$_\odot$) can be viewed as the mass range over which halos reacquire expelled baryons.
One may ask why the AGN feedback that is so effective in ejecting baryons in halos in the few $10^{12}$M$_\odot$  to $10^{13}$M$_\odot$ range seems to be incapable of doing so in more massive halos.
The answer likely lies in the combination of two factors.
One, as shown by the shut down of the star formation process (see Sect. \ref{sec:decoup} and refs. therein), the cold gas reservoir from which stars are formed and the black hole is fed, dries out at these scales never to be replenished. Two, as discussed in \cite{Gaspari:2014b}, the maximum energy that can be extracted from the central AGN is comparable to the binding energy of poor groups, thus, as halos grow in size, SMBH heating is restricted to progressively smaller self similar radii.
Although AGN feedback is incapable of baryon evacuation at the group and cluster scales, it does play a role in shaping the thermodynamics of these systems. At the group scale, where baryon decoupling is more substantial, it provides enough heat to reshape the thermodynamic profiles out to about a 1/5 of the virial radius (see Sect. \ref{sec:sub:mix}). At the cluster scale its effects are contained within the core. 

\section{Summary}\label{sec:summary}
In this paper, we present a semi-analytical model that builds on work from the early 2000s, it assumes  that the bulk of the entropy of the hot gas in massive halos is generated near the virial shock. We investigate how differences between entropy profiles predicted under gravitational collapse alone, known as self-similar profiles, and observed profiles, can be reconciled by incorporating feedback processes. After demonstrating that adding pre-heating to the entropy generation process reduces, but does not fully resolve, the discrepancy between the model and observed data, we explore an alternative form of feedback.
The core idea of our new approach is that by decoupling baryon accretion from dark matter accretion, we can reduce the density and increase the entropy of the accreted gas. We implement this "baryon decoupling" concept through a  parametric model for the gas fraction versus halo mass relation.
Our main finding can be summarized as follows.

\begin{itemize}
		
	\item Our model provides a simple explanation for the effectiveness of  gas fraction rescaling in reducing the scatter observed in entropy profiles of galaxy clusters.

	\item  The difference in slope between predicted, ${ {\rm d}\ln K / {\rm d}\ln r} \sim 1.2$, and observed, ${ {\rm d}\ln K / {\rm d}\ln r} \sim 0.8$, entropy profiles of massive clusters at large radii can be reconciled, for the better part, through the adoption of a heuristic model and the assumption that entropy is generated over a radial range rather than at just at the virial radius. Improved measurements of the entropy profile and the non-thermal pressure support in cluster outskirt are needed to further constrain the size of this region.

	\item Entropy and density profiles, generated from a single set of parameters describing the gas fraction versus halo mass relation, are in good agreement with observations from three different datasets that fully  sample the cluster mass scale.
	
	\item  Our model reproduces the growing departure from self-similarity observed in the data, as we move inward in individual profiles and down in mass, across different profiles. It is this result more than any other, that suggests that baryon decoupling captures a key feature of entropy generation.
	
	\item In our model, deviations from self-similarity in entropy at the cluster scale  stem from a phase of extremely low gas content centered around $10^{13}$M$_\odot$. We estimate that halos at this mass scale are missing between 50\% and 90\% of their baryons, corresponding to a gas fraction ranging between 2\% and 8\%.

	\item  Our model reproduces entropy profiles of X-ray bright groups, but not those of “quasi-isoentropic” groups, the latter clearly require late-time heating from the central AGN, which we do not include. Baryon decoupling does however reduce significantly the amount of AGN heating that is required. 
	
	\item The increase of scatter in the gas fraction, as we move down in halo mass, results from the reduction in the number of halos from which the re-accreted gas has previously been expelled. 
	
    \item To reproduce the scatter in entropy profiles at the cluster scale, the gas fraction at the group scale must exhibit large variations. This implies that X-ray selected samples are likely to be significantly biased against gas poor groups.

	\item  The "ejection" scale around $10^{13}$M$_\odot$, that we have identified by modeling the gas fraction versus halo mass relation, provides a physically meaningful divide between lower mass halos dominated by individual galaxies and higher mass multi-galaxy systems. Below this threshold, stellar mass builds up rapidly from the accretion of cold gas, above it, the accreted gas is predominantly hot and the central galaxy star formation rate is reduced to the point of almost being shut off.
	
	\item The AGN feedback, that is so effective in ejecting baryons in halos in the few $10^{12}$M$_\odot$  to $10^{13}$M$_\odot$ range, is incapable of doing so in more massive halos for two reasons: the cold gas reservoir from which stars are formed and the black hole is fed, dries out at these scales; the maximum energy that can be extracted from the central AGN is comparable to the binding energy of poor groups.
		
\end{itemize}

In this paper, we have conducted what might be described as a "thought experiment". We set out to determine whether a single modification: the decoupling of baryon accretion from dark matter accretion, could account for a specific phenomenon, the observed deviation of entropy profiles from the self-similar prediction. Remarkably, baryon decoupling proves effective in explaining much of the behavior we sought to understand. While certain aspects inevitably require more complex modifications, the simplicity of our model offers valuable insights and a clearer understanding of the underlying processes.
\begin{acknowledgements}
We are grateful to the referee, Mark Voit, for helpful comments that improved the quality of our manuscript.  S.M. thanks former Master student Giacomo Aprea for his contribution to this project.  We acknowledge financial support from INAF mainstream project No.1.05.01.86.13. M.G. acknowledges support from the ERC Consolidator Grant \textit{BlackHoleWeather} (101086804). L.L. acknowledges support from INAF grant 1.05.12.04.01. PT acknowledges support from the Next Generation European Union PRIN 2022 20225E4SY5 - "From ProtoClusters to Clusters in one Gyr".
\end{acknowledgements}

\bibliography{biblio_entropy}

\appendix
\section{Deriving ${\rm d}\ln K_{\rm v}/{\rm d}\ln M_{\rm h }$}
\label{sec:app}
In this appendix we derive the relation presented in Eq. \ref{eq:slope}. We start off with the expression for $K_{\rm v}$  in Eq. \ref{eq:entro_sm2}. As a first step we compute the natural logarithm of $K_{\rm v}$: 
\begin{equation}
	\ln K_{\rm v} =  \ln \left[ {1 \over 3} \, (\pi G^2)^{2/3} \,  f_{\rm b}^{-2/3} \right] \,
	+ \ln  \left( {M_{\rm h} \over \dot{M}_{\rm h}} \right)^{2/3} + \ln  M_{\rm h}^{2/3} \, .
	\label{eq:ap1:lnk}
\end{equation}
Next, by noting that ${M_{\rm h} / \dot{M}_{\rm h}}$ can be rewritten as  $t \; {\rm d}\ln t / {\rm d}\ln M_{\rm h}$ and defining $\alpha $ as: $\alpha \equiv {\rm d}\ln M_{\rm h } /{\rm d}\ln t $ we get:
\begin{equation}
	\ln K_{\rm v} =  \ln \left[ {1 \over 3} \, (\pi G^2)^{2/3} \,  f_{\rm b}^{-2/3} \right] \,
	+ {2 \over 3} \left(\ln t - \ln \alpha + \ln  M_{\rm h}\right)  \, .
	\label{eq:ap1:lnk2}
\end{equation}
Now we compute the derivative of $\ln K_{\rm v}$ with respect to $\ln M_{\rm h}$:
\begin{equation}
	{{\rm d}\ln K_{\rm v} \over {\rm d}\ln M_{\rm h }} =  {2 \over 3} \left({{\rm d}\ln t\over {\rm d}\ln M_{\rm h }} - {{\rm d}\ln \alpha \over {\rm d}\ln M_{\rm h }} +  {{\rm d}\ln M_{\rm h } \over {\rm d}\ln M_{\rm h }} \right)  \, ,
	\label{eq:ap1:dlnk}
\end{equation}
which simplifies to:
\begin{equation}
	{{\rm d}\ln K_{\rm v} \over {\rm d}\ln M_{\rm h }} =  {2 \over 3} \, \left( 1 + {1 \over \alpha} - { {\rm d}\ln \alpha \over {\rm d}\ln M_{\rm h }} \right) \, .
	\label{eq:ap1:dlnk2}
\end{equation}

\section{Applying hydrostatic equilibrium}
\label{sec:app2}
In this appendix we present the numerical method we have adopted to derive thermodynamic profiles under the condition of hydrostatic equilibrium.

\subsection{Thermodynamic variables at the virial radius}
We start by deriving expressions for thermodynamic variables at the shock radius, which is assumed to coincide with the virial radius, $R_{\rm v}$.
Assuming a strong shock, the following relation holds: 
\begin{equation}
	\rho_{\rm g,d} = 4  \rho_{\rm g,u}  ,
	\label{eq:ap2:rho1}
\end{equation}
where $\rho_{\rm g,u}$ and $\rho_{\rm g,d}$ are respectively  the gas density upstream and downstream of the shock. By combining Eq. \ref{eq:ap2:rho1} with equations describing the mass flow (Eq. \ref{eq:mdot}) and the velocity of the accreted gas (Eq. \ref{eq:vac}) we derive the following expression:
\begin{equation}
	\rho_{\rm g,d} \, = \, { 1 \over \pi G^{1/2} } \, R_{\rm v}^{-3/2} \,  M_{\rm h}^{-1/2} \dot{M}_{\rm g}  \, ,
	\label{eq:ap2:rho2}
\end{equation}
where  $M_{\rm h}$ is the halo mass and $\dot{M}_{\rm g}$ the gas accretion rate.
Combining energy conservation (Eq. \ref{eq:kt2}) with the equation describing the velocity of the accreted gas (Eq. \ref{eq:vac}) we derive the following expression for the gas temperature at the virial radius:
\begin{equation}
    k_{\rm b} T_{\rm v}  \, = \, {1 \over 3} \mu m_{\rm p}  {GM_{\rm h} \over R_{\rm v}}\, ,
	\label{eq:ap2:kt}
\end{equation}
where $k_{\rm b}$ is the Boltzmann constant, $\mu$  the mean molecular weight of the gas and $m_{\rm p}$  the proton mass.
Assuming our gas obeys the equation of state for a monoatomic gas: 
\begin{equation}
	P_{\rm g} \, = \, {\rho_{\rm g} \over \mu m_{\rm p}} \,  k_{\rm b} T  \, ,
	\label{eq:ap2:p1}
\end{equation}
and making use of Eqs. \ref{eq:ap2:rho2} and \ref{eq:ap2:kt}, we derive an expression for the gas pressure at the virial radius: 
\begin{equation}
	P_{\rm g,v} \, = \,  {G^{1/2} \over 3\pi}  \, R_{\rm v}^{-5/2} \,  M_{\rm h}^{1/2} \dot{M}_{\rm g} .
	\label{eq:ap2:p2}
\end{equation}
Similarly, starting from the definition of entropy:
\begin{equation}
	K \equiv {k_{\rm b}T \over \mu m_{\rm p} \rho^{2/3}} ,
	\label{eq:ap2:k}
\end{equation}
and making use of Eqs. \ref{eq:ap2:rho2} and \ref{eq:ap2:kt} we get the following expression for the entropy at the virial radius:
\begin{equation}
	K_{\rm v} =  {1 \over 3} \, (\pi G^2)^{2/3} \, { M_{\rm h}^{4/3} \over \dot{M}_{\rm g}^{2/3} } .
	\label{eq:ap2:k2}
\end{equation}

\subsection{Hydrostatic equilibrium equation}
The next step is to take the commonly adopted  form of the equation of hydrostatic equilibrium equation:
\begin{equation}
	{1 \over \rho_{\rm g}} \, {{\rm d} P_{\rm g}\over {\rm d} R} \, = \,- { G M_{\rm h}(<R)\over R^2}  \, ,
	\label{eq:ap2:hy1}
\end{equation}
and, in light of the power-law like nature of thermodynamic profiles, rewrite it as:
\begin{equation}
	{{\rm d} \ln P_{\rm g} \over {\rm d} \ln R} \, = \, -  { \mu m_{\rm p}\over k_{b}T } \; { G M_{\rm h}(<R)\over R}  \;   .
	\label{eq:ap2:hy2}
\end{equation}
Next, following \citep{Tozzi_Norman:2001}, we rewrite Eq. \ref{eq:ap2:hy2} by dividing each variable   by its value at the virial radius:
\begin{equation}
	{{\rm d} \ln p \over {\rm d} \ln x} \, = \, - C \, \tau^{-1} \, {  m(<x)\over x}  \, .
	\label{eq:ap2:hy3}
\end{equation}
where: 
\begin{multline}
	C = \, {G M_{\rm h}\over R_{\rm v} } \, { \mu m_{\rm p}\over k_{\rm b}T_{\rm v} } \;  , \; 
	\tau = T / T_{\rm v} \; ,  \; p = P_{\rm g}/P_{\rm g,v} \; , \\
	m = M_{\rm h}(R)/ M_{\rm h}(R_{\rm v}) \; , \; x = R/R_{\rm v} \; .
	\label{eq:ap2:c}
\end{multline}
By plugging Eq. \ref{eq:ap2:kt} into Eq. \ref{eq:ap2:c} we get:
\begin{equation}
	C = \, 3 \; .
	\label{eq:ap2:c2}
\end{equation}
An advantage of Eq. \ref{eq:ap2:hy3} is that it allows us to assign at a glance a physical interpretation to the numerical term, $C=3$. Since, for $x=1$, all terms on the right side of the equation are 1, $-3$ is the slope of the $\ln p$ versus $\ln x$ relation at the virial radius. 

Equations \ref{eq:ap2:rho1}, \ref{eq:ap2:kt}, \ref{eq:ap2:p1} , \ref{eq:ap2:k} completely describe physical conditions at the virial radius. By imposing hydrostatic equilibrium, we relate variations in thermodynamic quantities with the quantities themselves and, since these quantities are known at the virial radius, so are the variations which  relate  them. Furthermore, by rewriting the hydrostatic equilibrium equation in terms of ${{\rm d} \ln P_{\rm g} / {\rm d} \ln R}$ the two terms on the right hand side of Eq. \ref{eq:ap2:hy2} become proportional to one another when we evaluate Eq. \ref{eq:ap2:hy2} at the virial radius (see Eq. \ref{eq:ap2:kt}). This is what allows us to establish that  ${{\rm d} \ln P_{\rm g} / {\rm d} \ln R}$  computed at $R_{\rm v}$ is a constant. 

Finally, following \citep{Tozzi_Norman:2001}, we rewrite the normalized temperature in Eq. \ref{eq:ap2:hy3} in terms of normalized
pressure and entropy:  
\begin{equation}
	{{\rm d} \ln p \over {\rm d} \ln x} \, = \, - 3 \, p^{-2/5} \, k^{-3/5} \, {  m(<x)\over x}  \, ,
	\label{eq:ap2:hy5}
\end{equation}
where $ k = K/K_{\rm v}$.


\subsection{Gas mass equation}
The next step is to take the differential relation between gas mass within a given radius and the gas density at that radius  : 
\begin{equation}
	 {{\rm d}M_{\rm g} \over {\rm d}R} \, = \, 4 \pi \rho_{\rm g} R^2 \,  ,
	\label{eq:ap2:mgas_vs_rho}
\end{equation}
and subject it to the same sort of massaging we applied to the hydrostatic equilibrium equation.
We start by rewriting Eq. \ref{eq:ap2:mgas_vs_rho} through variables that are 1 at the virial radius:
\begin{equation}
	 {{\rm d}m_{\rm g} \over {\rm d}x} \, = \, C_{\rm g} \, \chi \, x^2 \,  ,
     \label{eq:ap2:mgas_vs_rho2}
\end{equation}
where: 
\begin{multline}
		C_{\rm g} = \, {4\pi R_{\rm v}^3 \rho_{\rm g,v}  \over M_{\rm g,v} }  \; \; ,  \; \;
	m_{\rm g} = M_{\rm g}(R)/ M_{\rm g,v}(R_{\rm v}) \; {\rm and} \\
	 \; \;  \chi = \rho_{\rm g}/\rho_{\rm g,v} \; .
	\label{eq:ap2:cg}
\end{multline}
We rewrite $C_{\rm g}$ in a more convenient form:
\begin{equation}
	C_{\rm g} \; = \; 3 \, {\rho_{\rm g,v}  \over  \langle{\rho_{\rm g,v}}\rangle}  \; \; {\rm where}   \; \; 
    \langle{\rho_{\rm g,v}}\rangle \; = \; { M_{\rm g,v} \over 4/3 \pi R_{\rm v}^3 } \; .
	\label{eq:ap2:cg2}
\end{equation}
Finally, as we have done for the pressure gradient, we rewrite Eq. \ref{eq:ap2:mgas_vs_rho2} in terms of natural logarithms and express density in terms of pressure and entropy:
\begin{equation}
	{{\rm d} \ln m_{\rm g} \over {\rm d} \ln x} \, = \, C_{\rm g} \, \left({p \over k}\right)^{3/5} \; { x^3 \over m_{\rm g}} \;  .
	\label{eq:ap2:mgas_vs_rho3}
\end{equation}
As for the hydrostatic equilibrium equation, see Eq. \ref{eq:ap2:hy3}, this form allows us to understand that $C_{\rm g}$ is the slope of the $\ln M_{\rm g}$ versus $\ln R$ relation at the virial radius:
\begin{equation}
	C_{\rm g} \; = \; \left({{\rm d} \ln M_{\rm g} \over {\rm d} \ln R}\right)_{R=R_{\rm v}} \; .
	\label{eq:ap2:cg3}
\end{equation}

\subsection{Entropy equation} \label{sec:sub:ent}
In our model, entropy generation occurs only at the virial (shock) radius, this implies that any further evolution of the gas will be adiabatic.
The gas may compress or expand to obey the hydrostatic equation, but it will not change its entropy. This implies that the entropy of a gas element within the halo can be expressed as a function of the halo mass and gas accretion rate at the time the element was shock heated, see Eq. \ref{eq:entro_sm}.
In differential terms, we may say that the infinitesimal increase in entropy is determined by the infinitesimal increase in gas mass. Thus, the variable ${\rm d} \ln K/{\rm d} \ln M_{\rm g}$ will not depend on the radius within the halo where the infinitesimal gas element  eventually resides but only on the halo mass at the time the gas was accreted. Furthermore, once a description of how the gas mass, $M_{\rm g}$, varies with halo is provided, $M_{\rm g}= M_{\rm g}(M_{\rm h})$, see Eq. \ref{eq:const_acc} for the self-similar case and Eq. \ref{eq:mdotg} for the baryon decoupling case,  ${\rm d} \ln K/{\rm d} \ln M_{\rm g}$ is completely specified and can be readily computed from $M_{\rm h}$, in mathematical terms:
\begin{equation}
	{{\rm d} \ln K \over {\rm d}\ln M_{\rm g}} \; = \; f(M_{\rm h}) \; .
	\label{eq:ap2:dk_dmg}
\end{equation} 

 
\subsection{Numerical solution}
The differential equations, Eqs. \ref{eq:ap2:hy5}, \ref{eq:ap2:mgas_vs_rho3} and \ref{eq:ap2:dk_dmg}, together with the boundary conditions provided in Eqs. \ref{eq:ap2:rho2}, \ref{eq:ap2:p2} and \ref{eq:ap2:k2}, make up a system that can be solved numerically.
To this end, we make use of the  finite differences method and rewrite Eqs. \ref{eq:ap2:hy5} , \ref{eq:ap2:mgas_vs_rho3} and \ref{eq:ap2:dk_dmg} as:
\begin{align}
	 \Delta \ln p  \;  & = \; - 3 \, p^{-2/5} \, k^{-3/5} \, {  m(<x)\over x}  \; \Delta \ln x \; , \label{eq:ap2:dp}\\
	 \Delta \ln m_{\rm g}  & = \; \; C_{\rm g} \;  \left({p \over k}\right)^{3/5} \;{ x^3 \over m_{\rm g}}  \; \Delta  \ln x \; , \label{eq:ap2:dmg}\\ 
	 \Delta \ln k  \;  & = \; \; {{\rm d} \ln k \over {\rm d}\ln m_{\rm g}}(m(<x)) \;  \Delta \ln m_{\rm g}	\; . \label{eq:ap2:dk}
\end{align}
As previously pointed out, we start from the virial radius where all normalized variables are 1 and $C_{\rm g}$ (see Eq. \ref{eq:ap2:cg2}) and ${{\rm d} \ln k / {\rm d} \ln m_{\rm g}}$ (see discussion in Sect. \ref{sec:sub:ent} and Eq. \ref{eq:ap2:dk_dmg}) are known. We move inwards by selecting a value for $\Delta \ln x$ that is sufficiently small to adequately sample gradients and, through Eqs. \ref{eq:ap2:dp} and \ref{eq:ap2:dmg}, compute respectively values for $\ln p$ and $\ln m_{\rm g}$ at $\ln x = 1 +  \Delta  \ln x$. From  $\Delta \ln m_{\rm g}$, through Eq. \ref{eq:ap2:dk},  we compute  $ \ln k$ at $\ln x = 1 +  \Delta  \ln x$. Next we insert the values of $p$, $m(<x)$, $m_{\rm g}$ and $k$ computed at $\ln x = 1 +  \Delta  \ln x$ in Eqs. \ref{eq:ap2:dp}, \ref{eq:ap2:dmg} and \ref{eq:ap2:dk} and proceed iteratively.

\begin{figure}
	\hspace{-0.5cm}
	\centerline{\includegraphics[angle=0,width=9.2cm]{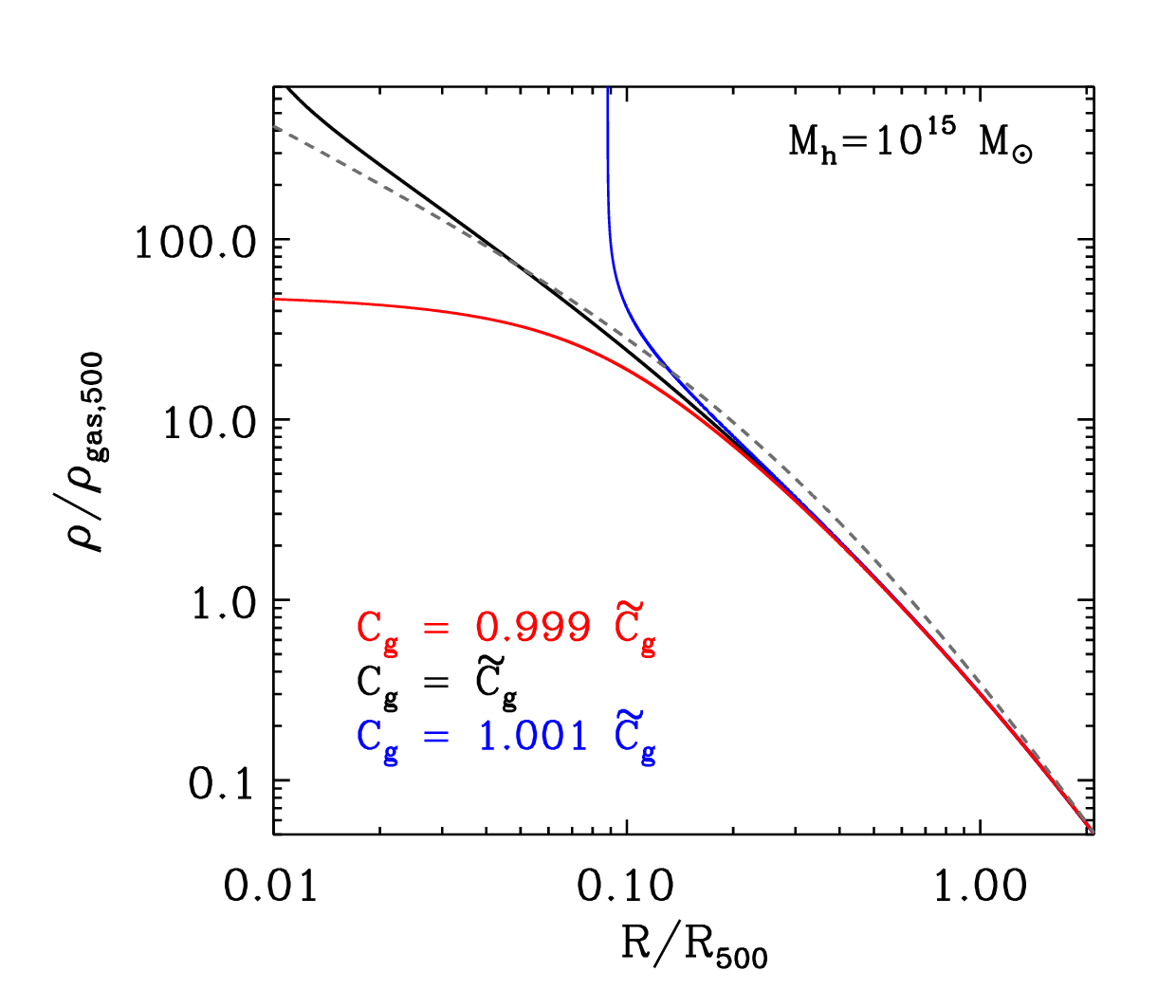}}
	\caption{Gas density profiles derived by solving for hydrostatic equilibrium for 3 different values of $C_{\rm g}$ as detailed in the label.  The dashed gray line shows the distribution of the gas as it is deposited in the gravitational well, without any redistribution.
	}
	\label{fig:ap2:cgs}
\end{figure}
As previously pointed out, the system is solved going inwards. Since gas at smaller radii is shock heated before gas at larger radii, in a sense, we solve the system by going back in time.  In other words, our scheme allows us to derive numerical solutions, which are not necessarily physical ones. This point is better understood by looking at Fig. \ref{fig:ap2:cgs}, where we show 3 different density profiles derived by solving for 3 distinct values of $C_{\rm g}$ that differ from one another by $\sim 0.1\%$.
The gas distribution derived from the $C_{\rm g} = \widetilde{C}_{\rm g}$ case (black solid line) differs by less than 10\% over most of the radial range with respect to the distribution obtained by assuming the gas does not move from the position it is deposited in from the accretion process (gray dashed line).
From Eq. \ref{eq:ap2:dmg} we see that, for $C_{\rm g} > \widetilde{C}_{\rm g}$  gas density will increase more rapidly as we go inwards with respect to the $C_{\rm g} =  \widetilde{C}_{\rm g}$. This effectively leads to gas being relocated to slightly larger radii, such a process cannot continue indefinitely because, sooner or later, we run out of gas. In the case reported here, where the variation in  $C_{\rm g}$ is modest, the run away occurs at relatively small radii, for larger values of $C_{\rm g}$ it is located at larger radii. Clearly this is an unphysical solution and we will not consider it further. 
The  $C_{\rm g} < \widetilde{C}_{\rm g}$ is more interesting, here Eq. \ref{eq:ap2:dmg} leads to a redistribution of the gas from larger to smaller radii. This is a mathematically viable solution, however from a physical point of view things are different. 

The broad agreement observed at large radii between entropy predicted under the assumption of full conversion of kinetic into thermal energy  and the observed one (see Fig. \ref{fig:k_vs_r_obs}) tells us that the redistribution, that follows the initial deposition of the gas and which is required to reach hydrostatic equilibrium, must be quite modest.
This is consistent with  findings based on our analysis, in Fig. \ref{fig:ap2:cgs} we see that, as we move away from  the virial/shock radius, $R_{\rm v} \sim 2.1 R_{500}$, the density derived by imposing hydrostatic equilibrium  remains close to the one derived by simply requiring that the gas remain where it is deposited by the accretion process over a large radial range. 
In summary we may say that, as the halo grows in size, a  redistribution process  sweeps outwards in the wake  of the expanding accretion shock, leaving behind it a gas distribution which is in hydrostatic equilibrium and close to the one derived by assuming the gas remains where it is deposited by the accretion process. 
This line of reasoning strongly suggests that the favored solution is the one with  $C_{\rm g} = \widetilde{C}_{\rm g}$.

We implement this solution in our numerical code through a recursive process that solves  Eqs. \ref{eq:ap2:dp}, \ref{eq:ap2:dmg} and \ref{eq:ap2:dk}  starting from the value of $C_{\rm g}$ derived from Eq. \ref{eq:ap2:cg3}. We then compare the density at  0.03 $R_{500}$ with the one derived assuming gas does not move from where it is deposited by the accretion process and modify $C_{\rm g}$ to reduce the difference. We iterate until the difference in density at  0.03 $R_{500}$ is smaller than 10\%. The search for $\widetilde{C}_{\rm g}$ is carried out through a standard bisection method.

\section{A derivation of the temperature correction}
\label{sec:app3}
In this appendix we derive Eq. \ref{eq:tred} presented in Sect. \ref{sec:sub:mix}.
We start by writing the gas mass falling onto the halo in a time interval $\Delta t$, $\dot{M}_{\rm g} \, \Delta t$, as  the sum of the masses of the infalling sub-halos: 
\begin{equation}
	\dot{M}_{\rm g}(M_{\rm h}^*) \, \Delta t \; = \; \int_{M_{\rm h}^{\rm min}}^{M_{\rm h}^*} \; M_{\rm g}(M_{\rm h}) \; { {\rm d}N\over {\rm d}M_{\rm h}}  \; {\rm d}M_{\rm h} \; ;
	\label{eq:ap3:dmg}
\end{equation} 
where ${\rm d}N / {\rm d}M_{\rm h}$ is the halo mass function, $M_{\rm h}^{\rm min}$ is the minimum mass over which the sum is extended and $M_{\rm h}^*$ is the mass of the halo on which the gas is accreting.
Note that this description of the process holds as long the timescale over which the halo mass function varies is much longer then $\Delta t$, the timescale over which thermalization occurs. 
The gravitational energy converted into thermal energy through shock heating in the same time interval may be written as:
\begin{equation}
	E_{\rm th}(M_{\rm h}^*) \; = \; {\dot{M}_{\rm g}(M_{\rm h}^*) \over \mu m_{\rm p}} \; k_{\rm b}T(M_{\rm h}^*) \; \Delta t  \; .
	\label{eq:ap3:eth1}
\end{equation} 
By substituting Eq. \ref{eq:ap3:dmg} into Eq. \ref{eq:ap3:eth1} we get:
\begin{equation}
	E_{\rm th}(M_{\rm h}^*) \; = \; { k_{\rm b}T(M_{\rm h}^*) \; \over \mu m_{\rm p}} \; \int_{M_{\rm h}^{\rm min}}^{M_{\rm h}^*} \; M_{\rm g}(M_{\rm h}) \; { {\rm d}N\over {\rm d}M_{\rm h}}  \; {\rm d}M_{\rm h}   \; .
	\label{eq:ap3:eth2}
\end{equation}
 
\begin{figure}
	\hspace{-0.5cm}
	\centerline{\includegraphics[angle=0,width=9.2cm]{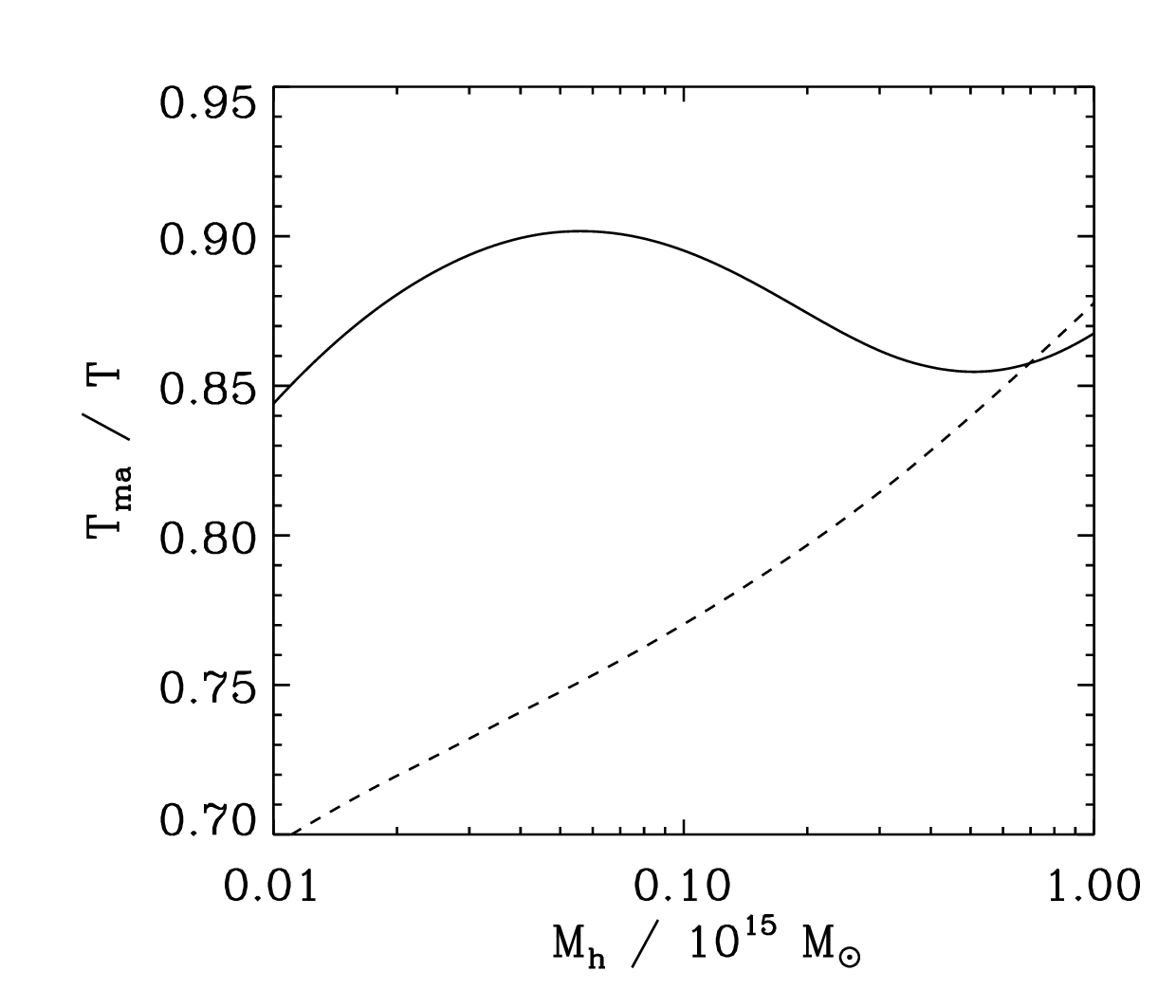}}
	\caption{Mixed accretion temperature, see text for definition, divided by temperature computed assuming all gas is outside  infalling sub-halos. The thick line shows the case of a gas fraction that varies with halo mass as in Fig. \ref{fig:f_gas}. The dashed line shows the case of a gas fraction that does not depend on halo mass.
	}
	\label{fig:ap3:tlos}
\end{figure}

 The binding energy  of the gas in the potential wells of the sub-halos falling onto the main halo in the time interval $\Delta t$, can be expressed as: 
 \begin{equation}
 	\widetilde{E_{\rm b}}(M_{\rm h}^*) \; = \;  \int_{M_{\rm h}^{\rm min}}^{M_{\rm h}^*} \; E_{\rm b}(M_{\rm h}) \; { {\rm d}N\over {\rm d}M_{\rm h}}  \; {\rm d}M_{\rm h}   \; ,
 	\label{eq:ap3:eb1}
 \end{equation}
 where $E_{\rm b}(M_{\rm h})$ is the binding energy of the gas within sub-halos of mass $M_{\rm h}$.
 We recall that accreting gas with no energy at infinity, once settled into hydrostatic equilibrium, has a  binding energy that equals its thermal energy \citep{Binney:1987}. 
 Thus the binding energy of the gas in a sub-halo of mass $M_{\rm h} $ can be expressed as the product of the number of gas particles multiplied by the temperature: 
 \begin{equation}
 E_{\rm b}(M_{\rm h}) \; = \;  {M_{\rm g}(M_{\rm h}) \over \mu m_{\rm p}}  \; k_{\rm b}T(M_{\rm h})  \; .
 	\label{eq:ap3:eb2}
 \end{equation}
 Plugging Eq. \ref{eq:ap3:eb2} into Eq. \ref{eq:ap3:eb1} we get:
  \begin{equation}
 	\widetilde{E_{\rm b}}(M_{\rm h}^*) \; = \;  \int_{M_{\rm h}^{\rm min}}^{M_{\rm h}^*} \; {M_{\rm g}(M_{\rm h}) \over \mu m_{\rm p}}  \; k_{\rm b}T(M_{\rm h}) \; { {\rm d}N\over {\rm d}M_{\rm h}}  \; {\rm d}M_{\rm h}   \; .
 	\label{eq:ap3:eb3}
 \end{equation}
 The next step is to compute the fraction of the energy made available by gravitational collapse that goes into unbinding the gas from sub-halos. This is:
  \begin{equation}
	{\widetilde{E_{\rm b}}(M_{\rm h}^*) \over E_{\rm th}(M_{\rm h}^*)} \; = \; { \displaystyle\int_{M_{\rm h}^{\rm min}}^{M_{\rm h}^*} \; f_{\rm g}(M_{\rm h}) \; M_{\rm h}  \;T(M_{\rm h}) \; { {\rm d}N\over {\rm d}M_{\rm h}}  \; {\rm d}M_{\rm h}  \over  T(M_{\rm h}^*)  \; \displaystyle\int_{M_{\rm h}^{\rm min}}^{M_{\rm h}^*} \; f_{\rm g}(M_{\rm h}) \; M_{\rm h} { {\rm d}N\over {\rm d}M_{\rm h}}  \; {\rm d}M_{\rm h} } \; ,
	\label{eq:ap3:eb_eth}
\end{equation}
 where we have made use of expressions for $\widetilde{E_{\rm b}}(M_{\rm h}^*)$ and $ E_{\rm th}(M_{\rm h}^*)$ provided respectively in Eqs. \ref{eq:ap3:eb3} and \ref{eq:ap3:eth2} and have substituted $M_{\rm g}(M_{\rm h})$ with $f_{\rm g}(M_{\rm h}) \; M_{\rm h}$.
 We rewrite  $T(M_{\rm h})$ as:
  \begin{equation}
      T(M_{\rm h}) \; = \;  T(M_{\rm h,o}) \left({M_{\rm h} \over M_{\rm h,o}}\right)^{2/3}  \;  ,
 	\label{eq:ap3:kt}
 \end{equation}
 where ${M_{\rm h,o}}$ is a reference halo mass  and $T(M_{\rm h,o})$ the value temperature takes on at ${M_{\rm h,o}}$. We  change integration variable from $M_{\rm h}$ to $m_{\rm h} = M_{\rm h}/M_{\rm h,o}$, and with a little algebra get:
   \begin{equation}
 	{\widetilde{E_{\rm b}}(M_{\rm h}^*) \over E_{\rm th}(M_{\rm h}^*)} \; = \; {m_{\rm h}^*}^{-2/3} \;{ \displaystyle\int_{m_{\rm h}^{\rm min}}^{m_{\rm h}^*} \; f_{\rm g}(m_{\rm h}) \; m_{\rm h}^{5/3}  \; { {\rm d}N\over {\rm d}m_{\rm h}}  \; {\rm d}m_{\rm h}  \over  \displaystyle\int_{m_{\rm h}^{\rm min}}^{m_{\rm h}^*} \; f_{\rm g}(m_{\rm h}) \; m_{\rm h} { {\rm d}N\over {\rm d}m_{\rm h}}  \; {\rm d}m_{\rm h} } \; .
 	\label{eq:ap3:eb_eth2}
 \end{equation}
The energy left for the thermalization of the gas is: $1 - {\widetilde{E_{\rm b}}(M_{\rm h}^*) / E_{\rm th}(M_{\rm h}^*)}$. Thus, if we define as mixed accretion temperature, $T_{\rm ma}$,  the gas temperature computed accounting for the energy invested in unbinding gas from infalling sub-halos, we find: 
   \begin{equation}
	  {T_{\rm ma}(M_{\rm h}^*)  \over T(M_{\rm h}^*)}\; = \; 1 - {m_{\rm h}^*}^{-2/3} \;{ \displaystyle\int_{m_{\rm h}^{\rm min}}^{m_{\rm h}^*} \; f_{\rm g}(m_{\rm h}) \; m_{\rm h}^{5/3}  \; { {\rm d}N\over {\rm d}m_{\rm h}}  \; {\rm d}m_{\rm h}  \over  \displaystyle\int_{m_{\rm h}^{\rm min}}^{m_{\rm h}^*} \; f_{\rm g}(m_{\rm h}) \; m_{\rm h} { {\rm d}N\over {\rm d}m_{\rm h}}  \; {\rm d}m_{\rm h} } \; ,
	\label{eq:ap3:tma_tsm}
\end{equation}
where $T(M_{\rm h}^*)$ is the temperature computed assuming all gas is outside  infalling sub-halos.
In Fig. \ref{fig:ap3:tlos}  we show $T_{\rm ma} / T$ as a function of halo mass.

\section{A derivation of the total gas energy}
\label{sec:app4}
In this appendix we derive the expression for the  thermal energy of the gas presented in Eq. \ref{eq:therm_ene}. In general terms, the thermal energy within a given volume, V, can be written as:
  \begin{equation}
     E_{\rm th} \; = \;  \displaystyle\int_{V} h  \; \rho_{\rm g}  \;  {\rm d}V \;  ,
	\label{eq:ap4:eth}
\end{equation}
where $h$, is the specific enthalpy (enthalpy per unit mass) and $\rho_{\rm g}$ is the gas mass density.
Under the standard assumption that gas in the halo can be described as an ideal monoatomic gas:
  \begin{equation}
	h \; = \;  {5 \over 2 } \; {k_{\rm b} T \over \mu m_{\rm p}} . 
	\label{eq:ap4:h}
\end{equation}
By assuming spherical symmetry and expressing the product of temperature with density as pressure we rewrite \ref{eq:ap4:eth} as:
  \begin{equation}
	E_{\rm th}(R) \; = \; 10 \, \pi \; \displaystyle\int_{0}^{R} p_{\rm g} \; r^2 {\rm d}r \;  .
	\label{eq:ap4:eth2}
\end{equation}
Next, we rescale pressure in units of $p_{500}$ and radius in units of $R_{500}$: 
\begin{equation}
	E_{\rm th}(x) = {15 \over 2} \; p_{500} V_{500} \; \int_0^x   \, \text{\cursive{p}} (\mathrm{x})  \, \mathrm{x}^2 {\rm d}\mathrm{x} \; ,
	\label{eq:ap4:eth3}
\end{equation}
where: $p_{500}$ is the characteristic pressure \citep[see][for a definition]{Arnaud:2010}, $V_{500} = 4/3 \pi R_{500}^3$, $x = R/R_{500}$ and $\text{\cursive{p}}(x) = p(x)/p_{500}$. Note also that within the integral we used the symbol $\mathrm{x}$ rather than $x$ to avoid confusion with the upper integration limit.
 
\end{document}